\documentclass{article}
\usepackage{bm,latexsym,amsmath,amssymb,amsfonts,fancyhdr,color,graphicx,multirow}
\usepackage[a4paper,bottom=3cm,top=3.3cm,head=5mm,width=18cm,dvipdfm]{geometry}
\usepackage[usenames,dvipsnames,svgnames,table]{xcolor}
\usepackage[ps2pdf,
            colorlinks=true,
            linkcolor=red,
            urlcolor=blue,
            citecolor=green,          
						bookmarks=true,
						bookmarksnumbered=true,
						breaklinks=true,
						pdfpagemode=Fullscreen,
						pdfstartview=FitBH]{hyperref}
\usepackage[dotinlabels]{titletoc}
\usepackage{titlesec}
\usepackage{authblk}

\numberwithin{equation}{section}
\allowdisplaybreaks[4]

\titlelabel{\thetitle.\quad \hspace{-0.8em}}
\titlecontents{section}
              [1.5em]
              {\vspace{4mm} \large \bf}
              {\contentslabel{1em}}
              {\hspace*{-1em}}
              {\titlerule*[.5pc]{.}\contentspage}
\titlecontents{subsection}
              [3.5em]
              {\vspace{2mm}}
              {\contentslabel{1.8em}}
              {\hspace*{.3em}}
              {\titlerule*[.5pc]{.}\contentspage}
\titlecontents{subsubsection}
              [5.5em]
              {\vspace{2mm}}
              {\contentslabel{2.5em}}
              {\hspace*{.3em}}
              {\titlerule*[.5pc]{.}\contentspage}

\hypersetup{ pdfauthor = {Shao-Feng Ge},
	     pdftitle = {A Novel Approach to Study Atmospheric Neutrino Oscillation}, % Insert title here!!!
	     pdfsubject = {}, % Insert subject here!!!
             pdfkeywords = {}, % Insert keywords here!!!
	     pdfcreator = {LaTeX with hyperref package},
	     pdfproducer = {dvips + ps2pdf} }

\definecolor{gesfpurple}{rgb}{0.47,0.19,0.42}

\definecolor{gesflanse}{rgb}{0.00,0.50,0.50}

\definecolor{gesfblue}{rgb}{0.08,0.42,0.76}

\definecolor{gesfred}{rgb}{1,0,0}

\definecolor{gesfwhite}{rgb}{1,1,1}

\definecolor{gesfblack}{rgb}{0,0,0}

\newcommand{\gsec}[1]{{\hypersetup{linkcolor=blue}Sec.~\ref{#1}\hypersetup{linkcolor=red}}}
\newcommand{\gfig}[1]{{\hypersetup{linkcolor=violet}Fig.~\ref{#1}\hypersetup{linkcolor=red}}}
\newcommand{\gtable}[1]{{\hypersetup{linkcolor=gesflanse}Table~\ref{#1}\hypersetup{linkcolor=red}}}

\newcommand{\gdelta}{{\delta'}}

\begin{document}

\title{\begin{flushright}
       \mbox{\normalsize KEK-TH-1662, arXiv:1309.3176}
       \end{flushright}
			 \vskip 20pt
       \textbf{\huge A Novel Approach to Study Atmospheric Neutrino Oscillation}} % Insert title here!!!
\author[1]{{\large Shao-Feng Ge} \footnote{gesf02@gmail.com}}
\author[2]{{\large Kaoru Hagiwara} \footnote{kaoru.hagiwara@kek.jp}}
\author[3]{{\large Carsten Rott} \footnote{carsten.rott@gmail.com}}
\affil[1]{\small KEK Theory Center, Tsukuba, 305-0801, Japan}
\affil[2]{KEK Theory Center and Sokendai, Tsukuba, 305-0801, Japan}
\affil[3]{Department of Physics, Sungkyunkwan University, Suwon 440-746, Korea}
%\affil[4]{Department of Physics and Center for Cosmology and Astro-Particle Physics, Ohio State University, Columbus, Ohio 43210, USA}
\date{\today}

\maketitle

\begin{abstract}
% Insert abstract here!!!
We develop a general theoretical framework to analytically disentangle the contributions of the neutrino mass hierarchy, the atmospheric mixing angle, and the CP phase, in neutrino oscillations. To illustrate the usefulness of this framework, especially that it can serve as a complementary tool to neutrino oscillogram in the study of atmospheric neutrino oscillations, we take PINGU as an example and compute muon- and electron-like event rates with event cuts on neutrino energy and zenith angle. Under the assumption of exact momentum measurements of neutrinos with a perfect e-$\mu$ identification and no backgrounds, we find that the PINGU experiment has the potential of resolving the neutrino mass hierarchy and the octant degeneracies within 1-year run, while the measurement of the CP phase is significantly more challenging. Our observation merits a serious study of the detector capability of estimating the neutrino momentum for both muon- and electron-like events.
\end{abstract}

%\hypersetup{linkcolor=black}
%\tableofcontents
%\hypersetup{linkcolor=red}

\section{Introduction}

%The parameters below are used as input in all subsequent calculations,
%\begin{subequations}
%\begin{eqnarray}
%&&
%  \sin^2 2 \theta_{\rm s} = 0.857 \pm 0.024 \,, \qquad
%  \sin^2 2 \theta_{\rm a} = 0.98  \pm 0.01 \,, \qquad
%  \sin^2 2 \theta_{\rm r} = 0.089 \pm 0.005 \,, \qquad
%\\
%&&
%  \delta m^2_{\rm s} = 7.6 \pm 0.2 \times 10^{-5} \mbox{eV}^2 \,, \qquad
%  \delta m^2_{\rm a} = 2.35 \pm 0.10 \times 10^{-3} \mbox{eV}^2 \,,
%\end{eqnarray}
%\end{subequations}
%%
%where the notations of mixing angles and mass squared differences have been 
%abbreviated according to their physical meanings: the solar mixing angle 
%$\theta_{\rm s} (\equiv \theta_{12})$, the atmospheric mixing angle 
%$\theta_{\rm a} (\equiv \theta_{23})$, the reactor mixing angle 
%$\theta_{\rm r} (\equiv \theta_{31})$, the mass squared difference effect in 
%solar neutrino oscillation $\delta m^2_{\rm s} (\equiv \delta m^2_{12})$,
%and the one effective in atmospheric neutrino oscillation
%$\delta m^2_{\rm a} (\equiv \delta m^2_{13})$. 

In the last two years, the field of neutrino physics has significantly advanced by constraining the reactor angle $\theta_{13}$. The T2K experiment~\cite{Abe:2011sj} was the first to report a hint of nonzero reactor angle, followed by MINOS~\cite{Adamson:2011qu} and Double CHOOZ~\cite{Abe:2011fz} which added up to a confidence level above 3 sigma.  It was measured accurately by Daya Bay~\cite{DayaBay1} and RENO~\cite{RENO} in March and April 2012, respectively, reaching 7.7~sigma~\cite{DayaBay2} by the October of the same year. 

The relatively large reactor angle opens up opportunities~\cite{Minakata:2012ue} for determining the mass hierarchy, the octant of the atmospheric mixing angle, and the CP phase.  The first could be achieved with a medium baseline reactor experiment~\cite{reactor,ge} and long baseline accelerator experiments~\cite{superbeam,verylong,accelerator,Blennow:2012gj} could measure all three of them. Atmospheric neutrino experiments~\cite{Abe:2011ts}--\cite{Blake:2012gn} could offer alternative ways to accomplish the same.

Recent studies have focused on magnetized detectors, which can distinguish neutrinos from antineutrinos~\cite{Blennow:2012gj,Samanta:2008af,Samanta:2009qw, Barger:2012fx}. Equipped with this capability, a detector of 50--100~Kton ($\sim 10^3$ tons) scale is enough to distinguish the mass hierarchy. Large volume water-Cherenkov or ice-Cherenkov detectors of tens of Mton ($\sim 10^6$ tons) scale could offer an alternative.  DeepCore~\cite{Collaboration:2011ym}, the existing in-fill to IceCube can reach down to energies of $\mathcal O$(10)~GeV and has recently reported the observation of muon neutrino disappearance oscillations~\cite{Aartsen:2013jza} and an electron neutrino flux consistent with expectations~\cite{:2012zk,:2012uu}, demonstrating the capabilities of an low-energy extension. DeepCore has also some sensitivity to neutrinos from the MSW resonance region~\cite{MSW} around $E_\nu \approx 5 \sim 10~\mbox{GeV}$. It can however only partially cover it~\cite{Mena:2008rh,FernandezMartinez:2010am} and to really exploit it a lower threshold detector would be needed.

There has been extensive interest~\cite{Akhmedov:2012ah,Agarwalla:2012uj,Ribordy:2013xea,Winter:2013ema,Blennow:2013vta,Tang11,discussions} recently by the IceCube Collaboration and theoretical community to extend the existing IceCube neutrino telescope~\cite{icecube} with an in-fill array called PINGU (Precision Icecube Next Generation Upgrade)~\cite{Koskinen:2011zz} that could detect neutrino events of $\mathcal O(1)~\mbox{GeV}$.  Such a detector opens up the opportunity of detecting more patterns of the atmospheric neutrino oscillation behavior, which is diluted at higher energy scale, especially due to matter effects~\cite{oscillogram}. A large benefit is the expected high event statistics at low energies. Event rates of $\mathcal O(100,000)$ per year from atmospheric neutrinos allow for measurements with small statistical uncertainty. During the preparation of this draft, a preliminary experimental study~\cite{IceCube:2013aaa} appeared. In Europe a similar detector to PINGU is being considered as part of the Km3NeT project. Our studies can be transferred to this ORCA - (Oscillation Research with Cosmics in the Abyss)~\cite{km3net}.

The expectation of a high statistics sample down to 1~GeV scale makes determining the mixing parameters with atmospheric neutrinos very promising~\cite{Mena:2008rh,FernandezMartinez:2010am, Akhmedov:2012ah,Ribordy:2013xea}.  The paper~\cite{Akhmedov:2012ah,Ribordy:2013xea,Winter:2013ema} adopts oscillograms~\cite{oscillogram} to depict the structure of oscillation resonances~\cite{MSW,parametric,length} when atmospheric neutrinos travel through the Earth. The ability to determine the mass hierarchy, the octant of the atmospheric mixing angle, and the CP phase is studied.  The event numbers and difference between normal hierarchy (NH) and inverted hierarchy (IH) are shown in oscillograms.  In~\cite{Blake:2012gn}, the Bayesian approach is explored in a generic way while the Toy Monte Carlo based on an extended unbinned likelihood ratio test statistic is implemented in~\cite{Franco:2013in}. When combined with accelerator experiments, the sensitivities on the octant of the atmospheric mixing angle~\cite{Abe:2011ts,Barger:2012fx,Akhmedov:2012ah,Chatterjee:2013qus} and the CP phase~\cite{Ghosh:2013yon} can be enhanced.

In \gsec{sec:propagation-basis}, we first develop a general framework of 
decomposing the neutrino oscillation probabilities and the event rates in 
the propagation basis, and apply it to the symmetric Earth matter profile in 
order to analytically disentangle the effects of the neutrino mass hierarchy, 
the atmospheric angle, and the CP phase. In 
\gsec{sec:rates}, we calculate and display the event rates that can be observed at
PINGU. Based on these results, we try to establish the potential 
of atmospheric neutrino measurement at PINGU in \gsec{sec:fit}, while its
dependence on the input values of the neutrino mass hierarchy, the atmospheric
mixing angle and the CP phase can be fully understood in our decomposition 
formalism. Finally, we summarize our conclusions in \gsec{sec:conclusions}. For 
more details about the basic inputs, including the atmospheric neutrino fluxes,
cross sections, effective fiducial volume of PINGU, and the Earth matter profile, 
as well as the numerical methods of evaluating the neutrino oscillation 
probabilities through Earth, please refer to \gsec{sec:setup}

\section{Disentangling Parameters in the Propagation Basis}
\label{sec:propagation-basis}

We first develop a general framework in the propagation basis
\cite{Akhmedov:1998xq,Yokomakura:2002av} for phenomenological study of 
neutrino oscillation.
It can analytically decompose the contributions of the neutrino mass hierarchy, 
the atmospheric mixing angle, and the CP phase. This decomposition
method can serve as a complementary tool to the neutrino oscillogram 
\cite{oscillogram} for the analysis of atmospheric neutrino oscillations, and can
apply generally to other types of neutrino oscillation experiments.

\subsection{Propagation Basis}

In the propagation basis, the atmospheric mixing angle 
$\theta_{23}$~\cite{Akhmedov:1998xq} and the CP phase $\delta$
\cite{Yokomakura:2002av} can be disentangled from the other mixing parameters 
as well as the Earth matter potential. 
This can be seen from the effective Hamiltonian,
\begin{equation}
  \mathcal H
=
  \frac 1 {2 E_\nu}
\left[
  U 
  \left\lgroup
  \begin{matrix}
    0 \\
  & \delta m^2_{\rm s} \\
  & & \delta m^2_{\rm a} 
  \end{matrix}
  \right\rgroup
  U^\dagger 
+
  \left\lgroup
  \begin{matrix}
    a(x) \\
  & 0 \\
  & & 0
  \end{matrix}
  \right\rgroup
\right] \,,
\label{eq:Ha}
\end{equation}
where 
\begin{equation}
  a(x) 
\equiv 2 E_\nu V(x)
= 2 \sqrt 2 E_\nu G_F N_{\rm e}(x) \,,
\end{equation}
represents the matter effect which is proportional to neutrino energy $E_\nu$
and the matter potential $V(x)$. The mass differences are denoted as,
\begin{subequations}
\begin{eqnarray}
&&  \delta m^2_{\rm s} \equiv m^2_2 - m^2_1 \,, \\
&&  \delta m^2_{\rm a} \equiv m^2_3 - m^2_1 \,.
\end{eqnarray}
\end{subequations}
The lepton-flavor mixing matrix $U$ relates the flavor basis 
($\nu_\alpha = \nu_{\rm e}, \nu_\mu, \nu_\tau$) and the mass eigenstates 
($m_{\nu_i} = m_i, i = 1, 2, 3$),
\begin{equation}
  \nu_\alpha 
=
  U_{\alpha i} \nu_i \,,
\end{equation}
and can be parametrized as 
$U \equiv O_{23}(\theta_{\rm a}) P_\delta O_{13}(\theta_{\rm r}) P^\dagger_\delta O_{12}(\theta_{\rm s})$,
\begin{equation}
  U
\equiv
  \left\lgroup
  \begin{matrix}
    1 \\
  & c_{\rm a} & s_{\rm a} \\
  &-s_{\rm a} & c_{\rm a}
  \end{matrix}
  \right\rgroup
  \left\lgroup
  \begin{matrix}
    1 \\
  & 1 \\
  & & e^{i \delta}
  \end{matrix}
  \right\rgroup
  \left\lgroup
  \begin{matrix}
    c_{\rm r} &   & s_{\rm r} \\
        & 1 &     \\
  - s_{\rm r} &   & c_{\rm r}
  \end{matrix}
  \right\rgroup
  \left\lgroup
  \begin{matrix}
    1 \\
  & 1 \\
  & & e^{- i \delta}
  \end{matrix}
  \right\rgroup
  \left\lgroup
  \begin{matrix}
    c_{\rm s} & s_{\rm s} \\
  - s_{\rm s} & c_{\rm s} \\
    & & 1
  \end{matrix}
  \right\rgroup \,,
\label{eq:U}
\end{equation}
where $c_\alpha \equiv \cos \theta_\alpha$ and 
$s_\alpha \equiv \sin \theta_\alpha$. 
The solar, the atmospheric, and the reactor mixing angles are labelled as,
\begin{equation}
  (s,a,r) \equiv (12,23,13) \,,
\end{equation} 
according to how they were measured. For convenience, we denote the three rotation matrices in (\ref{eq:U}) from the left to the right as $O_{23}$, $O_{13}$, and $O_{12}$ respectively. 

The 2--3 mixing matrix $O_{23}$ and the rephasing matrix $P_\delta$ can be 
extracted out as overall matrices~\cite{Yokomakura:2002av},
%With a close look at the Hamiltonian (\ref{eq:Ha}), we would find that the 
%potential is degenerate between $\nu_\mu$ and $\nu_\tau$.
%\begin{equation}
%  \mathcal H
%=
%  \frac 1 {2 E}
%\left[
%  (O_{23} P_\delta) (O_{31} P^\dagger_\delta O_{12})
%  \left\lgroup
%  \begin{matrix}
%    0 \\
%  & \delta m^2_{\rm s} \\
%  & & \delta m^2_{\rm a} 
%  \end{matrix}
%  \right\rgroup
%  (O_{31} P^\dagger_\delta O_{12})^\dagger (O_{23} P_\delta)^\dagger
%+
%  (O_{23} P_\delta)
%  \left\lgroup
%  \begin{matrix}
%    a(x) \\
%  & 0 \\
%  & & 0
%  \end{matrix}
%  \right\rgroup
%  (O_{23} P_\delta)^\dagger
%\right] \,.
%\end{equation} 
%
%As $P^\dagger_\delta$ commutes with $O_{12}$, it's equivalent to place it in the
%inner side of $O_{12}$. Since it can further commute with the diagonal mass 
%matrix, the two inner rephasing matrices $P_\delta$ and $P^\dagger_\delta$ on 
%the both sides can combine can hence cancel with each other,
\begin{equation}
  \mathcal H
=
  \frac 1 {2 E_\nu} (O_{23} P_\delta) 
\left[
  (O_{13} O_{12})
  \left\lgroup
  \begin{matrix}
    0 \\
  & \delta m^2_{\rm s} \\
  & & \delta m^2_{\rm a} 
  \end{matrix}
  \right\rgroup
  (O_{13} O_{12})^\dagger
+
  \left\lgroup
  \begin{matrix}
    a(x) \\
  & 0 \\
  & & 0
  \end{matrix}
  \right\rgroup
\right]
  (O_{23} P_\delta)^\dagger \,.
\label{eq:Hprime}
\end{equation} 
In this way, $O_{23}$ and $P_\delta$ are separated from the neutrino mass
hierarchy, which is encoded in the first term inside the square bracket, as
well as the matter effect, represented by the second term. In other words, 
the atmospheric mixing angle $\theta_{\rm a}$ and the CP phase $\delta$ are 
disentangled from the remaining mixing parameters, analytically. 
This is a significant simplification in the analysis of neutrino oscillation,
especially the atmospheric neutrino oscillation that suffers from complicated 
matter profile, as described in \gsec{sec:PREM}. 

To make it explicit, the original Hamiltonian $\mathcal H$ can be rotated to 
the equivalent $\mathcal H'$ in the propagation basis through a similar
transformation,
\begin{equation}
  \mathcal H'
=
  \frac 1 {2 E_\nu}
\left[
  (O_{13} O_{12})
  \left\lgroup
  \begin{matrix}
    0 \\
  & \delta m^2_{\rm s} \\
  & & \delta m^2_{\rm a} 
  \end{matrix}
  \right\rgroup
  (O_{13} O_{12})^\dagger
+
  \left\lgroup
  \begin{matrix}
    a(x) \\
  & 0 \\
  & & 0
  \end{matrix}
  \right\rgroup
\right]
=
 (O_{23} P_\delta)^\dagger \mathcal H (O_{23} P_\delta) \,.
\label{eq:H'}
\end{equation}
There are only four mixing parameters involved in $\mathcal H'$, the two 
mass squared differences $\delta m^2_{\rm s}$ and $\delta m^2_{\rm a}$, 
the solar mixing angle $\theta_{\rm s}$, and the reactor mixing angle 
$\theta_{\rm r}$. Correspondingly, we can define a {\it propagation basis} 
($\nu'_i$)~\cite{Akhmedov:1998xq,Yokomakura:2002av} that is related to the 
flavor basis ($\nu_\alpha$) and the mass eigenstates ($\nu_i$) as follows:
\begin{eqnarray}
  \nu_\alpha
=
  [O_{23}(\theta_{\rm a}) P_\delta]_{\alpha i} \nu'_i \,.
\end{eqnarray}
The transformed Hamiltonian $\mathcal H'$ is the effective Hamiltonian
defined in the propagation basis. Once the neutrino oscillation amplitudes,
\begin{equation}
  S'_{\rm ij} 
\equiv
  \langle \nu'_j | S' | \nu'_i \rangle
\end{equation}
are calculated with the Hamiltonian $\mathcal H'$ in the propagation basis, the oscillation amplitudes in the flavor basis,
\begin{equation}
  S_{\beta \alpha}
\equiv
  \langle \nu_\beta | S | \nu_\alpha \rangle
\end{equation}
are simply obtained by the unitarity transformation,
\begin{equation}
  S
=
  (O_{23} P_\delta) S' (O_{23} P_\delta)^\dagger
\equiv
  (O_{23} P_\delta) 
  \left\lgroup
  \begin{matrix}
    S'_{11} & S'_{12} & S'_{13} \\
    S'_{21} & S'_{22} & S'_{23} \\
    S'_{31} & S'_{32} & S'_{33}
  \end{matrix}
  \right\rgroup
  (O_{23} P_\delta)^\dagger \,,
\label{eq:SS'}
\end{equation}
This makes the formalism much simpler.

We find the propagation basis very useful in the phenomenological study of neutrino oscillation. It allows us to analytically factor out $\theta_{\rm a}$~\cite{Akhmedov:1998xq} and $\delta$~\cite{Yokomakura:2002av} from the numerical evaluation of the oscillation amplitudes that involves many factors and can be very complicated. The contributions of the still unknown neutrino mass hierarchy, the octant of the atmospheric mixing angle $\theta_{\rm a}$, and the CP phase $\delta$ are now disentangled from each other. A general formalism based on this feature can help to reveal the pictures behind neutrino oscillation phenomena. This is especially important when the three unknown parameters are under close investigations at current and future neutrino experiments.

\subsection{Oscillation Probabilities}

The oscillation probabilities are measured in the flavor basis. It is necessary to explicitly express the flavor basis amplitude matrix $S$ in terms of its counterpart $S'$ in the propagation basis by the unitary transformation with $O_{23} P_\delta$, namely to expand (\ref{eq:SS'}). According to the definition of the mixing matrix in (\ref{eq:U}) , $O_{23} P_\delta$ can be explicitly written as,
\begin{equation}
  O_{23} P_\delta
=
  \left\lgroup
  \begin{matrix}
    1 \\
  & c_{\rm a} & s_{\rm a} e^{i \delta} \\
  & - s_{\rm a} & c_{\rm a} e^{i \delta} 
  \end{matrix}
  \right\rgroup \,,
\qquad
  (O_{23} P_\delta)^\dagger
=
  \left\lgroup
  \begin{matrix}
    1 \\
  & c_{\rm a} & - s_{\rm a} \\
  & s_{\rm a} e^{- i \delta} &   c_{\rm a} e^{- i \delta} 
  \end{matrix}
  \right\rgroup \,.
\label{eq:similarT}
\end{equation}
The mixing from the propagation to the flavor basis occurs between the second and the third indices. We can expect the first element of $S'$ to be unaffected when (\ref{eq:similarT}) is combined with (\ref{eq:SS'})~\cite{Yokomakura:2002av},
\begin{subequations}
\begin{eqnarray}
  S_{\rm e e}
& = &
  S'_{11} \,,
\\
  S_{\rm e \mu}
& = &
  c_{\rm a} S'_{12} + s_{\rm a} e^{- i \delta} S'_{13} \,, 
\\
  S_{\mu \rm e}
& = &
  c_{\rm a} S'_{21} + s_{\rm a} e^{+ i \delta} S'_{31} \,,
\\
  S_{\mu \mu}
& = &
  c^2_{\rm a} S'_{22} + c_{\rm a} s_{\rm a} (e^{- i \delta} S'_{23} + e^{+ i \delta} S'_{32}) + s^2_{\rm a} S'_{33} \,.
\label{eq:Smumu}
\end{eqnarray}
\label{eq:Sij}
\end{subequations}
\hspace{-3mm}
Note that only the elements among $e$ and $\mu$ flavors are shown since they are sufficient to derive all the flavor basis oscillation probabilities,
\begin{equation}
  P_{\alpha \beta}
\equiv
  P(\nu_\alpha \rightarrow \nu_\beta)
=
  | \langle \nu_\beta | S | \nu_\alpha \rangle |^2
=
  |S_{\beta \alpha}|^2 \,,
\end{equation}
\hspace{-2mm}
from $\nu_{\rm e}$ and $\nu_\mu$ (as well as from $\bar \nu_{\rm e}$ and $\bar \nu_\mu$, as shown below). Explicitly we find,
\begin{subequations}
\begin{eqnarray}
  P_{\rm ee}
\equiv
  |S_{\rm ee}|^2
\label{eq:Pee}
& = &
  |S'_{11}|^2 \,,
\\
  P_{\rm e \mu}
\equiv
  |S_{\mu \rm e}|^2
& = &
  c^2_{\rm a} |S'_{12}|^2
+ s^2_{\rm a} |S'_{13}|^2
+ 2 c_{\rm a} s_{\rm a} (\cos \delta \mathbb R + \sin \delta \mathbb I) (S'_{12} S'^*_{13}) \,,
\\
  P_{\mu \rm e}
\equiv
  |S_{\rm e \mu}|^2
& = &
  c^2_{\rm a} |S'_{21}|^2 + s^2_{\rm a} |S'_{31}|^2
+ 2 c_{\rm a} s_{\rm a} (\cos \delta \mathbb R - \sin \delta \mathbb I) (S'_{21} S'^*_{31}) \,,
\\
  P_{\mu \mu}
\equiv
  |S_{\mu \mu}|^2
& = &
  c^4_{\rm a} |S'_{22}|^2 + s^4_{\rm a} |S'_{33}|^2 + 2 c^2_{\rm a} s^2_{\rm a} \mathbb R (S'_{22} S'^*_{33})
\nonumber
\\
& + &
  c^2_{\rm a} s^2_{\rm a} \left[ |S'_{23}|^2 + 2 (\cos 2 \delta \mathbb R + \sin 2 \delta \mathbb I) (S'_{23} S'^*_{32}) + |S'_{32}|^2 \right]
\nonumber
\\
& + &
  2 c_{\rm a} s_{\rm a} \cos \delta \mathbb R [(c^2_{\rm a} S'_{22} + s^2_{\rm a} S'_{33}) (S'_{23} + S'_{32})^*]
\nonumber
\\
& + &
  2 c_{\rm a} s_{\rm a} \sin \delta \mathbb I [(c^2_{\rm a} S'_{22} + s^2_{\rm a} S'_{33}) (S'_{32} - S'_{23})^*] \,,
\end{eqnarray}
\label{eq:Pij}
\end{subequations}
\hspace{-2.5mm}
where $\mathbb R$ and $\mathbb I$ gives the real and imaginary parts, respectively. The dependence on the atmospheric mixing angle $\theta_{\rm a}$ and the CP phase $\delta$ can be clearly seen in the above expressions. The transition probability into $\nu_\tau$ are then obtained by unitarity conditions,
\begin{subequations}
\begin{eqnarray}
  P_{\rm e \tau}
& = &
  1 - P_{\rm ee} - P_{\rm e \mu} \,,
\\
  P_{\mu \tau}
& = &
  1 - P_{\mu \rm e} - P_{\mu \mu} \,,
\end{eqnarray}
\end{subequations}
\hspace{-2.5mm}
while we neglect contributions from tiny components of $\nu_\tau$ and $\bar \nu_\tau$ flux in the atmospheric neutrinos~\cite{Athar:2012it}.

The oscillation probabilities for antineutrinos are then obtained simply as,
\begin{equation}
  \overline P_{\alpha \beta}
\equiv
  P(\bar \nu_\alpha \rightarrow \bar \nu_\beta)
=
  P_{\alpha \beta} (a(x) \rightarrow - a(x), \delta \rightarrow - \delta) \,,
\label{eq:Pbar}
\end{equation}
\hspace{-2mm}
by reversing the sign of the matter potential in the Hamiltonian (\ref{eq:Ha}) and the CP phase $\delta$ in the neutrino mixing matrix (\ref{eq:U}), which is identical to the parametrization adopted in Review of Particle Physics~\cite{PDG2012}.

\subsection{Simplifications with Symmetric Matter Profile}

The expressions in (\ref{eq:Pij}) can be significantly simplified in the approximation of the symmetric or reversible matter profile along the baseline, such as those of atmospheric neutrinos in the earth whose matter profile is approximately spherically symmetric as in PREM~\cite{Dziewonski:1981xy} adopted in our study. It has been known that~\cite{Akhmedov:2001kd} the oscillation amplitude matrix after experiencing a reversible matter profile is symmetric in the absence of CP violation. This is indeed the case for the oscillation amplitudes through the Earth in the propagation basis, giving,
\begin{equation}
  S'_{\rm ij} = S'_{\rm ji} \,.
\end{equation}

Based on the above observation, the atmospheric neutrino oscillation amplitudes (\ref{eq:Sij}) can be further simplified,
\begin{subequations}
\begin{eqnarray}
  S_{\rm ee}
& = &
  S'_{11} \,,
\\
  S_{\rm e \mu}
& = &
  c_{\rm a} S'_{12} + s_{\rm a} e^{- i \delta} S'_{13} \,, 
\\
  S_{\mu \rm e}
& = &
  c_{\rm a} S'_{12} + s_{\rm a} e^{+ i \delta} S'_{13} \,,
\\
  S_{\mu \mu}
& = &
  c^2_{\rm a} S'_{22} + s^2_{\rm a} S'_{33} + 2 c_{\rm a} s_{\rm a} \cos \delta S'_{23} \,.
\label{eq:Smumu2}
\end{eqnarray}
\end{subequations}
As a convention, we adopt those elements $S'_{\rm ij}$ with $i \leq j$. It is now manifest that the flavor oscillation amplitudes $S_{\rm e \mu}$ and $S_{\mu \rm e}$ differ only by the CP phase and the expression for $S_{\mu \mu}$ (\ref{eq:Smumu}) is greatly simplified in (\ref{eq:Smumu2}). The oscillation probabilities now read,
\begin{subequations}
\begin{eqnarray}
&&
  P_{\rm ee}
\equiv
  |S_{\rm ee}|^2
=
  |S'_{11}|^2 \,,
\\
&&
  P_{\rm e\mu}
\equiv
  |S_{\mu \rm e}|^2
=
  c^2_{\rm a} |S'_{12}|^2
+ s^2_{\rm a} |S'_{13}|^2
+ 2 c_{\rm a} s_{\rm a} (\cos \delta \mathbb R + \sin \delta \mathbb I) (S'_{12} S'^*_{13}) \,,
\\
&&
  P_{\mu \rm e}
\equiv
  |S_{\rm e \mu}|^2
=
  c^2_{\rm a} |S'_{12}|^2 + s^2_{\rm a} |S'_{13}|^2
+ 2 c_{\rm a} s_{\rm a} (\cos \delta \mathbb R - \sin \delta \mathbb I) (S'_{12} S'^*_{13}) \,,
\\
&&
  P_{\mu\mu}
\equiv
  |S_{\mu \mu}|^2
=
  |c^2_{\rm a} S'_{22} + s^2_{\rm a} S'_{33}|^2
+ 4 c^2_{\rm a} s^2_{\rm a} \cos^2 \delta |S'_{23}|^2
+ 4 c_{\rm a} s_{\rm a} \cos \delta \mathbb R [(c^2_{\rm a} S'_{22} + s^2_{\rm a} S'_{33})S'^*_{23}] \,.
\end{eqnarray}
\label{eq:Pij-2}
\end{subequations}

Throughout our studies in this report we adopt the expression (\ref{eq:Pij-2}) for computing the oscillation probabilities in our numerical calculation, which are exact in the limit of the symmetric earth matter profile PREM~\cite{Dziewonski:1981xy} and neglecting the depth of the detector beneath the earth surface as compared to the baseline lengths. The oscillation probabilities for antineutrinos $\overline P_{\alpha \beta}$ are then computed as in (\ref{eq:Pbar}).

%\subsection{Asymmetric Contribution of Atmospheric Angle $\theta_{\rm a}$}
%\label{sec:expansion}

\subsection{Expansion of Oscillation Probabilities with respect to $x_{\rm a} = \cos 2 \theta_{\rm a}$ and $\delta m^2_{\rm s}$}

Although the expressions (\ref{eq:Pij-2}) for the oscillation probabilities $P_{\alpha \beta}$, and $\overline P_{\alpha \beta}$ via (\ref{eq:Pbar}), are simple enough to perform numerical analysis efficiently, we can obtain further insight by keeping only the leading terms of the following two small parameter of the three neutrino model,
\begin{subequations}
\begin{eqnarray}
  x_{\rm a}
& \equiv &
  \cos 2 \theta_{\rm a}
=
  \sqrt{1 - \sin^2 2 \theta_{\rm a}}
=
  0.21^{+0.06}_{-0.10} \,,
\label{eq:xa}
\\
  \frac {\delta m^2_{\rm s}}{|\delta m^2_{\rm a}|}
& = &
  0.032 \pm 0.002\,,
\label{eq:massratio}
\end{eqnarray}
\end{subequations}
whose numerical values are constrained from the data~\cite{PDG2012,Machado:2011ar,MINOS}, as summarized below in (\ref{eq:inputs}).

First, by expanding $c_{\rm a}$ and $s_{\rm a}$ in terms of $x_{\rm a}$:
\begin{equation}
  c^2_{\rm a}
=
  \frac 1 2 (1 + x_{\rm a}) \,,
\qquad
  s^2_{\rm a}
=
  \frac 1 2 (1 - x_{\rm a}) \,,
\qquad
  c^2_{\rm a} s^2_{\rm a}
=
  \frac 1 4 (1 - x^2_{\rm a}) \,,
\label{eq:expansion}
\end{equation}
the oscillation probabilities $P_{\alpha \beta}$ (\ref{eq:Pij-2}) are expanded as,
\begin{subequations}
\begin{eqnarray}
&&
  P_{\rm ee}
=
  |S'_{11}|^2 \,,
\\
&&
  P_{\rm e \mu}
=
  \frac 1 2 \left( 1 - |S'_{11}|^2 \right)
+ \frac {x_{\rm a}} 2 (|S'_{12}|^2 - |S'_{13}|^2)
+ (\cos \gdelta \mathbb R + \sin \gdelta \mathbb I) (S'_{12} S'^*_{13})
+ \mathcal O(x^4_{\rm a}) \,,
\\
&&
  P_{\mu \rm e}
=
  \frac 1 2 \left( 1 - |S'_{11}|^2 \right)
+ \frac {x_{\rm a}} 2 (|S'_{12}|^2 - |S'_{13}|^2)
+ (\cos \gdelta \mathbb R - \sin \gdelta \mathbb I) (S'_{12} S'^*_{13})
+ \mathcal O(x^4_{\rm a}) \,,
\\
&&
  P_{\mu\mu}
=
  \frac 1 4 |S'_{22} + S'_{33}|^2
+ \frac {x_{\rm a}} 2 (|S'_{22}|^2 - |S'_{33}|^2)
+ \cos \gdelta \mathbb R [(S'_{22} + S'_{33})S'^*_{23}]
\nonumber
\\
&&
\phantom{P_{\mu\mu}}
+ x_{\rm a} \cos \gdelta \mathbb R [S'_{23} (S'_{22} - S'_{33})^*]
+ \frac 1 4 |S'_{22} - S'_{33}|^2 x^2_{\rm a}
+ \cos^2 \gdelta |S'_{23}|^2
+ \mathcal O(x^4_{\rm a}) \,.
\label{eq:Pmumu-3}
\end{eqnarray}
\label{eq:Pij-3}
\end{subequations}
\hspace{-2.5mm} We can clearly identify the linear terms of $x_{\rm a}$ in $P_{\rm e \mu}$ and $P_{\mu \rm e}$, which are identical, and also in $P_{\mu \mu}$. In the above expansion, we keep the terms of order $x^2_{\rm a}$, which turn out to have significant impacts in the measurement of $x_a$ despite the smallness of $x^2_{\rm a} \lesssim 0.05$ at 90\% confidence level. Furthermore, we introduce a short-hand notation,
%the expansion of $x_{\rm a}$ up to $\mathcal O(x^3_{\rm a})$, as it may happen that both $|S'_{22} + S'_{33}|^2$ and $|S'_{22}|^2 - |S'_{33}|^2$ vanish simultaneously with $S'_{22} = - S'_{33}$. Under this circumstance, the $x^2_{\rm a}$ term would dominate. It is interesting to note that $\cos \delta$ and $\sin \delta$ are modulated by a common factor $1-x^2_{\rm a}/2$ originating from the prefactor $\sqrt{2 c_{\rm a} s_{\rm a}}$ always associated with $\cos \delta$ and $\sin \delta$ back in (\ref{eq:Pij-2}). The effect of CP phase will be suppressed by a large deviation of the atmospheric mixing angle from its maximal value $\theta_{\rm a} = 45^\circ$. For convenience, we define
\begin{equation}
  \cos \gdelta 
\equiv 
  2 c_{\rm a} s_{\rm a} \cos \delta 
\approx
  \sqrt{ 1 - x^2_{\rm a} } \cos \delta \,,
\qquad  
  \sin \gdelta 
\equiv
  2 c_{\rm a} s_{\rm a} \sin \delta
\approx
  \sqrt{ 1 - x^2_{\rm a} } \sin \delta \,.
\end{equation}
in (\ref{eq:Pij-3}), without expanding the factor $\sqrt{1 - x^2_{\rm a}}$, since all the $\delta$-dependence in the transition probabilities (\ref{eq:Pij-2}) are functions of $2 c_{\rm a} s_{\rm a} \cos \delta$ and $2 c_{\rm a} s_{\rm a} \sin \delta$. The uncertainty of the $\delta$-measurement should be modulated by the factor $1/\sqrt{1 - x^2_{\rm a}}$.

We find it quite useful to express the oscillation probabilities $P_{\alpha \beta}$ in (\ref{eq:Pij-3}) and the corresponding antineutrino oscillation probabilities $\overline P_{\alpha \beta}$ as,
\begin{subequations}
\begin{eqnarray}
  P_{\alpha \beta}
& \equiv &
  P^{(0)}_{\alpha \beta}
+ P^{(1)}_{\alpha \beta} x_{\rm a}
+ P^{(2)}_{\alpha \beta} \cos \gdelta
+ P^{(3)}_{\alpha \beta} \sin \gdelta
+ P^{(4)}_{\alpha \beta} x_{\rm a} \cos \gdelta
+ P^{(5)}_{\alpha \beta} x^2_{\rm a}
+ P^{(6)}_{\alpha \beta} \cos^2 \gdelta \,,
\\
  \overline P_{\alpha \beta}
& \equiv &
  \overline P^{(0)}_{\alpha \beta}
+ \overline P^{(1)}_{\alpha \beta} x_{\rm a}
+ \overline P^{(2)}_{\alpha \beta} \cos \gdelta
+ \overline P^{(3)}_{\alpha \beta} \sin \gdelta
+ \overline P^{(4)}_{\alpha \beta} x_{\rm a} \cos \gdelta
+ \overline P^{(5)}_{\alpha \beta} x^2_{\rm a}
+ \overline P^{(6)}_{\alpha \beta} \cos^2 \gdelta \,,
\label{eq:barPij}
\end{eqnarray}
\label{eq:Pij-decomposition}
\end{subequations}
\hspace{-3mm} where $P^{(0)}_{\alpha \beta}$ and $\overline P^{(0)}_{\alpha \beta}$ are the leading terms, while $P^{(k)}_{\alpha \beta}$ and $\overline P^{(k)}_{\alpha \beta}$ with $k = 1, \cdots, 6$ are the coefficients of corresponding terms linear in $x_{\rm a}$, $\cos \gdelta$, $\sin \gdelta$, $x_{\rm a} \cos \gdelta$, $x^2_{\rm a}$, and $\cos^2 \gdelta$, respectively. The magnitude of these coefficients determines the experimental sensitivity of measuring the two mixing parameters, $x_{\rm a}$ and $\delta$. The coefficients $P^{(k)}_{\alpha \beta}$ of (\ref{eq:Pij-3}) are shown in the following table.
\begin{equation}
\begin{tabular}{c|cccc}
 & $P^{(k)}_{\rm ee}$ & $P^{(k)}_{\rm e\mu}$ & $P^{(k)}_{\mu \rm e}$ & $P^{(k)}_{\mu \mu}$ \\[1mm]
\hline
(0) & $|S'_{11}|^2$ & $\frac 1 2 (1 - |S'_{11}|^2)$ & $\frac 1 2 (1 - |S'_{11}|^2)$ & $\frac 1 4 |S'_{22} + S'_{33}|^2$ \\[1mm]
(1) & 0 & $\frac 1 2 (|S'_{12}|^2 - |S'_{13}|^2)$ & $\frac 1 2 (|S'_{12}|^2 - |S'_{13}|^2)$ & $\frac 1 2 (|S'_{22}|^2 - |S'_{33}|^2)$ \\[1mm]
(2) & 0 & $\mathbb R(S'_{12} S'^*_{13})$ & $\mathbb R(S'_{12} S'^*_{13})$ & $\mathbb R[S'_{23}(S'_{22} + S'_{33})^*]$ \\[1mm]
(3) & 0 & $\mathbb I(S'_{12} S'^*_{13})$ & $-\mathbb I(S'_{12} S'^*_{13})$ & 0 \\[1mm]
(4) & 0 & 0 & 0 & $\mathbb R[S'_{23}(S'_{22} - S'_{33})^*]$ \\[1mm]
(5) & 0 & 0 & 0 & $\frac 1 4 |S'_{22} - S'_{33}|^2$ \\[1mm]
(6) & 0 & 0 & 0 & $|S'_{23}|^2$ 
\end{tabular}
\label{eq:Ps}
\end{equation}
\hspace{-2mm} %The coefficients $\overline P^{(k)}_{\alpha \beta}$ for the antineutrino oscillations are obtained simply by replacing the propagation basis amplitudes $S'_{\rm ij}$ by $\overline S'_{\rm ij}$ ($a(x) \rightarrow - a(x)$), and by reversing the sign of the coefficients of $\sin \delta$, that is $\overline P^{(3)}_{\alpha \beta}$ for $(\alpha \beta) = (e \mu)$ and $(\mu e)$ in (\ref{eq:Ps}).
It is clearly seen from (\ref{eq:Ps}) that $P_{\rm e e} = P(\nu_{\rm e} \rightarrow \nu_{\rm e})$ has no dependence on $\theta_{\rm a}$ and $\delta$, all the other oscillation probabilities have terms $P^{(1)}_{\alpha \beta}$ linear in $x_{\rm a}$, the coefficients of $\cos \gdelta$ are the same for $P_{\rm e \mu}$ and $P_{\mu \rm e}$, those of $\sin \gdelta$ have the same magnitude but the opposite sign between $P_{\rm e \mu}$ and $P_{\mu \rm e}$, while $P_{\mu \mu}$ has no dependence on $\sin \gdelta$. Most of these properties of the oscillation probabilities are expected from theoretical considerations, while they are made explicit in (\ref{eq:Pij-decomposition}) and (\ref{eq:Ps}). 
The corresponding coefficients for the antineutrino oscillations, $\overline P^{(k)}_{\alpha \beta}$ in (\ref{eq:barPij}) are obtained from $P^{(1)}_{\alpha \beta}$ in (\ref{eq:Ps}) as follows:
\begin{subequations}
\begin{eqnarray}
  \overline P^{(k)}_{\alpha \beta} 
& = &
  P^{(k)}_{\alpha \beta} (S'_{\rm ij} \rightarrow \overline S'_{\rm ij}) 
\qquad 
  \mbox{for } k = 0,1,2,4,5,6 \,,
\\
  \overline P^{(3)}_{\alpha \beta}
& = &
- P^{(3)}_{\alpha \beta} (S'_{\rm ij} \rightarrow \overline S'_{\rm ij}) \,,
\end{eqnarray}
\label{eq:nu-antinu}
\end{subequations}
\hspace{-2mm} where $\overline S'_{\rm ij}$ are the oscillation amplitudes in the propagation basis which are obtained from $S'_{\rm ij}$ by reversing the sign of the matter potential $a(x)$:
\begin{equation}
  \overline S'_{\rm ij} = S'_{\rm ij} (a(x) \rightarrow - a(x)) \,.
\end{equation}
The relation (\ref{eq:Pbar}) between the $\nu$ and $\bar \nu$ oscillation probabilities, $P_{\alpha \beta}$ and $\overline P_{\alpha \beta}$, respectively, is simplified significantly in the propagation basis where the matter dependence and the $\delta$ dependence of the oscillation amplitudes are factorized.

The parameter dependences of the oscillation probabilities $P_{\alpha \beta}$ and $\overline P_{\alpha \beta}$ are further simplified significantly when we take account of the smallness of the mass squared difference $\delta m^2_{\rm s}$ as compared to $|\delta m^2_{\rm a}|$, (\ref{eq:massratio}). We note in the propagation basis Hamiltonian (\ref{eq:H'}) that if we set $\delta m^2_{\rm s} \equiv \delta m^2_{12} = 0$, then the oscillation occurs only between $\nu'_1$ and $\nu'_3$, and hence the transitions between $\nu'_1$ and $\nu'_2$, and those between $\nu'_2$ and $\nu'_3$ should be suppressed,
\begin{equation}
  |S'_{12}|, \quad |S'_{23}| 
=
  \mathcal O \left( \frac {\delta m^2_{\rm s}}{\delta m^2_{\rm a}} \right) \,,
\label{eq:order}
\end{equation}
in the propagation basis. In our numerical study of the atmospheric neutrino oscillations in the energy range $2~\mbox{GeV} < E_\nu < 20~\mbox{GeV}$, we find $|S'_{12}| < 0.15$, and $|S'_{23}| < 0.06$. We therefore obtain the following approximation by dropping all the terms of order $(\delta m^2_{\rm s} / \delta m^2_{\rm a})^2$:
\begin{subequations}
\begin{eqnarray}
&&
  P_{\rm e e}
=
  |S'_{11}|^2 \,,
\\
&&
  P_{\rm e \mu}
=
  \frac {1 - x_{\rm a}} 2 (1 - |S'_{11}|^2)
+ (\cos \gdelta \mathbb R + \sin \gdelta \mathbb I) (S'_{12} S'^*_{13})
+ \mathcal O \left( x^4_{\rm a}, \left( \frac {\delta m^2_{\rm s}}{\delta m^2_{\rm a}} \right)^2 \right) \,,
\\
&&
  P_{\mu \rm e}
=
  \frac {1 - x_{\rm a}} 2 (1 - |S'_{11}|^2)
+ (\cos \gdelta \mathbb R - \sin \gdelta \mathbb I) (S'_{12} S'^*_{13})
+ \mathcal O \left( x^4_{\rm a}, \left( \frac {\delta m^2_{\rm s}}{\delta m^2_{\rm a}} \right)^2 \right) \,,
\\
&&
  P_{\mu \mu}
=
  \frac 1 4 |S'_{22} + S'_{33}|^2
+ \frac {x_{\rm a}} 2 (1 - |S'_{11}|^2)
+ \frac 1 4 x^2_{\rm a} |S'_{22} - S'_{33}|^2
\nonumber
\\
&&
\phantom{P_{\mu \mu}}
- \cos \gdelta \mathbb R(S'_{12} S'^*_{13})
+ x_{\rm a} \cos \gdelta \mathbb R [S'_{23} (S'_{22} - S'_{33})^*] 
+ \mathcal O \left( x^4_{\rm a}, \left( \frac {\delta m^2_{\rm s}}{\delta m^2_{\rm a}} \right)^2 \right) \,, \qquad \qquad
\end{eqnarray}
\label{eq:Pij-simplified}
\end{subequations}
and (\ref{eq:Ps}) is further simplified as follows,
\begin{equation}
\begin{tabular}{c|cccc}
 & $P^{(k)}_{\rm ee}$ & $P^{(k)}_{\rm e\mu}$ & $P^{(k)}_{\mu \rm e}$ & $P^{(k)}_{\mu \mu}$ \\[1mm]
\hline
(0) & $|S'_{11}|^2$ & $\frac 1 2 (1 - |S'_{11}|^2)$ & $\frac 1 2 (1 - |S'_{11}|^2)$ & $\frac 1 4 |S'_{22} + S'_{33}|^2$ \\[1mm]
(1) & 0 &-$\frac 1 2 (1 - |S'_{11}|^2)$ &-$\frac 1 2 (1 - |S'_{11}|^2)$ & $\frac 1 2 (1 - |S'_{11}|^2)$ \\[1mm]
(2) & 0 & $\mathbb R(S'_{12} S'^*_{13})$ & $\mathbb R(S'_{12} S'^*_{13})$ & $-\mathbb R(S'_{12} S'^*_{13})$ \\[1mm]
(3) & 0 & $\mathbb I(S'_{12} S'^*_{13})$ & $-\mathbb I(S'_{12} S'^*_{13})$ & 0 \\[1mm]
(4) & 0 & 0 & 0 & $\mathbb R[S'_{23}(S'_{22} - S'_{33})^*]$ \\[1mm]
(5) & 0 & 0 & 0 & $\frac 1 4 |S'_{22} - S'_{33}|^2$ \\[1mm] 
(6) & 0 & 0 & 0 & 0 
\end{tabular}
\label{eq:Ps2}
\end{equation}
In this approximation, there are only 6 independent oscillation factors in the propagation basis, of which $|S'_{11}|^2$ determines the overall rates $P^{(0)}_{\rm ee}$, $P^{(0)}_{\rm e \mu}$, $P^{(0)}_{\mu \rm e}$, as well as all the coefficients of $x_{\rm a}$, $P^{(1)}_{\rm e \mu}$, $P^{(1)}_{\mu \rm e}$, and $P^{(1)}_{\mu \mu}$. The overall rate $P^{(0)}_{\mu \mu}$ is governed by $|S'_{22} + S'_{23}|^2$. The coefficients of all the $\sin \gdelta$ terms are $P^{(3)}_{\rm e \mu} = - P^{(3)}_{\mu \rm e} = \mathbb I(S'_{12} S'^*_{13})$, and the coefficients of the $\cos \gdelta$ terms are $P^{(2)}_{\rm e \mu} = P^{(2)}_{\mu \rm e} = - P^{(2)}_{\mu \mu} = \mathbb R (S'_{12} S'^*_{13})$. The remaining cross term $x_{\rm a} \cos \gdelta$ is governed by $\mathbb R [S'_{23} (S'_{22} - S'_{33})^*]$ in $P^{(4)}_{\mu \mu}$. For the $x^2_{\rm a}$ term, its coefficient $|S'_{22} - S'_{33}|^2$ is also independent. The coefficient $P^{(6)}_{\mu \mu}$ of $\cos^2 \gdelta$ in (\ref{eq:Pij-decomposition}) and (\ref{eq:Ps}) is $(\delta m^2_s / \delta m^2_a)^2$ order, and hence is dropped in (\ref{eq:Pij-simplified}) and (\ref{eq:Ps2}). 

\begin{figure}[h!]
\centering
\vspace{1mm}
\includegraphics[height=8.8cm,width=7cm,angle=-90]{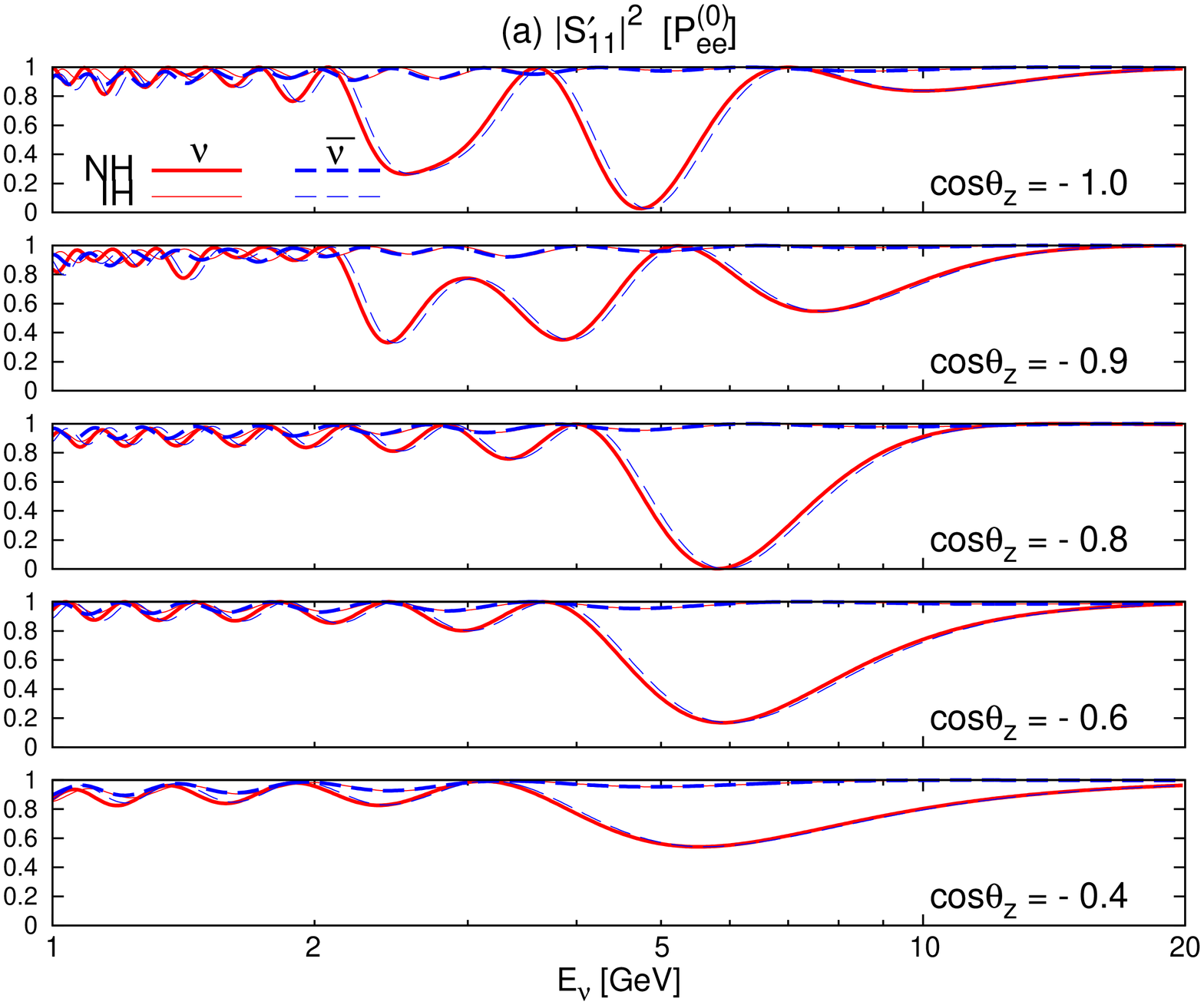}
\includegraphics[height=8.8cm,width=7cm,angle=-90]{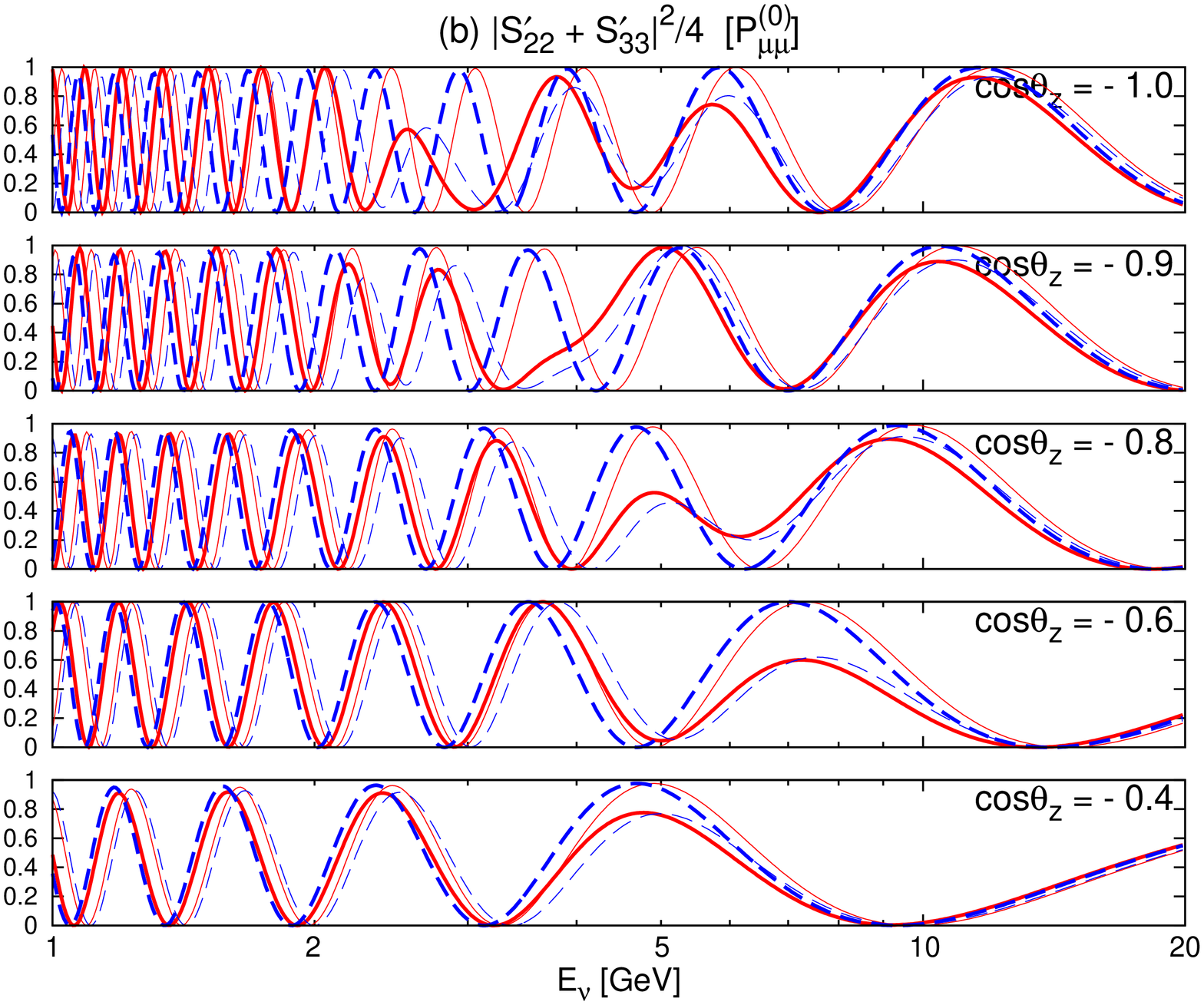}
\vspace{3mm}
\caption{(a) The oscillation probabilities $P^{(0)}_{\rm ee} = |S'_{11}|^2$ [solid lines], and $\overline P^{(0)}_{\rm ee} = |\overline S'_{11}|^2$ [dashed lines] plotted against the $\nu$ and $\bar \nu$ energies, respectively, for the zenith angles $\cos \theta_{\rm z} = -1, -0.9, -0.8, -0.6, -0.4$. The thick curves are for NH, while the thin curves are for IH. (b) The same as (a), but for the coefficient $P^{(0)}_{\mu \mu}$ ($\nu$) and $\overline P^{(0)}_{\mu \mu}$ ($\bar \nu$).}
%\caption{Zeroth-order coefficients of oscillation probabilities 
%         $|S'_{11}|^2$ for $P^{(0)}_{\rm ee}$ [Left] and
%         $|S'_{22} + S'_{33}|^2/4$ for $P^{(0)}_{\mu \mu}$ [Right].
%         Both NH [thick] and IH [thin] are plotted for neutrino [red solid]
%         and antineutrino [blue dash].}
\label{fig:S0}
\end{figure}

\begin{figure}[h]
\vspace{3mm}
\centering
\includegraphics[height=8.8cm,width=7cm,angle=-90]{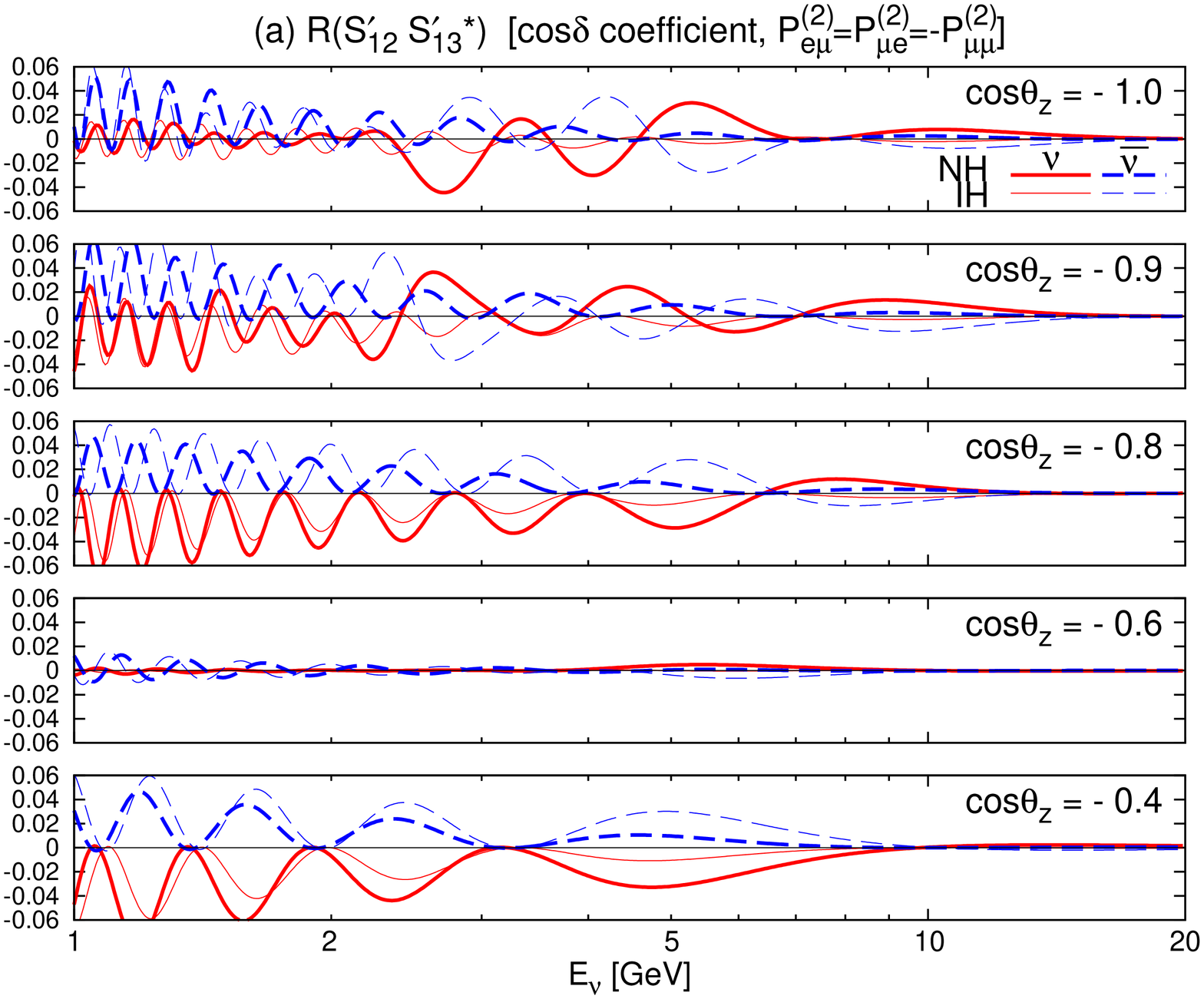}
\includegraphics[height=8.8cm,width=7cm,angle=-90]{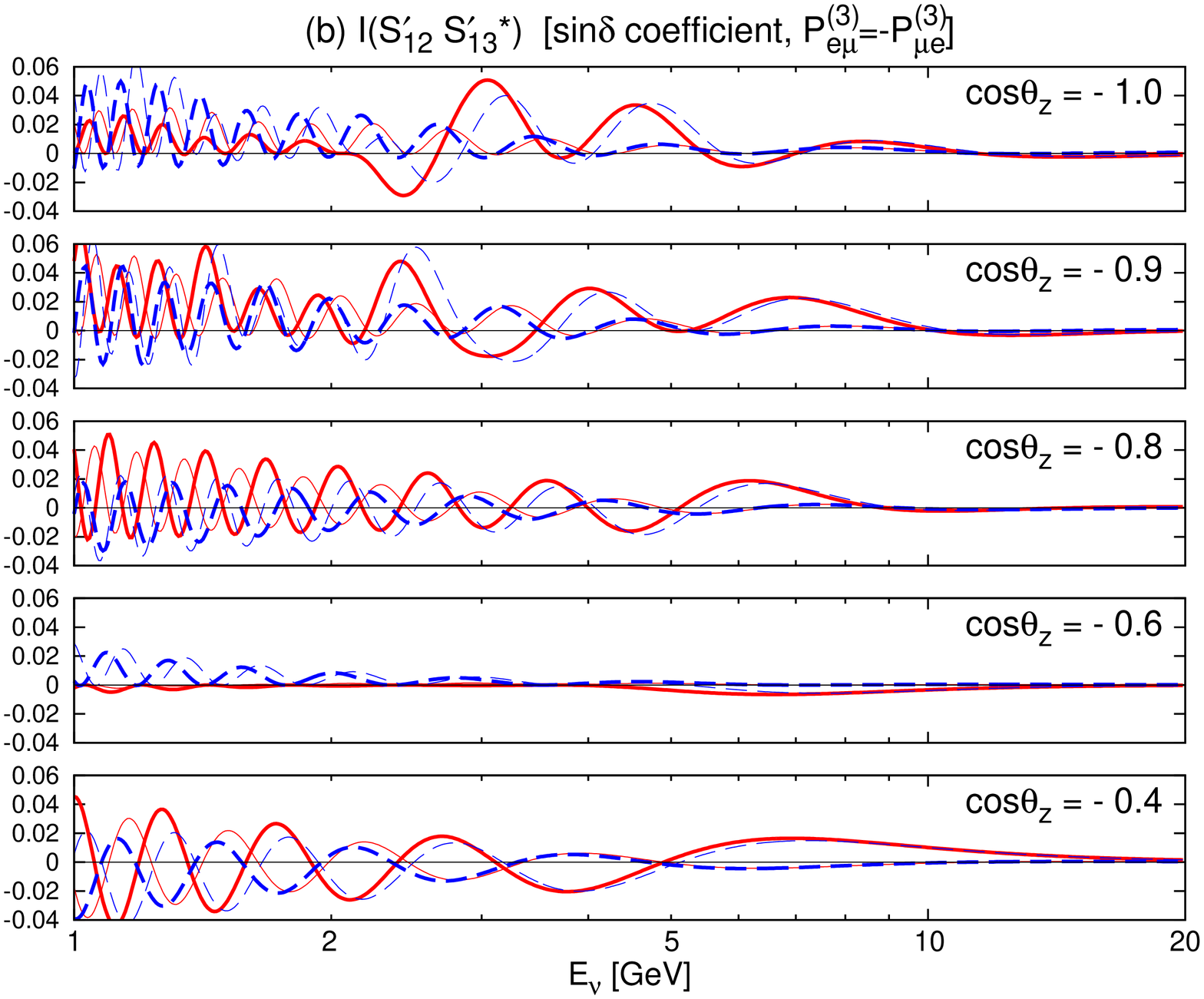}
\vspace{3mm}
\caption{(a) The coefficients of $\cos \gdelta = \sqrt{1 - x^2_{\rm a}} \cos \delta$, $\mathbb R(S'_{12} S'^*_{13}) = P^{(2)}_{\rm e \mu} = P^{(2)}_{\mu \rm e} = - P^{(2)}_{\mu \mu}$ and $\mathbb R(\overline S'_{12} \overline S'^*_{13}) = \overline P^{(2)}_{\rm e \mu} = \overline P^{(2)}_{\mu \rm e} = - \overline P^{(2)}_{\mu \mu}$, and (b) the coefficients of $\sin \gdelta = \sqrt{1 - x^2_{\rm a}} \sin \delta$, $\mathbb I(S'_{12} S'^*_{13}) = P^{(3)}_{\rm e \mu} = - P^{(3)}_{\mu \rm e}$ and $\mathbb I(\overline S'_{12} \overline S'^*_{13}) = - \overline P^{(3)}_{\rm e \mu} = \overline P^{(3)}_{\mu \rm e}$, are plotted against the neutrino energy $E_\nu$ for $\cos \theta_{\rm z} = -1, -0.9, -0.8, -0.6, -0.4$. The thick and thin lines are for NH and IH, respectively, for $\nu$ [solid lines] and $\bar \nu$ [dashed lines] oscillations.}
\label{fig:Sd}
\end{figure}

We examine the energy and zenith angle dependence of these 6 oscillation factors in \gfig{fig:S0} and \gfig{fig:Sd}. Shown in \gfig{fig:S0}(a) and (b) are the $E_\nu$ dependence of the coefficients $P^{(0)}_{\rm ee} = |S'_{11}|^2$ ($\overline P^{(0)}_{\rm ee} = |\overline S'_{11}|^2$) and $P^{(0)}_{\mu \mu} = |S'_{22} + S'_{33}|^2/4$ ($\overline P^{(0)}_{\mu \mu} = |\overline S'_{22} + \overline S'_{33}|^2$), respectively, for the baseline along five zenith angles, $\cos \theta_{\rm z} = -1, -0.9, -0.8, -0.6$ and $-0.4$. In each panel, the solid and dashed curves are for $\nu$ and $\bar \nu$ oscillations, respectively, shown by the thick lines for NH and by the thin lines for IH. It should be noted that the coefficient $|S'_{11}|$ in \gfig{fig:S0} not only determines $P^{(0)}_{\rm ee}$, as in (\ref{eq:Pee}), but also governs all the coefficients of $x_{\rm a}$, $P^{(1)}_{\rm e \mu} = P^{(1)}_{\mu \rm e} = P^{(1)}_{\mu \mu}$ in the approximation (\ref{eq:Pij-simplified}).

We immediately notice in \gfig{fig:S0}(a) the absence of the significant oscillation in $P^{(0)}_{\rm ee}$ for the $\nu$ in IH [solid-thin lines] and for the $\bar \nu$ ($\overline P^{(0)}_{\rm ee}$) in NH [dashed-thick lines]. Likewise, in \gfig{fig:S0}(b) for $P^{(0)}_{\mu\mu}$ (IH) and $\overline P^{(0)}_{\mu\mu}$ (NH), the oscillation curves for the same contributions, solid-thin and dashed-thick lines, show the vacuum-oscillation like pattern. They are consequences of the absence of the MSW resonance in these cases, as explained in \gsec{sec:MSW}. Conversely, the strong oscillation pattern for $P_{\rm ee}$ (NH) and $\overline P_{\rm ee}$ (IH) in \gfig{fig:S0}(a) and the significant deviation from the vacuum oscillation pattern for $P^{(0)}_{\mu \mu}$ (NH) and $\overline P^{(0)}_{\mu \mu}$ (IH) in \gfig{fig:S0}(b) are both consequences of the MSW resonance at $E_\nu \sim 6$~GeV for the earth matter density of $\rho \sim 5 \mbox{g}/\mbox{cm}^2$ along the baseline with $\cos \theta_{\rm z} < - 0.6$; see \gfig{fig:Pee:MH}. More generally, we find
\begin{subequations}
\begin{eqnarray}
  \overline P^{(k)}_{\alpha \beta} ({\rm NH}) 
& \approx &
  P^{(k)}_{\alpha \beta} ({\rm IH})
\qquad \mbox{for} \qquad k = 0,1,3,5,6 \,,
\\
  \overline P^{(k)}_{\alpha \beta} ({\rm NH})
& \approx &
- P^{(k)}_{\alpha \beta} ({\rm IH})
\qquad \mbox{for} \qquad k = 2,4 \,,
\end{eqnarray}
\label{eq:Pk-approximate}
\end{subequations}
\hspace{-2mm} and vice versa for $\overline P^{(k)}_{\alpha \beta} ({\rm IH})$. The relative minus signs for $k=2,3,4$, as compared to the relations (\ref{eq:nu-antinu}) within the same hierarchy, are consequences of the extra minus sign in $S'_{13}$ and $S'_{22} - S'_{33}$, when both $\delta m^2_{\rm a}$ and $a(x)$ reverse signs in the limit of vanishing $\delta m^2_{\rm s}$. %This is different from the case in (\ref{eq:nu-antinu}) where the minus sign between $P^{(3)}_{\alpha \beta}$ and $\overline P^{(3)}_{\alpha \beta}$ comes from reversing the CP phase $\delta$. Here, another extra sign appears in $S'_{13}$ and $S'_{22} - S'_{33}$. This is due to the fact that both $\delta m^2_{\rm a}$ and $a(x)$ receive a minus sign. In the limit of vanishing $\delta m^2_{\rm s}$, the oscillation happens effectively between $1$ and $3$. The mixing term $S'_{13} = 2 i \sin (\phi/2) e^{i \phi/2}$ where $\phi$ is the oscillation phase which is proportional to the effective mass square difference and hence would receive a minus. This would lead directly to a minus, which comes from $\sin(\phi/2)$, to $S'_{13}$. The minus sign in the exponential would lead to a small phase shift as shown in \gfig{fig:Sd}. This also applies to $S'_{22} - S'_{33}$.

Therefore, if we observe the presence or absence of the MSW resonance effects in $\nu$ and $\bar \nu$ oscillations, we can determine the neutrino mass hierarchy. However, as shown clearly in Figs.~\ref{fig:S0}(a) and (b), the oscillation probabilities of $\nu$ in one mass hierarchy are very similar to those of $\bar \nu$ in the other mass hierarchy. Therefore, the capability of an atmospheric neutrino detector that cannot distinguish particle charges depend critically on the difference in the flux times cross section products of $\nu$ and $\bar \nu$, as shown in \gfig{fig:flux}.

In \gfig{fig:Sd}, we show the coefficients of $\cos \gdelta = \sqrt{1 - x^2_{\rm a}} \cos \delta$ and $\sin \gdelta = \sqrt{1 - x^2_{\rm a}} \sin \delta$, which determines the sensitivity of the neutrino oscillation among the $e$ and $\mu$ flavors on the CP phase $\delta$. The real part $\mathbb R(S'_{12} S'^*_{13})$ shown in \gfig{fig:Sd}(a) governs all the coefficients of $\cos \gdelta$, see (\ref{eq:Ps2}), whereas the imaginary part $\mathbb I(S'_{12} S'^*_{13})$ in \gfig{fig:Sd}(b) dictates the $\sin \gdelta$ coefficients of $P^{(3)}_{\rm e \mu} = - P^{(3)}_{\mu \rm e}$ and $\overline P^{(3)}_{\rm e \mu} = - \overline P^{(3)}_{\mu \rm e}$. In all the cases, we confirm the vacuum oscillation like patterns for $\nu$ in IH [solid-thin lines] and for $\bar \nu$ in NH [dashed-thick lines], and significantly different patterns for $\nu$ in NH [solid-thick lines] and for $\bar \nu$ in IH [dashed-thin lines]. The approximate relations (\ref{eq:Pk-approximate}) between the $\nu$ oscillation in IH and the $\bar \nu$ oscillation in NH, and {\it vice versa}, between the $\nu$ oscillation in NH and the $\bar \nu$ oscillation in IH also holds rather well, despite small phase-shifts due to oscillations in $\delta m^2_{\rm s}$. In addition, we note the smallness of their magnitudes, typically at the level of $3\%$ for $\cos \theta_{\rm z} < -0.8$, being terms of order $\delta m^2_{\rm s} / \delta m^2_{\rm a}$. Note that they are larger at lower energies, $E_\nu \lesssim 6~\mbox{GeV}$. Consequently, the measurement of the CP phase $\delta$ may require sensitivity to the $\nu_\mu \leftrightarrow \nu_{\rm e}$ and $\bar \nu_\mu \leftrightarrow \bar \nu_{\rm e}$ oscillations at relatively low energies.
 
\begin{figure}[h!]
\vspace{3mm}
\centering
\includegraphics[height=8.8cm,width=7cm,angle=-90]{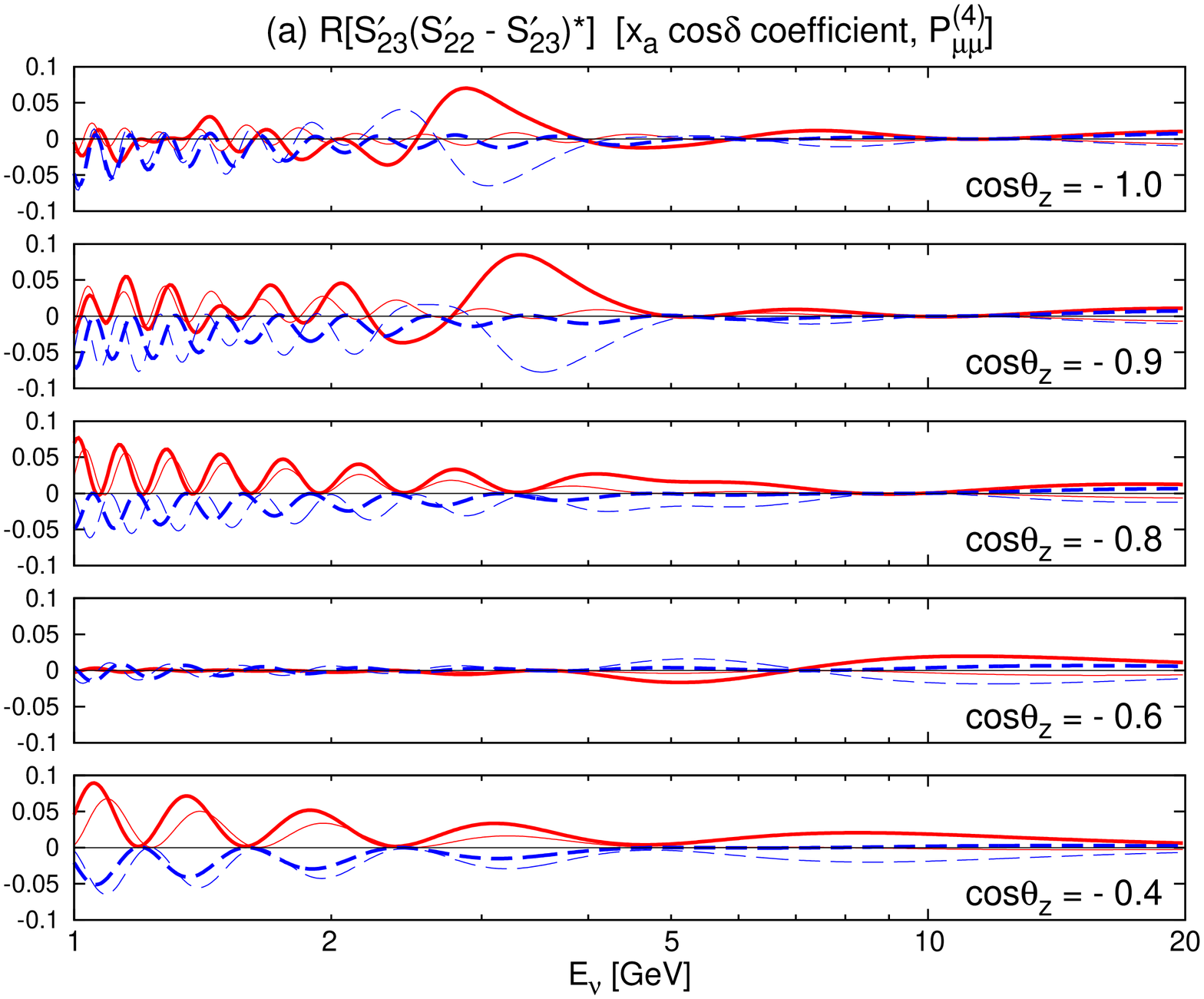}
\includegraphics[height=8.8cm,width=7cm,angle=-90]{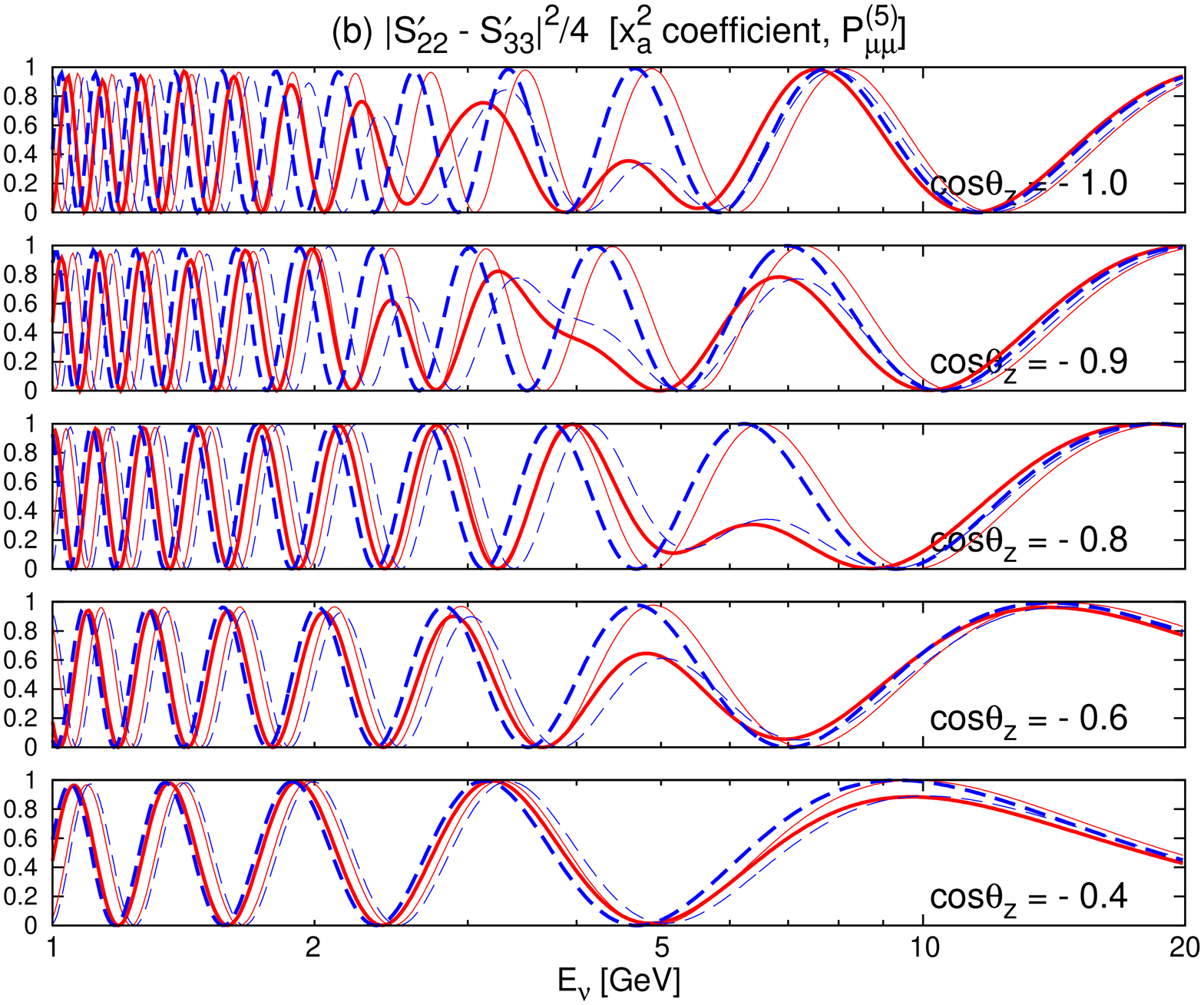}
\vspace{3mm}
\caption{(a) The coefficients $\mathbb R [S'_{23} (S'_{22} - S'_{33})^*] = P^{(4)}_{\mu \mu}$ and $\mathbb R [\overline S'_{23} (\overline S'_{22} - \overline S'_{33})^*] = \overline P^{(4)}_{\mu \mu}$ of the cross term $x_{\rm a} \cos \gdelta = x_{\rm a} \sqrt{1 - x^2_{\rm a}} \cos \delta$, and (b) the coefficients $\frac 1 4 |S'_{22} - S'_{33}|^2 = P^{(5)}_{\mu \mu}$ and $\frac 1 4 |\overline S'_{22} - \overline S'_{33}|^2 = \overline P^{(5)}_{\mu \mu}$ of the quadratic term $x^2_{\rm a}$, are plotted against the neutrino energy $E_\nu$ for $\cos \theta_{\rm z} = -1, -0.9, -0.8, -0.6, -0.4$. The thick and thin lines are for NH and IH, respectively, for $\nu$ [solid lines] and $\bar \nu$ [dashed lines] oscillations.}
\label{fig:Sxa2}
\end{figure}

Finally, in \gfig{fig:Sxa2}(a) we show the coefficients of the cross term $x_{\rm a} \cos \gdelta$, which appears only in $P^{(4)}_{\mu \mu}$ and $\overline P^{(4)}_{\mu \mu}$, and in \gfig{fig:Sxa2}(b) the coefficients $P^{(5)}_{\mu \mu}$ and $\overline P^{(5)}_{\mu \mu}$ of the quadratic term $x^2_{\rm a}$. We should note that $P^{(4)}_{\alpha \beta}$ is of the same order as $P^{(2)}_{\alpha \beta}$ and $P^{(3)}_{\alpha \beta}$, and also satisfy the same features described in the last paragraph. Generally speaking, the coefficient of the $x_{\rm a} \cos \delta$ mixing term is small in magnitude as compared to those of $x_{\rm a}$, $P^{(1)}_{\rm e \mu} = P^{(1)}_{\mu \rm e} = - P^{(1)}_{\mu \mu} = - \frac 1 2 (1 - |S'_{11}|^2)$ as can be inferred from the $|S'_{11}|^2$ plots of \gfig{fig:S0}(a), especially at high energies of $E_\nu \gtrsim 5~\mbox{GeV}$. Therefore, we expect little $\delta$-dependence in the $x_{\rm a}$ measurement. On the other hand, the coefficients of $x^2_{\rm a}$, $P^{(5)}_{\mu \mu}$ and $\overline P^{(5)}_{\mu \mu}$ shown in \gfig{fig:Sxa2}(b), are large in magnitudes and can dominate the terms linear in $x_{\rm a}$, even for $x^2_{\rm a} \sim 0.04$, especially for the neutrino oscillation in IH and the antineutrino oscillation in NH where $\frac 1 2 (1 - |S'_{11}|^2)$ has very small magnitude at $E_\nu \gtrsim 4~\mbox{GeV}$, see the thin-solid and thick-dashed curves in \gfig{fig:S0}(a). Consequently, their contributions can be significant in the measurement of $x_a$. %For the quadratic term, $P^{(5)}_{\alpha \beta}$ is as large as $P^{(0)}_{\alpha \beta}$ and $P^{(1)}_{\alpha \beta}$. We should note that $P^{(0)}_{\mu \mu}$ and $P^{(1)}_{\mu \mu}$ vanish around $E_\nu \sim 8 \mbox{GeV}$ while $P^{(5)}_{\mu \mu}$ reaches its maximum. At this point, the quadratic term dominates.

\section{Event Rates at a Charge-Blind Detector}
\label{sec:rates}

We are now ready to study systematically the event rate distributions of atmospheric neutrino observation at PINGU. It cannot distinguish particle charges, but has the capability of resolving high energy $\mu^\pm$ tracks from $e^\pm$ and/or hadronic showers~\cite{Aartsen:2013jza,:2012zk,:2012uu}.

Out from the four components in the atmospheric neutrino flux, namely the fluxes of electron- and muon-neutrino/antineutrino, as shown in \gfig{fig:flux}, the PINGU detector~\cite{Koskinen:2011zz} is assumed to observe both electron-like and muon-like events \cite{Cowen},
\begin{subequations}
\begin{eqnarray}
  \frac {d N_{\rm e}}{d E_\nu d \cos \theta_{\rm z}}
& \hspace{-3mm} = \hspace{-3mm} &
\left\{
[ \phi_{\nu_{\rm e}}(E_\nu, \cos \theta_{\rm z}) P_{\rm ee}(E_\nu, \cos \theta_{\rm z})
+ \phi_{\nu_\mu}(E_\nu, \cos \theta_{\rm z}) P_{\mu \rm e}(E_\nu, \cos \theta_{\rm z})] \sigma_{\nu_{\rm e}}(E_\nu) 
\right.
\nonumber
\\
&&
\left.
+ [\phi_{\bar \nu_{\rm e}}(E_\nu, \cos \theta_{\rm z}) \overline P_{\rm ee}(E_\nu, \cos \theta_{\rm z}) 
+ \phi_{\bar \nu_\mu}(E_\nu, \cos \theta_{\rm z}) \overline P_{\mu \rm e}(E_\nu, \cos \theta_{\rm z})] \sigma_{\bar \nu_{\rm e}}(E_\nu)
\right\} \rho V_{{\rm eff}}(E_\nu) \,,
\\
  \frac {d N_\mu}{d E_\nu d \cos \theta_{\rm z}}
& \hspace{-3mm} = \hspace{-3mm} &
\left\{
[ \phi_{\nu_{\rm e}}(E_\nu, \cos \theta_{\rm z}) P_{\rm e\mu}(E_\nu, \cos \theta_{\rm z})
+ \phi_{\nu_\mu}(E_\nu, \cos \theta_{\rm z}) P_{\mu \mu}(E_\nu, \cos \theta_{\rm z})] \sigma_{\nu_\mu}(E_\nu) 
\right.
\nonumber
\\
&&
\left.
+ [\phi_{\bar \nu_{\rm e}}(E_\nu, \cos \theta_{\rm z}) \overline P_{\rm e \mu}(E_\nu, \cos \theta_{\rm z}) 
+ \phi_{\bar \nu_\mu}(E_\nu, \cos \theta_{\rm z}) \overline P_{\mu \mu}(E_\nu, \cos \theta_{\rm z})] \sigma_{\bar \nu_\mu}(E_\nu)
\right\} \rho V_{{\rm eff}}(E_\nu) \,,
\end{eqnarray}
\end{subequations}
\hspace{-1.5mm} by summing over contributions from charged-current (CC) $e^\pm$ and $\mu^\pm$ production events. Since the fiducial volume is universal for both $e^\pm$ and $\mu^\pm$ channels, as explained in \gsec{sec:FXV}, it serves as an overall factor. For each flavor, neutrino and antineutrino contribute with the corresponding CC cross sections. It should be noted that we neglect the contributions from tau-neutrino/antineutrino since the tau-neutrino flux is very small~\cite{Athar:2012it} and also because the charged-current $\tau^\pm$ production events followed by their pure-leptonic decays contribute mainly to events with low observable energies, which may not contribute much due to the smaller fiducial volume. In addition, the tau neutrino cross section is small compared to the CC electron and muon neutrino cross sections \cite{Jeong:2010nt}. Contributions from the charged-current $\tau^\pm$ production events as well as neutral-current events will be studied elsewhere. 

Since the number of signal events depends on the oscillation probabilities linearly, which have been decomposed into six terms in (\ref{eq:Pij-decomposition}), the event rates can also be decomposed accordingly,
\begin{equation}
  \frac {d N_\alpha}{d E_\nu d \cos \theta_{\rm z}}
\equiv
  N^{(0)}_\alpha
+ N^{(1)}_\alpha x_{\rm a}
+ N^{(2)}_\alpha \cos \gdelta
+ N^{(3)}_\alpha \sin \gdelta
+ N^{(4)}_\alpha x_{\rm a} \cos \gdelta
+ N^{(5)}_\alpha x^2_{\rm a} \,
+ N^{(6)}_\alpha \cos^2 \gdelta \,.
\label{eq:Ni-decomposition}
\end{equation}
\hspace{-1.8mm} 
By combining with the explicit expressions of the decomposed oscillation probabilities in (\ref{eq:Ps}), the coefficients for electron-like event number rates are,
\begin{subequations}
\begin{eqnarray}
  N^{(0)}_{\rm e} 
& = &
\left\{
  \left[ \phi_{\nu_{\rm e}} |S'_{11}|^2
  + \phi_{\nu_\mu}
  \frac 1 2 (1 - |S'_{11}|^2) \right]
 \sigma_{\nu_{\rm e}}
+ \left[ \phi_{\bar \nu_{\rm e}} |\overline S'_{11}|^2 
  + \phi_{\bar \nu_\mu}
  \frac 1 2 \left( 1 - |\overline S'_{11}|^2 \right) \right]
  \sigma_{\bar \nu_{\rm e}}
\right\} \rho V_{{\rm eff}} \,,
\qquad
\\
  N^{(1)}_{\rm e} 
& = &
\left\{
- \phi_{\nu_\mu} \frac 1 2 (1 - |S'_{11}|^2) \sigma_{\nu_{\rm e}}
- \phi_{\bar \nu_\mu}
  \frac 1 2 \left( 1 - |\overline S'_{11}|^2 \right)
  \sigma_{\bar \nu_{\rm e}}
\right\} \rho V_{{\rm eff}} \,, \qquad
\\
  N^{(2)}_{\rm e} 
& = &
\left[
  \phi_{\nu_\mu} \mathbb R \left( S'_{12} S'^*_{13} \right)
  \sigma_{\nu_{\rm e}}
+ \phi_{\bar \nu_\mu}
  \mathbb R \left( \overline S'_{12} \overline S'^*_{13} \right)
  \sigma_{\bar \nu_{\rm e}}
\right] \rho V_{{\rm eff}} \,,
\\
  N^{(3)}_{\rm e} 
& = &
\left[
- \phi_{\nu_\mu} \, \mathbb I \left( S'_{12} S'^*_{13} \right)
  \sigma_{\nu_{\rm e}}
+ \phi_{\bar \nu_\mu} \,
  \mathbb I \left( \overline S'_{12} \overline S'^*_{13} \right)
  \sigma_{\bar \nu_{\rm e}}
\right] \rho V_{{\rm eff}} \,,
\\
  N^{(4)}_{\rm e} 
& = &
  N^{(5)}_{\rm e} 
=
  N^{(6)}_{\rm e} 
=
  0 \,.
\end{eqnarray}
\end{subequations}
For brevity, the arguments $E_\nu$ and $\cos \theta_{\rm z}$ have been omitted.  Note that there is no term with $x_{\rm a} \cos \gdelta$, $x^2_{\rm a}$ or $\cos^2 \gdelta$ dependence for electron-like events since $P^{(4)}_{\rm ee} = P^{(4)}_{\mu \rm e} = P^{(5)}_{\rm ee} = P^{(5)}_{\mu \rm e} = P^{(6)}_{\rm ee} = P^{(6)}_{\mu \rm e} = 0$ and the same for antineutrinos as shown in (\ref{eq:Ps}) and (\ref{eq:Ps2}). In other words, the atmospheric angle $\theta_{\rm a}$ and the CP phase $\delta$ are naturally disentangled in the electron-like events, which depend
on $\theta_{\rm a}$ through $N^{(1)}_{\rm e}$ while the dependence on
the CP phase $\delta$ comes from $N^{(2)}_{\rm e}$ and $N^{(3)}_{\rm e}$. 

For the muon-like events, we find,
\begin{subequations}
\begin{eqnarray}
  N^{(0)}_\mu
& \hspace{-1mm} = \hspace{-1mm} &
\left\{
  \left[ \phi_{\nu_{\rm e}} \frac 1 2 (1 - |S'_{11}|^2) 
       + \phi_{\nu_\mu} \frac 1 4 |S'_{22} + S'_{33}|^2 \right]
  \sigma_{\nu_\mu}
+
  \left[ \phi_{\bar \nu_{\rm e}} \frac 1 2 (1 - |\overline S'_{11}|^2) 
       + \phi_{\bar \nu_\mu} \frac 1 4 |\overline S'_{22} + \overline S'_{33}|^2 \right] \sigma_{\bar \nu_\mu} \right\} \rho V_{{\rm eff}} \,,
\qquad
\\
  N^{(1)}_\mu
& \hspace{-1mm} = \hspace{-1mm} &
\left\{
  (\phi_{\nu_\mu} - \phi_{\nu_{\rm e}}) \frac 1 2 (1 - |S'_{11}|^2) \sigma_{\nu_\mu} 
+ (\phi_{\bar \nu_\mu} - \phi_{\bar \nu_{\rm e}}) \frac 1 2 (1 - |\overline S'_{11}|^2) \sigma_{\bar \nu_\mu}
\right\} \rho V_{{\rm eff}} \,,
\\
  N^{(2)}_\mu
& \hspace{-1mm} = \hspace{-1mm} &
\left\{
  \left( \phi_{\nu_{\rm e}} - \phi_{\nu_\mu} \right) \mathbb R (S'_{12} S'^*_{13}) 
  \sigma_{\nu_\mu}
+ \left( \phi_{\bar \nu_{\rm e}} - \phi_{\bar \nu_\mu} \right) \mathbb R(\overline S'_{12} \overline S'^*_{13}) 
  \sigma_{\bar \nu_\mu} 
\right\} \rho V_{{\rm eff}} \,,
\\
  N^{(3)}_\mu
& \hspace{-1mm} = \hspace{-1mm} &
\left\{
  \phi_{\nu_{\rm e}} \mathbb I (S'_{12} S'^*_{13}) \sigma_{\nu_\mu}
- \phi_{\bar \nu_{\rm e}} \mathbb I (\overline S'_{12} \overline S'^*_{13}) 
  \sigma_{\bar \nu_\mu}
\right\} \rho V_{{\rm eff}} \,,
\\
  N^{(4)}_\mu
& \hspace{-1mm} = \hspace{-1mm} &
\left\{
  \phi_{\nu_\mu} \mathbb R \left[ S'_{23} (S'_{22} - S'_{33})^* \right] \sigma_{\nu_\mu}
+ \phi_{\bar \nu_\mu} \mathbb R \left[ \overline S'_{23} (\overline S'_{22} - \overline S'_{33})^* \right] \sigma_{\bar \nu_\mu}
\right\} \rho V_{{\rm eff}} \,,
\\
  N^{(5)}_\mu
& \hspace{-1mm} = \hspace{-1mm} &
  \left\{ \phi_{\nu_\mu} \frac 1 4 |S'_{22} - S'_{33}|^2 \sigma_{\nu_\mu}
        + \phi_{\bar \nu_\mu} \frac 1 4 |\overline S'_{22} - \overline S'_{33}|^2 \sigma_{\bar \nu_\mu} 
  \right\} \rho V_{{\rm eff}} \,,
\\
  N^{(6)}_\mu
& \hspace{-1mm} = \hspace{-1mm} &
  \left\{ \phi_{\nu_\mu} |S'_{23}|^2 \sigma_{\nu_\mu}
        + \phi_{\bar \nu_\mu} |\overline S'_{23}|^2 \sigma_{\bar \nu_\mu} 
  \right\} \rho V_{{\rm eff}}  \,.
\end{eqnarray}
\end{subequations}
Note that for muon-like events, the crossing term $x_{\rm a} \cos \gdelta$ has nonvanishing coefficient $N^{(4)}_\mu$. Consequently, with muon-like events included, we should observe some correlation between the measurements of the atmospheric mixing angle $x_{\rm a}$ and the CP phase $\delta$, as will be described in \gsec{sec:fit}. In addition, a nonzero $N^{(6)}_\mu$, which is one further order of magnitude smaller than $N^{(4)}_\mu$, is kept according to (\ref{eq:Ps}) just to show its magnitude. %This is different from the case of electron-like events where $N^{(5)}_\mu$ is zero at the first place before expanding w.r.t $\delta m^2_{\rm s} / \delta m^2_{\rm a}$. 
%\begin{figure}[h]
%\centering
%\vspace{1mm}
%\includegraphics[height=8.9cm,width=10cm,angle=-90]{N0dN_m2.eps}
%\includegraphics[height=8.9cm,width=10cm,angle=-90]{N0dN_e2.eps}
%\vspace{2mm}
%\caption{Decomposition coefficients $N^{(1)}_\alpha$ [for $\cos \delta$, green
%         solid], $N^{(2)}_\alpha$ [for $\sin \delta$, blue dashed], 
%         $N^{(3)}_\alpha$ [for $\cos^2 \delta$, red short dashed],
%         and $N^{(5)}_\alpha$ [for $x \cos \delta$, black dotted] of the 
%         muon-like [Left] and electron-like [Right] events for NH [thick]
%         and IH [thin].}
%\label{fig:Ni2}
%\end{figure}

\begin{figure}[h!]
\centering
\vspace{4mm}
\includegraphics[height=8.9cm,width=7cm,angle=-90]{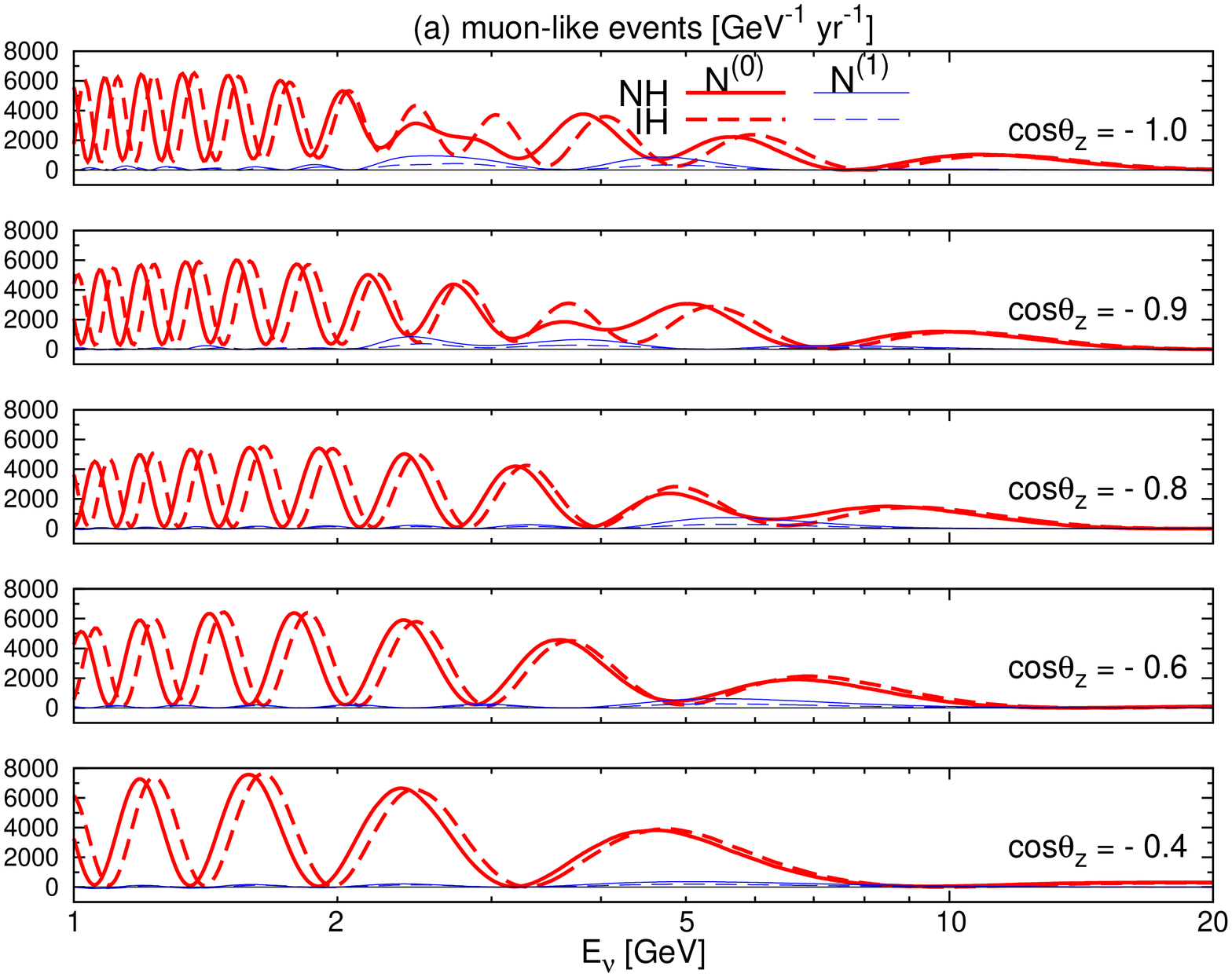}
\includegraphics[height=8.9cm,width=7cm,angle=-90]{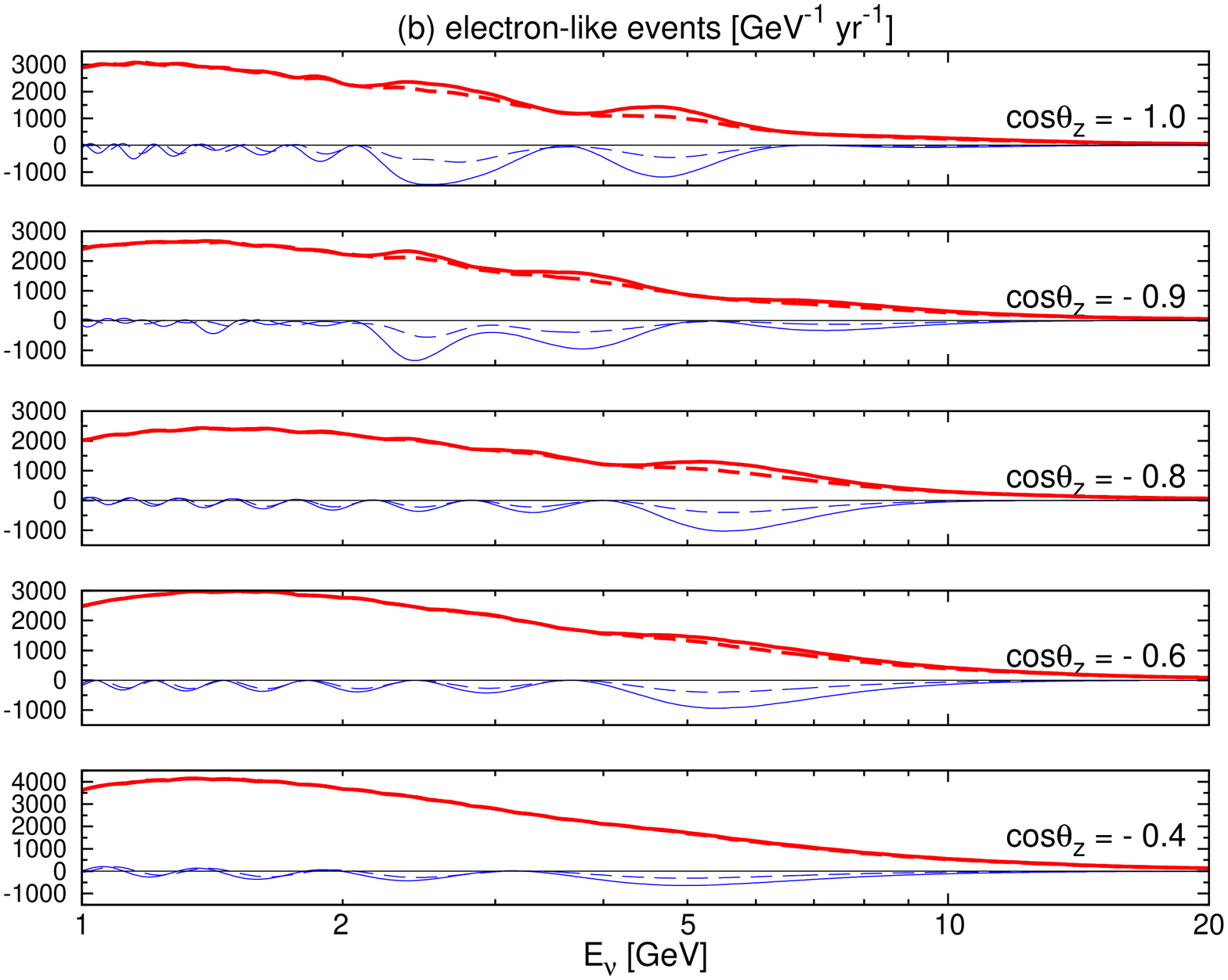}
\vspace{2mm}
\caption{The overall rates $N^{(0)}_\alpha$ [red-thick lines] and the coefficients $N^{(1)}_\alpha$ of $x_{\rm a}$ [blue-thin lines] are plotted against $E_\nu$ at $\cos \theta_{\rm z} = -1, -0.9, -0.8, -0.6, -0.4$, for the muon-like [$\alpha = \mu$] events (a) and the electron-like [$\alpha = e$] events (b). The solid curves are for NH, while the dashed curves are for IH. The vertical scale gives the number of events per GeV in one year.}
\label{fig:Ni1}
\end{figure}

\begin{figure}[h!]
\centering
\vspace{1mm}
\includegraphics[height=8.9cm,width=7cm,angle=-90]{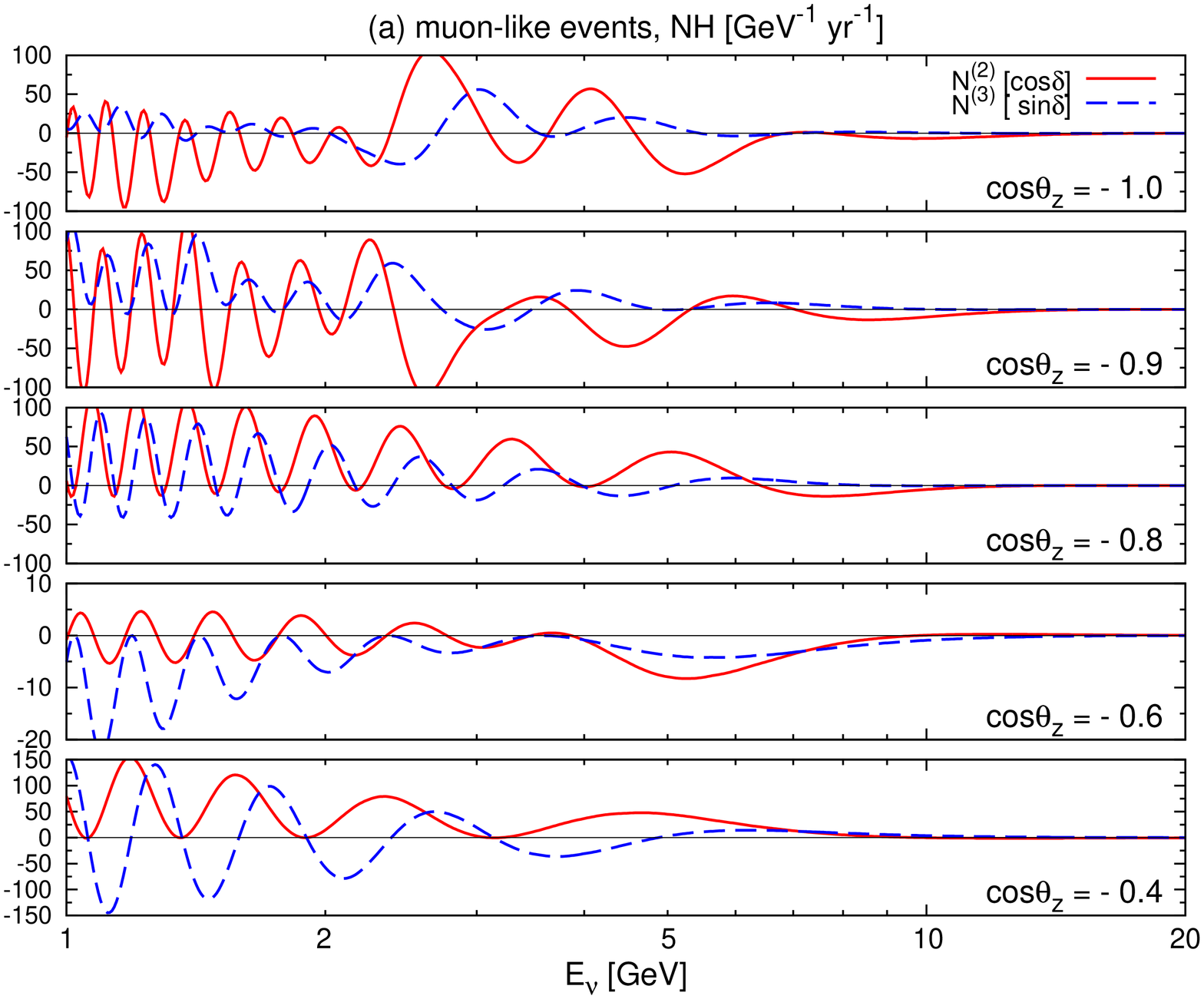}
\includegraphics[height=8.9cm,width=7cm,angle=-90]{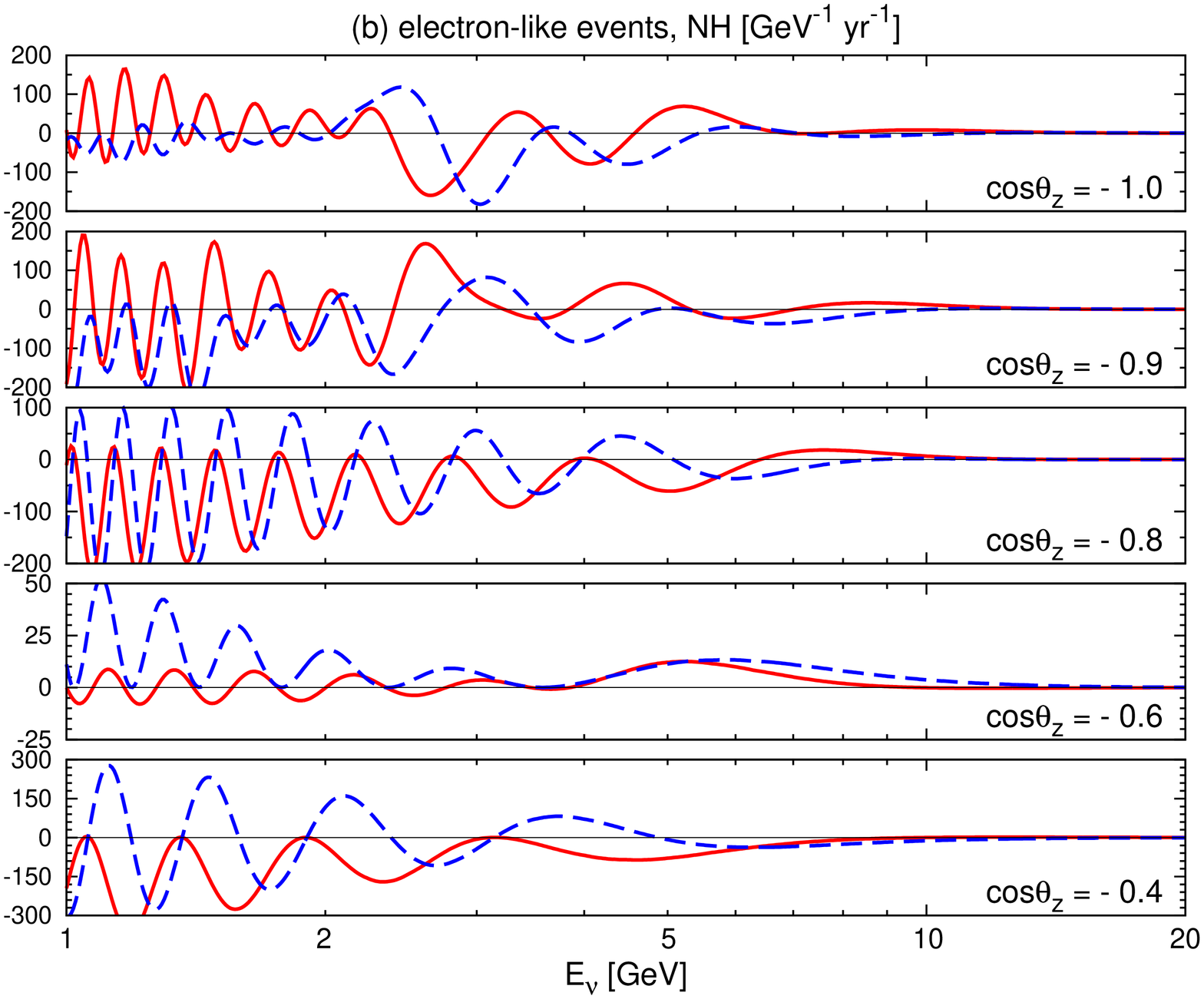}
\vspace{2mm}
\caption{The coefficients $N^{(2)}_\alpha$ of $\cos \gdelta = \sqrt{1 - x^2_{\rm a}} \cos \delta$ [red-solid lines] and $N^{(3)}_\alpha$ of $\sin \gdelta = \sqrt{1 - x^2_{\rm a}} \sin \delta$ [blue-dashed lines] for the muon-like ($\alpha = \mu$) events (a) and electron-like events (b) for NH.}
\label{fig:Ni2N}
\vspace{3mm}
\end{figure}

\begin{figure}[h!]
\centering
\vspace{1mm}
\includegraphics[height=8.9cm,width=7cm,angle=-90]{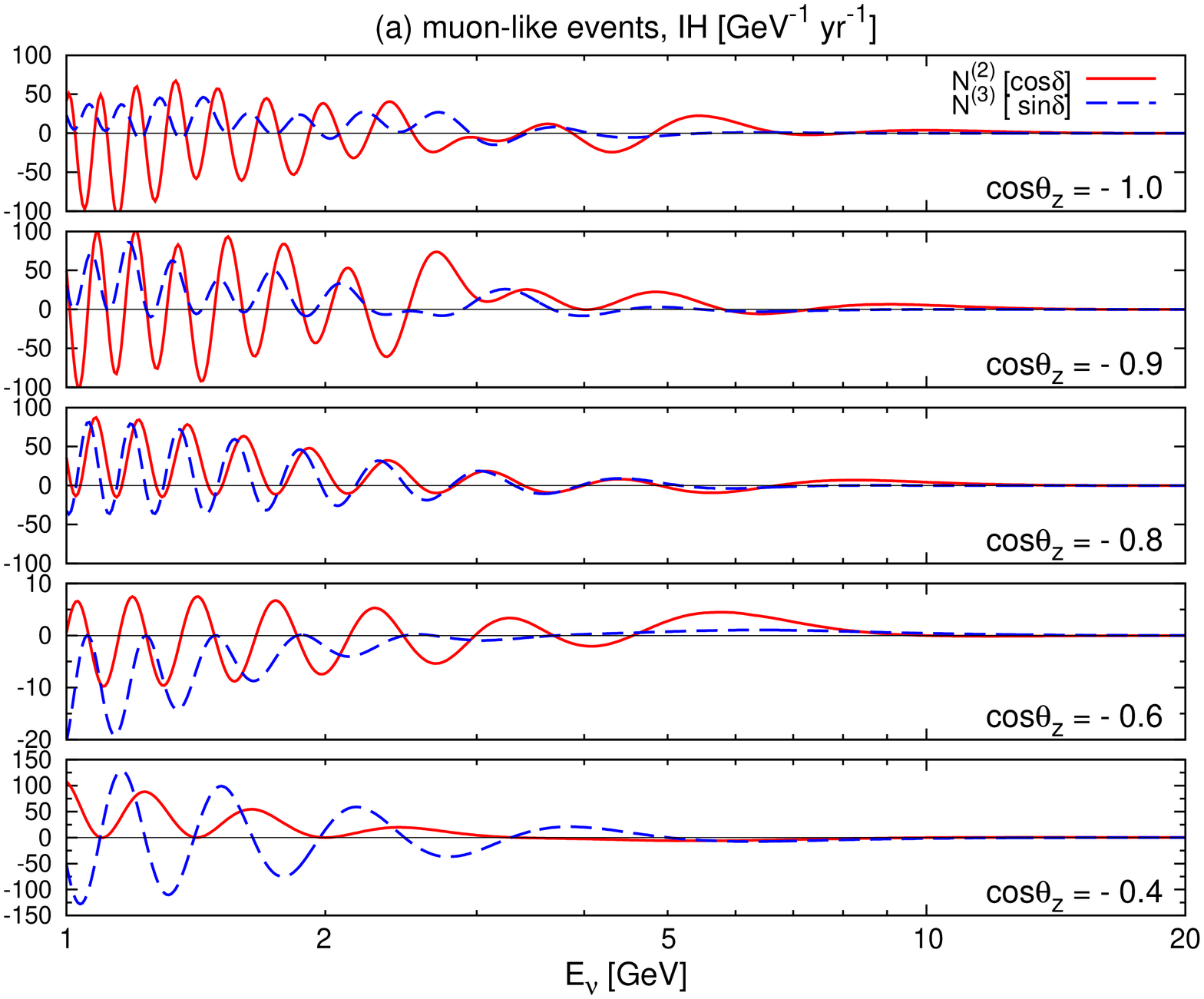}
\includegraphics[height=8.9cm,width=7cm,angle=-90]{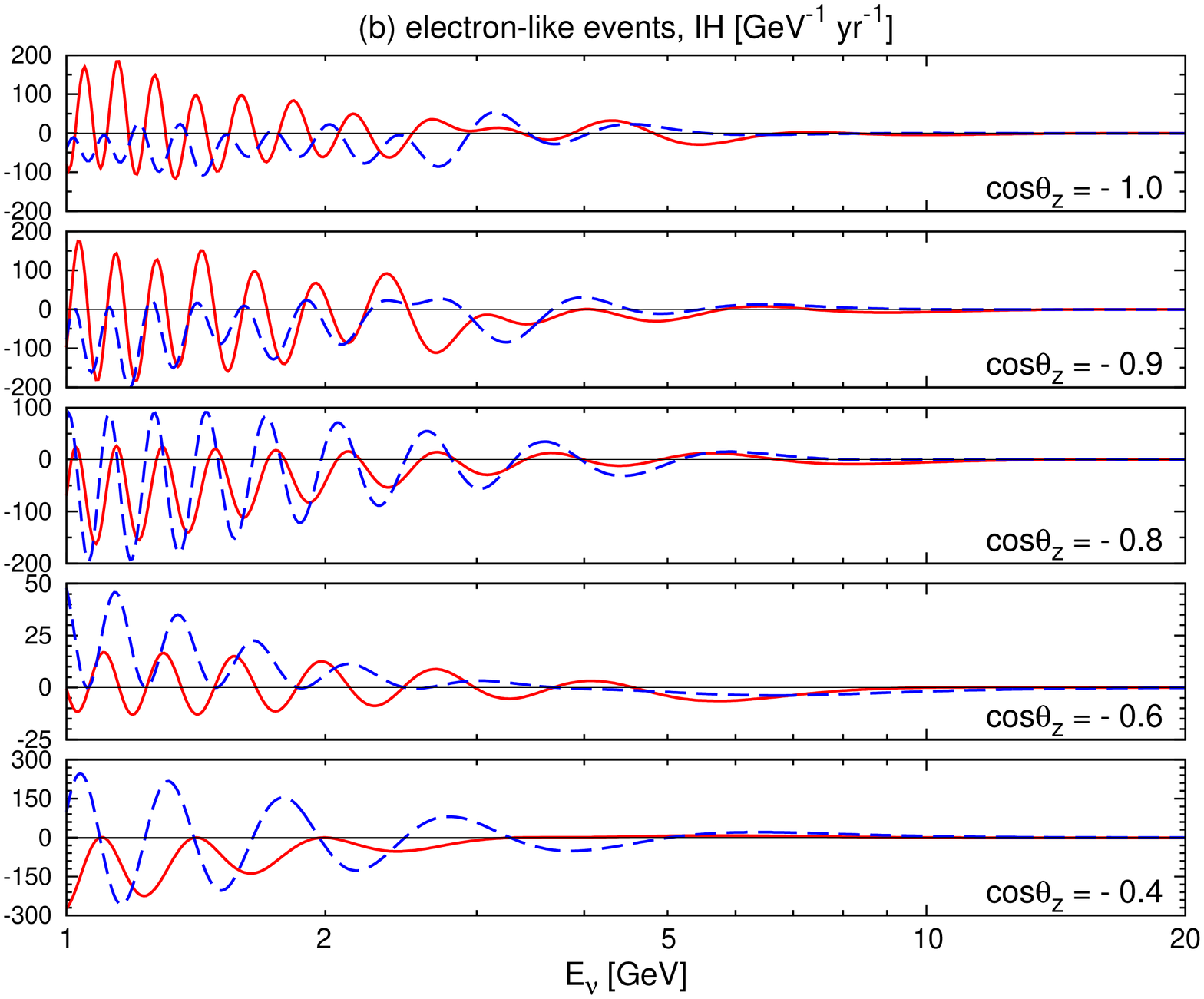}
\vspace{2mm}
\caption{The coefficients $N^{(2)}_\alpha$ of $\cos \gdelta = \sqrt{1 - x^2_{\rm a}} \cos \delta$ [solid red lines] and $N^{(3)}_\alpha$ of $\sin \gdelta = \sqrt{1 - x^2_{\rm a}} \sin \delta$ [dashed blue lines] for the muon-like ($\alpha = \mu$) events (a) and electron-like events (b) for IH.}
\label{fig:Ni2I}
\end{figure}

\begin{figure}[h!]
\centering
\vspace{4mm}
\includegraphics[height=8.9cm,width=7cm,angle=-90]{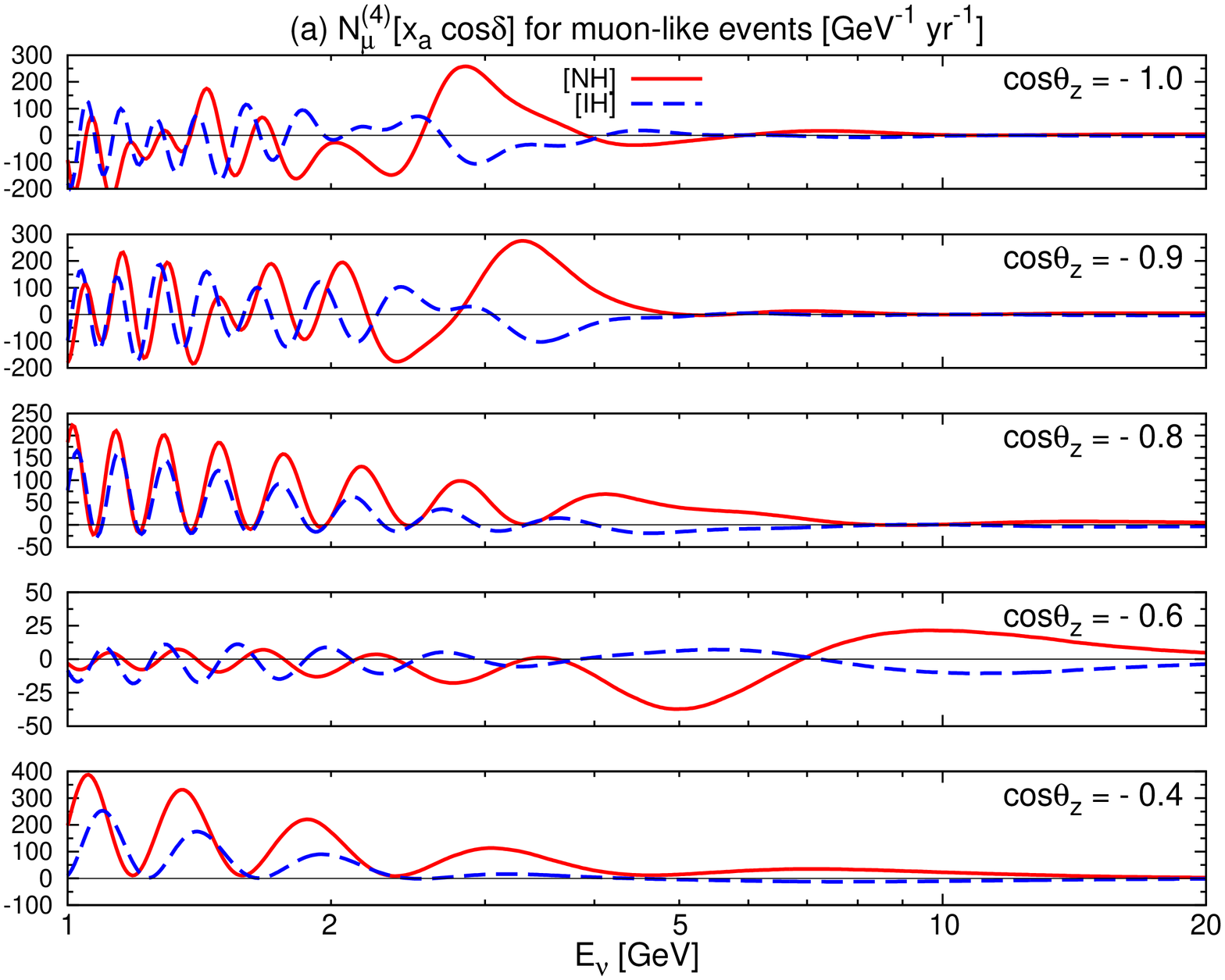}
\includegraphics[height=8.9cm,width=7cm,angle=-90]{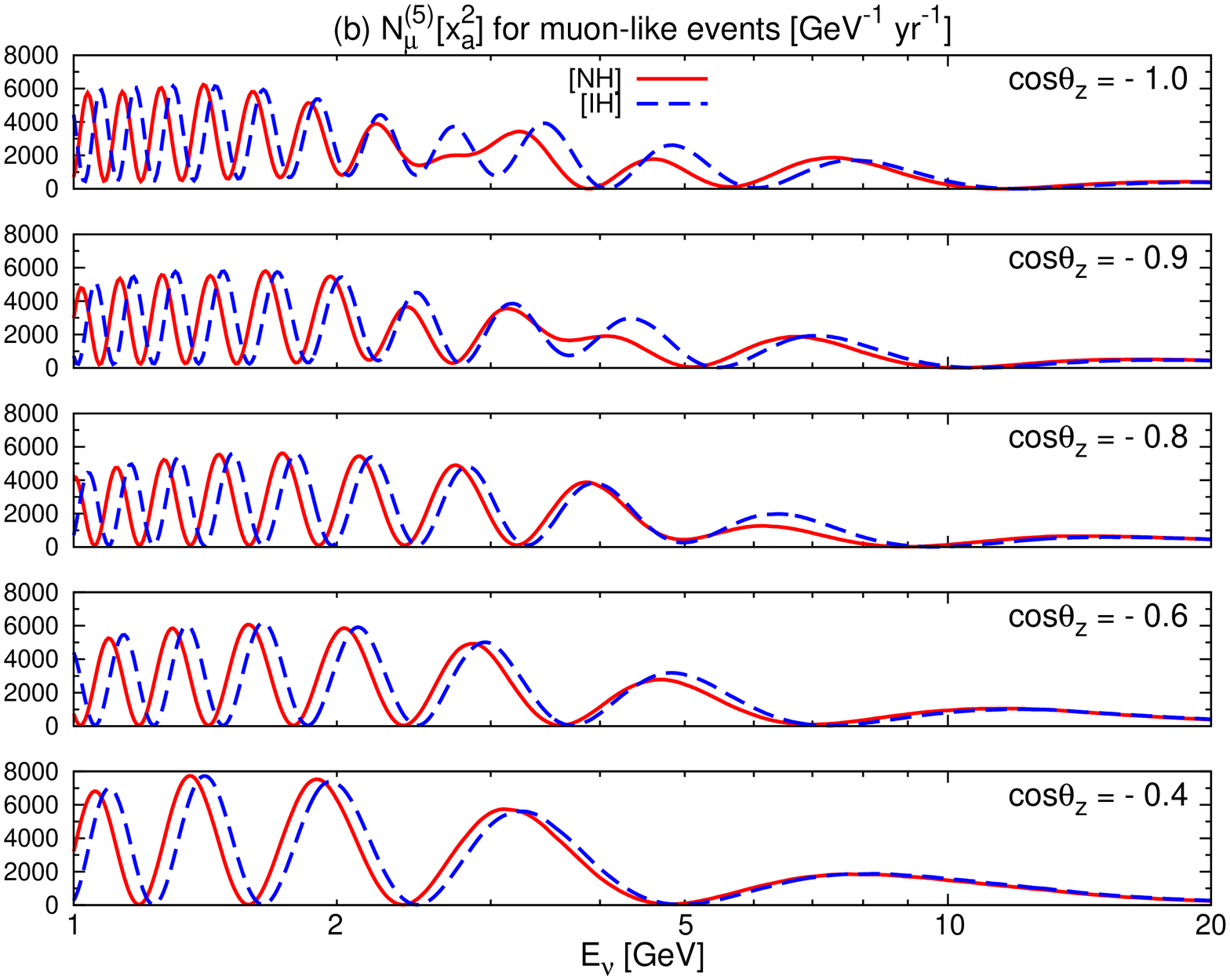}
\vspace{2mm}
\caption{(a) The coefficients $N^{(4)}_\mu$ of $x_{\rm a} \cos \gdelta = x_{\rm a} \sqrt{1 - x^2_{\rm a}} \cos^2 \delta$, and (b) $N^{(5)}_\mu$ of $x^2_{\rm a}$ for both NH [solid lines] and IH [dashed lines].}
\label{fig:N45}
\end{figure}

By combining everything together, the atmospheric neutrino flux, cross section and effective fiducial volume of \gfig{fig:flux} in \gsec{sec:FXV}, and the oscillation probabilities discussed in \gsec{sec:propagation-basis}, the energy and the zenith angle dependences of the coefficients for muon- and electron-like event rates are shown in \gfig{fig:Ni1}, \gfig{fig:Ni2N}, \gfig{fig:Ni2I}, and \gfig{fig:N45}. Note the different scales of the plots, which are adjusted to show the structure of the coefficients.

In \gfig{fig:Ni1}, we show the overall rates $N^{(0)}_\alpha$ in red-thick lines and the coefficients $N^{(1)}_\alpha$ of $x_{\rm a}$ in blue-thin lines, as functions of $E_\nu$ at several $\cos \theta_{\rm z}$, for muon-like [$\alpha = \mu$] events (a) and for electron-like [$\alpha = e$] events (b). The solid curves are for the normal hierarchy (NH), while the dashed curves are for the inverted hierarchy (IH). The muon-like event rates $N^{(0)}_\mu$ in the left panel \gfig{fig:Ni1}(a) show significant oscillatory behavior for both NH (thick-solid lines) and IH (thick-dashed lines). However, the huge hierarchy dependences in the $\nu_\mu \rightarrow \nu_\mu$ oscillation (MSW \cite{MSW} resonance only for NH) and in the $\bar \nu_\mu \rightarrow \bar \nu_\mu$ oscillation (MSW resonance only for IH) as shown in \gfig{fig:S0}(b) diminish significantly because of the cancellation between the $\nu_\mu$ and $\bar \nu_\mu$ contributions. Because the flux times cross section for $\nu_\mu$ is a factor of about three larger than that for $\bar \nu_\mu$ as shown in \gfig{fig:flux}(b), the hierarchy dependence of the $\nu_\mu \rightarrow \nu_\mu$ oscillation survives, resulting in the smaller rate for IH at the MSW resonant energy of $\sim 6~\mbox{GeV}$ at $\cos \theta_{\rm z} ~ - 0.8$. Especially at $\cos \theta_{\rm z} \lesssim - 0.9$, shown in the top two panels of \gfig{fig:Ni1}(a), the nearly maximal resonant oscillation of $\nu_\mu \rightarrow \nu_\mu$ for NH at $E_\nu \sim 4~\mbox{GeV}$ shown by the thick-red curves in the top two panels of \gfig{fig:S0}(b) gives rise to the significant difference in the muon-like event rate in the $3 \sim 5~\mbox{GeV}$ region due to the so-called parametric resonance \cite{parametric, length}. Because of the large event numbers, there is a possibility that these differences can be identified in experiments and that the neutrino mass hierarchy is determined. We note, however, that the finite energy and angular resolution of real experiments may make it difficult to identify differences which depend strongly on the energy, such as those in the oscillation phase observed at $E_\nu \lesssim 4~\mbox{GeV}$ at all $\cos \theta_{\rm z}$. On the other hand, the hierarchy dependence of the electron-like event rate $N^{(0)}_{\rm e}$, shown also by thick-red lines in \gfig{fig:Ni1}(b), has little dependence on $\cos \theta_{\rm z}$ and does not oscillate in $E_\nu$. Although both the overall rate and the difference is small, the event is consistently higher for NH than IH in the broad energy range of $2 \sim 10~\mbox{GeV}$, reflecting the MSW and parametric enhancements of the $\nu_\mu \rightarrow \nu_{\rm e}$ oscillation, $P^{(0)}_{\mu \rm e} = \frac 1 2 (1 - |S'_{11}|^2)$, for NH; see (\ref{eq:Ps2}) and \gfig{fig:S0}(a). Such moderate $E_\nu$ and $\cos \theta_{\rm z}$ dependences of the electron-like event rate on the mass hierarchy may allow actual experiments to identify the difference.

Let us now examine the coefficients $N^{(1)}_\mu$ and $N^{(1)}_{\rm e}$ of $x_{\rm a}$, which are shown by thin-blue lines in \gfig{fig:Ni1} (a) and (b), respectively, also in solid for NH and in dashed for IH. Note that $N^{(1)}_\mu$ is positive definite while $N^{(1)}_{\rm e}$ tends to be negative at high energies ($E_\nu \gtrsim 2~\mbox{GeV}$). The coefficients $N^{(1)}_\mu$ and $N^{(1)}_{\rm e}$ are both proportional to $\frac 1 2 (1 - |S'_{11}|^2)$ in the approximation of (\ref{eq:Ps2}), and hence the energy-angular dependences are mild especially at high energy region of $4 \sim 10~\mbox{GeV}$, just like the electron-like event rate $N^{(0)}_{\rm e}$. This will help experiments to measure $x_{\rm a}$. Because of the positive sign of $N^{(1)}_\mu$, the mass hierarchy determination by using only the muon-like events should be easier for $x_{\rm a} < 0$ ($\sin^2 \theta_{\rm a} < 0.5$) than for $x_{\rm a} > 0$ ($\sin^2 \theta_{\rm a} > 0.5$). The trend can be reversed when the electron-like events are also included because $N^{(1)}_{\rm e}$ has negative sign and has relatively larger magnitude. Likewise, $x_{\rm a}$ will be measured more accurately for NH than for IH when only the muon-like events are studied, whereas the measurement for IH can be significantly improved by including the electron-like events in the analysis. % and the opposite with electron-like events also included, since the magnitudes of $N^{(1)}_\alpha$ are larger for NH than those for IH.

The dependence on the CP phase $\delta$ is shown in \gfig{fig:Ni2N} and \gfig{fig:Ni2I}, respectively, for NH and IH. In both figures, the $N^{(2)}_\alpha$ [red-solid lines] coefficient of $\sqrt{1 - x^2_{\rm a}} \cos \delta$ and $N^{(3)}_\alpha$ [blue-dashed lines] of $\sqrt{1 - x^2_{\rm a}} \sin \delta$ are shown for the muon-like events in the left (a) and for the electron-like events in the right (b) panels. %We can see that they vanish in the energy range $E_\nu \gtrsim 10 \mbox{GeV}$ and have rich pattern in the lower end. Due to detector energy resolution, patterns with $E_\nu \lesssim 4 \mbox{GeV}$ are not expected to hold much distinguishing power. It comes to the same conclusion that for measuring the CP phase $\delta$, the most promising range is $4 \mbox{GeV} \lesssim E_\nu \lesssim 10 \mbox{GeV}$. It should be noted that in this range $N^{(2)}_\mu$ has larger amplitude than $N^{(3)}_\mu$ while $N^{(2)}_{\rm e}$ can be comparable with $N^{(3)}_{\rm e}$. If only muon-like events are used for determining the CP phase $\delta$, $\cos \delta$ would receive stronger constraint than $\sin \delta$. This can be improved if electron-like events are also included in the analysis. All these feature apply for both NH and IH cases. The biggest difference is for IH the coefficients are systematically smaller than the NH case. If the neutrino mass hierarchy is inverted, it would be extremely difficult to determine the CP phase. 
Let us first examine the $\delta$-dependence in the NH case as shown in \gfig{fig:Ni2N}. We first note the significantly smaller magnitudes of the coefficients $N^{(2)}_\alpha$ and $N^{(3)}_\alpha$, which are typically $100/\mbox{GeV}$, as compared to $N^{(0)}_\alpha$ and $N^{(1)}_\alpha$ which are measured in unit of $1000/\mbox{GeV}$ as shown in \gfig{fig:Ni1}. If we restrict our attention to the higher energy region of $E_\nu > 4~\mbox{GeV}$ which is less sensitive to the experimental energy-angular smearing effects, the electron-like events in \gfig{fig:Ni2N}(b) have higher sensitivity to both $\cos \delta$ [red-solid lines] and $\sin \delta$ [blue-dashed lines] than the muon-like events in \gfig{fig:Ni2N}(a). All the four coefficients $N^{(2)}_\alpha$ and $N^{(3)}_\alpha$ for $\alpha = \mu$ and $\alpha = e$ have larger magnitudes at lower energies, $E_\nu \lesssim 3~\mbox{GeV}$, although they oscillate rapidly with $E_\nu$. The expected energy resolution of the PINGU detector may smear out those rapid oscillation. However, in a certain $\cos \theta_{\rm z}$ region the coefficients tend to have a definite sign which may survive after the energy smearing. For instance, let us examine the $\sin \gdelta$ measurement by using the coefficients $N^{(3)}_\alpha$ shown by blue-dashed lines in \gfig{fig:Ni2N}. The average of $N^{(3)}_\mu$ at $\cos \theta_{\rm z} = -0.9$ is clearly positive in the whole energy range shown in the figures, whereas that of $N^{(3)}_{\rm e}$ tends to be negative in the whole region. They tend to oscillate about zero at $\cos \theta_{\rm z} = - 1.0$, and the sign reverses at $\cos \theta_{\rm z} = - 0.6$. Therefore, if the angular resolution of experiments can resolve $\cos \theta_{\rm z} = -0.9$ ($\theta_{\rm z} \sim 154^\circ$) from $\cos \theta_{\rm z} = -0.6$ ($\theta_{\rm z} \sim 127^\circ$), then it might be possible to measure $\sin \gdelta$ by using the total number of events including the low energy region. The same applies for the $\cos \gdelta$ measurements, for which the coefficient $N^{(2)}_\mu$ [red-solid lines] tends to be positive at around $\cos \theta_{\rm z} = - 0.8$ and at $-0.4$, while the opposite trend is expected for $N^{(2)}_{\rm e}$. Although a quantitative study with realistic event simulation is beyond the scope of the present paper, probability of using the low energy data for measuring $\delta$ may worth serious studies. 

In case of IH, the coefficients $N^{(2)}_\alpha$ and $N^{(3)}_\alpha$ behave as in \gfig{fig:Ni2I} (a) and (b), respectively, for muon-like ($\alpha = \mu$) and electron-like ($\alpha = e$) events. Although the general trend of the oscillation patterns looks very similar between \gfig{fig:Ni2N} for NH and \gfig{fig:Ni2I} for IH, the magnitude of the coefficients at high energies ($E_\nu > 4~\mbox{GeV}$) are significantly smaller for IH than those for NH. We should therefore expect that the $\delta$ measurement is more difficult for IH than for NH, when higher energy data are used. On the other hand, the $\cos \theta_{\rm z}$ dependence of the sign and the magnitude of the coefficients integrated over the low energy region below $4~\mbox{GeV}$ are similar to those of NH for $\cos \theta_{\rm z} \gtrsim -0.6$. Dedicated studies with realistic energy and angular resolution may reveal the possibility of measuring $\delta$ even in the IH case.

We show the coefficient $N^{(4)}_\mu$ of the cross term $x_{\rm a} \cos \gdelta = x_{\rm a} \sqrt{1 - x^2_{\rm a}} \cos \delta$  and the coefficient $N^{(5)}_\mu$ of the quadratic term $x^2_{\rm a}$ in \gfig{fig:N45}(a) and (b), respectively, for NH [red-solid lines] and IH [blue-dashed lines]. %Both coefficients have small contribution. Although $N^{(4)}_\mu$ can be comparable with $N^{(2)}_\alpha$ and $N^{(3)}_\alpha$ in \gfig{fig:Ni2N} and \gfig{fig:Ni2I}, it is always associated with $x_{\rm a}$ which introduces at least 20\% suppression while $N^{(5)}_\mu$ is itself small already. This $N^{(4)}_\mu$ would lead to correlation between the atmospheric mixing angle $\theta_{\rm a}$ and the CP phase $\delta$ which will be made explicit with $\chi$ analysis in \gsec{sec:fit}. The nonlinear effect induced by $N^{(5)}_\mu$ can be just neglected without affecting the physical picture. 
The magnitude of the coefficient $N^{(4)}_\mu$ is similar to those of $N^{(2)}_\alpha$ and $N^{(3)}_\alpha$ for $\cos \delta$ and $\sin \delta$ in \gfig{fig:Ni2N} and \gfig{fig:Ni2I}, as expected from an order of magnitude estimates in (\ref{eq:order}). Also, its energy dependence is similar to that of $\cos \delta$ coefficient $N^{(2)}_\mu$. We therefore expect that the uncertainty of the $\cos \delta$ measurement depends on the sign of $x_{\rm a}$ when the muon-like events are used in the analysis.

The magnitude of the coefficient $N^{(5)}_\mu$ of $x^2_{\rm a}$ in \gfig{fig:N45}(b) is of the same order of magnitude as that of $N^{(1)}_\alpha$ for the linear term in \gfig{fig:Ni1}. Note that the maximum value of the coefficient $N^{(1)}_\mu$ of $x_{\rm a}$ in \gfig{fig:Ni1}(a) is around $140$, $250$, $690$, and $570$ for $\cos \theta_{\rm z} = -1$, $-0.9$, $-0.8$, and $-0.6$, respectively, for NH, whereas those of the IH are smaller. On the other hand, the maximal values of the $x^2_{\rm a}$ coefficient $N^{(5)}_\mu$ in \gfig{fig:N45}(b) are $1800$, $1870$, $1264$, and $1050$, for the same region. Around the peak of $N^{(5)}_\mu$, its magnitude is typically more than one order of magnitude larger than that of $N^{(1)}_\mu$, the term $N^{(5)}_\mu x^2_{\rm a}$ can be as sensitive to $x_{\rm a}$ as the term $N^{(1)}_\mu x_{\rm a}$ even when $x_{\rm a} \sim 0.2$.

\section{A Simple $\chi^2$ Analysis}
\label{sec:fit}

In this section, we examine the potential sensitivity of the PINGU experiment to the three unknown neutrino oscillation parameters at the neutrino event level. In other words, we assume that both the neutrino energy ($E_\nu$) and its momentum direction ($\cos \theta_{\rm z}$) are measured exactly for each event, and by ignoring uncertainties in the neutrino flux, cross sections and effective fiducial volume, the probability to misidentify $\mu^\pm$ and $e^\pm$ events, as well as backgrounds from $\tau$-decay and neutral current events. Although these assumptions are far from reality, the results are still useful in identifying the maximum information hidden in the data, motivating and directing studies with full detector simulations. Especially, it can demonstrate that the decomposition method in the propagation basis is extremely powerful to reveal the hidden patterns behind the neutrino oscillogram.

\subsection{$\chi^2$ Function}
\label{sec:chi2}

We introduce a conventional $\chi^2$ technique to investigate experimental sensitivities on the neutrino mass hierarchy, the atmospheric mixing angle $\theta_{\rm a}$ and its octant, as well as the CP phase $\delta$. Note that the result of this method, $\chi^2_{\rm min}$ corresponds to the so-called ``average experiment''~\cite{Schwetz:2006md} or ``Asimov data set''~\cite{Asimov}. The uncertainty from statistical fluctuation can be easily estimated to be $\Delta(\chi^2_{\rm min}) \approx 2 \sqrt{\Delta(\chi^2_{\rm min})}$~\cite{uncertainty}. It not only applies to the case of discrete variables such as the neutrino mass hierarchy and the octant, but actually applies generally as long as the binned event number is large enough such that statistical fluctuation can be approximated by Gaussian distribution. Based on these two key parameters, $\chi^2_{\rm min}$ and its variation $\Delta(\chi^2_{\rm min}$, statistical interpretation can be made.

The $\chi^2$ function receives contributions from the statistical uncertainty of the event numbers as functions of the neutrino energy $E_\nu$ and its momentum direction $\cos \theta_{\rm z}$,
\begin{equation}
  \chi^2
\equiv
  \sum_\alpha
  \int d E_\nu d \cos \theta_{\rm z} 
\left[
  \dfrac { \left( \dfrac {dN_\alpha}{d E_\nu d \cos \theta_{\rm z}} \right)^{\rm{th}}
         - \left( \dfrac {dN_\alpha}{d E_\nu d \cos \theta_{\rm z}} \right)^{\rm{obs}}}
    {\sqrt{ \left( \dfrac {dN_\alpha}{d E_\nu d \cos \theta_{\rm z}} \right)^{\rm{obs}}}}
\right]^2
+ \chi^2_{\rm{para}} \,,
\label{eq:chi2-1}
\end{equation}
as well as external constraint on neutrino oscillation parameters which has been denoted as $\chi^2_{\rm{para}}$,
\begin{eqnarray}
  \chi^2_{\rm{para}}
& = &
  \left[ \frac {(\delta m^2_{\rm a})^{\rm{fit}} - \overline{\delta m^2_{\rm a}}}{\Delta \delta m^2_{\rm a}} \right]^2
+ \left[ \frac {(\delta m^2_{\rm s})^{\rm{fit}} - \overline{\delta m^2_{\rm s}}}{\Delta \delta m^2_{\rm s}} \right]^2
\nonumber
\\
& + &
  \left[ \frac {(\sin^2 2 \theta_{\rm r})^{\rm{fit}} - \overline{\sin^2 2 \theta_{\rm r}}}{\Delta \sin^2 2 \theta_{\rm r}} \right]^2
+ \left[ \frac {(\sin^2 2 \theta_{\rm s})^{\rm{fit}} - \overline{\sin^2 2 \theta_{\rm s}}}{\Delta \sin^2 2 \theta_{\rm s}} \right]^2
+ \left[ \frac {(\sin^2 2 \theta_{\rm a})^{\rm{fit}} - \overline{\sin^2 2 \theta_{\rm a}}}{\Delta \sin^2 2 \theta_{\rm a}} \right]^2 \,,
\label{eq:chi2para}
\end{eqnarray}
with the current central values and expected uncertainties in the near future~\cite{PDG2012,Machado:2011ar,MINOS,global},
\begin{subequations}
\begin{eqnarray}
&&
  \delta m^2_{\rm a} = 2.35 \pm 0.1 \times 10^{-3} \mbox{eV}^2 \,,
\qquad
  \delta m^2_{\rm s} = 7.50 \pm 0.2 \times 10^{-5} \mbox{eV}^2 \,,
\qquad
  \sin^2 2 \theta_{\rm s} = 0.857 \pm 0.024 \,,
\\
&&
  \sin^2 2 \theta_{\rm r} = 0.098 \pm 0.005 \,,
\qquad
  \sin^2 2 \theta_{\rm a} = 0.957 \pm 0.030 \,.
\end{eqnarray}
\label{eq:inputs}
\end{subequations}
\hspace{-2mm} We generate events with the mean values of the parameters in (\ref{eq:inputs}) for one of the mass hierarchies, except for $x_{\rm a} = \cos^2 \theta_{\rm a} - \sin^2 \theta_{\rm a}$, for which we examine three input values $\pm 0.2$ and $0$, which are consistent with the present constraints. As for the CP phase $\delta$, we examine four cases, $0$, $\pi$, and $\pm \pi/2$. We then use MINUIT~\cite{minuit} to find the minimum of the $\chi^2$ function by varying all the six parameters, $\delta m^2_{\rm a}$, $\delta m^2_{\rm s}$, $\sin^2 2 \theta_{\rm s}$, $\sin^2 2 \theta_{\rm r}$, $x_{\rm a}$, and $\delta$.  %Since $\delta m^2_{\rm s}$ and $\theta_{\rm s}$ do not have a significant effect on atmospheric neutrino oscillations through Earth, they would not affect the $\chi^2$ minimum by much. Nevertheless, for completeness, the following minimizations are carried out over all these five variables by MINUIT~\cite{minuit}.

Not all the information in atmospheric neutrino mixing pattern can be retrieved after reconstructing the events.  The mixing pattern with low energy and small $|\cos \theta_{\rm z}|$ will be lost due to smearing and detector resolution.  To see how this would affect the result, we will apply simple event cuts on neutrino energy $E_\nu$ and the zenith angle $\theta_{\rm z}$ when presenting the results below. 

The dependence on the true values of the neutrino oscillation parameters is consistent with those in the previous studies \cite{Abe:2011ts,Ghosh:2012px,Winter:2013ema,Blennow:2013vta}, except for the $x_{\rm a}$-dependence of the hierarchy sensitivity as explained in \gsec{sec:mass}.

\subsection{The Mass Hierarchy}
\label{sec:mass}

As shown in \gfig{fig:S0}, the neutrino mass hierarchy can be determined by observing the MSW resonances due to the Earth matter effect which occurs only for NH in neutrino oscillations and for IH in antineutrino oscillations. Although the differences are partially cancelled for a detector like PINGU which is incapable of distinguishing neutrino from antineutrino, it is still possible to determine the neutrino mass hierarchy with atmospheric neutrino oscillation because of incomplete cancellation. The hierarchy can be defined as,
\begin{equation}
  \Delta \chi^2_{\tiny \rm MH}
\equiv
  | \chi^2_{\rm min} (NH) - \chi^2_{\rm min} (IH)| \,,
\end{equation}
where the $\chi^2$ minimum is obtained by setting the neutrino mass hierarchy
to be normal (NH) or inverted (IH). 

\begin{table}[h]
\centering
\begin{tabular}{c|c||ccc|ccc}
\multicolumn{2}{c||}{\multirow{2}{*}{$\Delta \chi^2_{\tiny \rm MH}$}}
& \multicolumn{3}{c|}{NH} & \multicolumn{3}{c}{IH} \\
\multicolumn{2}{c||}{} & $\bar x_{\rm a} = - 0.2$ & $\bar x_{\rm a} = 0.0$ & $\bar x_{\rm a} = + 0.2$ & $\bar x_{\rm a} = - 0.2$ & $\bar x_{\rm a} = 0.0$ & $\bar x_{\rm a} = + 0.2$ \\
\hline\hline
\multirow{4}{*}{$\mu$-like} & $\bar \delta = 0^\circ$   & 50.5 & 49.5 & 40.4        & 51.0       & 60.6 & 52.0 \\
                            & $\bar \delta = 90^\circ$  & 48.5 & 51.3 & 39.2        & 50.1       & 60.8 & 52.0 \\
                            & $\bar \delta = 180^\circ$ & 44.5 & 44.1 & {\bf 35.4}  & {\bf 48.2} & 57.8 & 50.4 \\
                            & $\bar \delta = 270^\circ$ & 46.3 & 44.6 & 36.2        & 49.0       & 58.0 & 50.4 \\
\hline
\multirow{4}{*}{$\mu$+e-like} & $\bar \delta = 0^\circ$ & 112.1 & 78.7 & 52.7        & 79.0 & 82.7 & 62.2 \\
                            & $\bar \delta = 90^\circ$  & 102.8 & 73.0 & 50.1        & 78.2 & 81.0 & 60.2 \\
                            & $\bar \delta = 180^\circ$ &  98.6 & 68.6 & {\bf 43.4}  & 74.2 & 78.2 & {\bf 58.4} \\
                            & $\bar \delta = 270^\circ$ & 107.6 & 74.1 & 45.4        & 75.0 & 80.0 & 60.2 \\
\end{tabular}
\caption{The dependence of hierarchy sensitivity $\Delta \chi^2_{\tiny \rm MH}$ on the input values of the atmospheric angle's deviation, $\bar x_{\rm a}$, and the CP phase, $\bar \delta$, with 1-year running of PINGU. The cases of both NH and IH, muon- and electron-like events have been considered with event cut $E_\nu > 6~\mbox{GeV}$ and $\cos \theta_{\rm z} < -0.4$.}
\label{tab:true}
\end{table}
 
Since the CP phase $\delta$ and the atmospheric mixing angle $\theta_{\rm a}$ have not been pinned down yet, their values would affect the distinguishability of the mass hierarchy. The dependence of the hierarchy sensitivity $\Delta \chi^2_{\tiny \rm MH}$ on the input values of $\delta$ and the parameter $x_{\rm a} = \cos^2 \theta_{\rm a} - \sin^2 \theta_{\rm a}$ is summarized in \gtable{tab:true}. Four typical cases of $\bar \delta = \frac n 2 \pi$ with n = 0, 1, 2, 3 respectively and three possibilities for $\bar x_{\rm a} = \pm 0.2, 0.0$ have been shown for both NH in the left and IH in the right. Since muon-like events are easier to be measured, we first show the results with only muon-like events in the upper part and then also include the electron-like events in the lower part. The results in \gtable{tab:true} are obtained with the event cuts $E_\nu > 6~\mbox{GeV}$ and $\cos \theta_{\rm z} < - 0.4$ which will be detailed below. Even with this limited parameter space, the hierarchy sensitivity $\Delta \chi^2_{\tiny \rm MH}$ is sizable, being larger than $35$ for NH and $48$ for IH for all cases of input values for $\delta$ and $\theta_{\rm a}$, under the assumption of a perfect detector. The smallest value of $\Delta \chi^2_{\tiny \rm MH}$ in each block has been marked as black bold numbers. 

With only muon-like events, the hierarchy sensitivity is comparable between NH
and IH. This is because in the considered energy range, $E_\mu > 6 \, \mbox{GeV}$,
the zeroth order event rates $N^{(0)}_\mu$ alternate without preference for 
neither NH or IH. If the
energy cut is lowered down to $E_\nu > 4 \, \mbox{GeV}$, the event rate for IH
dominates over the one with NH, rendering larger hierarchy sensitivity for NH.
The situation does not change much when the electron-like events are included.
This can be seen by comparing the results with $\bar x_{\rm a} = 0$ where 
complication for the $x_{\rm a}$ terms are absent. 

For all four blocks, the hierarchy sensitivity is larger with 
$\bar \delta = 0^\circ$ than the result with $\bar \delta = 180^\circ$. 
This is because the in the considered
energy range $E_\nu > 6 \, \mbox{GeV}$, the coefficient $N^{(2)}_\mu$ of 
$\cos \delta'$ dominates. In addition, it is mainly negative for NH and 
positive for IH, as shown in \gfig{fig:Ni2N}. 
With a positive $\cos \delta'$, the difference between NH 
and IH is enlarged when $N^{(2)}_\mu \cos \delta'$ is added to the zeroth order
$N^{(0)}_\mu$.

The dependence on $x_{\rm a}$ is a little more complicated due to the presence 
of both the linear and quadratic terms. Let us first compare the results with
$\bar x_{\rm a} = \pm 0.2$ which have the same contribution from the quadratic 
term. For NH, the hierarchy sensitivity is larger for negative $\bar x_{\rm a}$.
This is because in most part of the considered parameter region, especially in
the regions $\cos \theta_{\rm z} \gtrsim -0.7$ and 
$\cos \theta_{\rm z} \lesssim -0.9$, negative 
$x_{\rm a}$ makes the difference between NH and IH larger. This trend remains
when the energy cut is lowered down to $E_\nu > 4 \, \mbox{GeV}$ and is 
enhanced when electron-like events are included.
The contribution from the quadratic term $N^{(5)}_\mu$ is always positive and
its hierarchy dependence is the opposite to that of $N^{(0)}_\mu$, leading to
a negative effect. If only linear term of $x_{\rm a}$ is present, the 
dependence is monotonically decreasing and is lifted by the quadratic term
when $\bar x_{\rm a}$ vanishes. For IH, the linear term coefficient $N^{(1)}_\mu$
is much smaller reducing the difference between $\bar x_{\rm a} = \pm 0.2$
and making the increase at $\bar x_{\rm a} = 0$ prominent. Note that the 
dependence on $\bar x_{\rm a}$ is different from the results in 
\cite{Abe:2011ts,Ghosh:2012px,Winter:2013ema,Blennow:2013vta} due to detector responses.

In the upper-left block of \gtable{tab:true} for NH with only muon-like events, the dependence on $\delta$ is smaller than that on $x_{\rm a}$. For each row with fixed input value of $\delta$, the variation is around $9 \sim 12$, while for each column with fixed input value of $x_{\rm a}$, it is around $5 \sim 6$. This property applies for all the other three blocks. In the upper-right block for IH with only muon-like events, the variant in rows is around $9 \sim 10$, but the variant in columns is much smaller being around $2 \sim 3$. Such trends are expected since the magnitude of the observable coefficients of $x_{\rm a}$, namely $N^{(1)}_\mu$, is much larger than those of $\cos \gdelta$ and $\sin \gdelta$, $N^{(2)}_\mu$ and $N^{(3)}_\mu$, respectively. The variation in $x_{\rm a}$ further increases after electron-like events are included since the ratio of coefficients $N^{(1)}_{\rm e}/N^{(0)}_{\rm e}$ is more significant than  $N^{(1)}_\mu/N^{(0)}_\mu$ as shown in \gfig{fig:Ni1}. The $x_{\rm a}$-dependence of $\chi^2_{\tiny \rm MH}$ is consistently larger for NH than for IH, because the coefficient $N^{(1)}_\alpha$ of $x_{\rm a}$ is larger for NH than for IH, as shown in \gfig{fig:Ni1}. It is remarkable that when electron-like events are included in the analysis, the hierarchy distinguishing power increases significantly for $x_{\rm a} = - 0.2$, but not much for $x_{\rm a} = +0.2$. This is because of the negative sign of $N^{(1)}_{\rm e}$, shown in \gfig{fig:Ni1}(b), which enlarges the hierarchy dependence of the event rate for negative $x_{\rm a}$. 

Although the absolute magnitude of $\Delta \chi^2_{\tiny \rm MH}$ in \gtable{tab:true} for a perfect detector without systematic uncertainty do not have much significance, the relative importance of electron-like events and possible impacts of the $x_{\rm a}$ value in the hierarchy determination may want further studies. Note that the neutrino mass hierarchy can be resolved no matter what true values of the atmospheric angle and the CP phase can be, in contrast to the CP-hierarchy and octant-hierarchy degeneracies from which accelerator based neutrino experiments suffer \cite{degeneracy}.

\begin{table}[h]
\centering
\begin{tabular}{c|c||ccc|ccc}
\multicolumn{2}{c||}{\multirow{2}{*}{$\Delta \chi^2_{\tiny \rm MH}$}}
& \multicolumn{3}{c|}{NH}
& \multicolumn{3}{c}{IH} \\
\multicolumn{2}{c||}{} & $\cos \theta_{\rm z} \hspace{-1mm} < \hspace{-1mm} - 0.2$ & $\cos \theta_{\rm z} \hspace{-1mm} < \hspace{-1mm} - 0.4$ & $\cos \theta_{\rm z} \hspace{-1mm} < \hspace{-1mm} - 0.6$ & $\cos \theta_{\rm z} \hspace{-1mm} < \hspace{-1mm} - 0.2$ & $\cos \theta_{\rm z} \hspace{-1mm} < \hspace{-1mm} - 0.4$ & $\cos \theta_{\rm z} \hspace{-1mm} < \hspace{-1mm} - 0.6$ \\
\hline\hline
\multirow{4}{*}{$\mu$-like} & $E_\nu \hspace{-1mm} > \hspace{-1mm} 2~\mbox{GeV}$ 
& 259.8 & 255.6      & 249.2 & 131.2 & 129.7       & 123.6 \\
                             & $E_\nu \hspace{-1mm} > \hspace{-1mm} 4~\mbox{GeV}$ 
& 155.5 & 152.4      & 132.2 &  98.9 &  98.1        & 91.2 \\
                             & $E_\nu \hspace{-1mm} > \hspace{-1mm} 6~\mbox{GeV}$ 
&  39.2 & {\bf 35.4} & 28.5  &  49.9 &  {\bf 48.2} &  40.7 \\
                             & $E_\nu \hspace{-1mm} > \hspace{-1mm} 8~\mbox{GeV}$ 
&   9.3 & 8.3        & 4.4   &  15.4 &  13.2       &   6.2 \\
\hline
\multirow{4}{*}{$\mu$+e-like} & $E_\nu \hspace{-1mm} > \hspace{-1mm} 2~\mbox{GeV}$ 
& 281.4 & 277.0      & 267.6 & 142.5 & 139.1       & 136.2 \\
                             & $E_\nu \hspace{-1mm} > \hspace{-1mm} 4~\mbox{GeV}$  
& 176.1 & 171.6      & 150.1 & 117.6 & 115.6       & 112.6 \\
                             & $E_\nu \hspace{-1mm} > \hspace{-1mm} 6~\mbox{GeV}$ 
&  51.8 & {\bf 43.4} & 34.7  & 61.1  & {\bf 58.4}  & 53.0 \\
                             & $E_\nu \hspace{-1mm} > \hspace{-1mm} 8~\mbox{GeV}$  
&  11.6 & 9.5        & 4.7   & 20.7  &  18.8       &  13.5 \\
\end{tabular}
\caption{The dependence of the hierarchy sensitivity $\Delta \chi^2_{\tiny \rm MH}$ on event selection cuts for 1-year running of PINGU. The cases of NH and IH, $\mu$ and e+$\mu$-like events respectively have been considered. The input values of $x_{\rm a}$ and $\delta$ are taken as those in Table~\ref{tab:true} that produces the smallest value of $\Delta \chi^2_{\tiny \rm MH}$. Namely, $(x_{\rm a}, \delta) = (-0.2, 180^\circ)$ for $\mu$-like events with IH, and $(x_{\rm a}, \delta) = (+0.2, 180^\circ)$ for the others.}
\label{tab:cuts}
\end{table}

Those events at low energy and/or small $|\cos \theta_{\rm z}|$ will be largely smeared out, and oscillating features may be averaged out. This is because the energy smearing mainly comes from neutrino scattering, which is expected to scale as a linear function $\delta E \propto E$, and statistical fluctuation, which scales as $\delta E \propto \sqrt E$. On the other hand, the neutrino oscillation period shrinks quickly at low energy, approximately as a quadratic function $\Delta E \propto E^2$. No oscillation signal can be expected to survive below some energy threshold. For angular resolution, $\delta \theta_{\rm z}$, it should be roughly a constant in the neutrino frame. When converted to the parameter in phase space, the resolution, $\delta(\cos \theta_{\rm z}) = \sin \theta_{\rm z} \delta \theta_{\rm z}$, is much larger for horizontal events, $\sin \theta_{\rm z} \approx 1$. Hence, these regions may not contribute much in more realistic studies, and can be omitted by simply applying event selection cuts $E_\nu > E^{\rm{cut}}_\nu$ and $\cos \theta_{\rm z} < \cos \theta^{\rm{cut}}_{\rm z}$.  

The dependences on $E^{\rm{cut}}_\nu$ and $\cos\theta_{\rm z}^{\rm{cut}}$ are shown in \gtable{tab:cuts} where the input values of $\delta$ and $x_{\rm a}$ are chosen corresponding to the smallest value of $\Delta \chi^2_{\tiny \rm MH}$ in each block of \gtable{tab:true} respectively.  We observe that the results depend strongly on the range of $E_\nu$ and $\cos \theta_{\rm z}$, which can be effectively analyzed in real experiments. For instance, when $E^{{\rm cut}}_\nu$ is raised from $6~\mbox{GeV}$ to $8~\mbox{GeV}$, $\Delta \chi^2_{\tiny \rm MH}$ drops by nearly a factor of $3 \sim 7$, while for $E_\nu > 8~\mbox{GeV}$, changing of the zenith angle coverage from $\cos \theta_{\rm z} < -0.4$ to $\cos \theta_{\rm z} < -0.6$ reduces $\Delta \chi^2_{\tiny \rm MH}$ by further factor of $2$. It is therefore very important to have low energy threshold of the detector and to study the smearing effects in detail.

\subsection{The Atmospheric Angle and Its Octant}
\label{sec:xa}

Once the mass hierarchy is determined, $\chi^2$ fit can be performed with the correct mass hierarchy as an input, and the same $\chi^2$ function (\ref{eq:chi2-1}) can be used to measure the atmospheric mixing angle and its octant discussed here, as well as the CP phase $\delta$ which will be discussed in \gsec{sec:CP}. 
In order to make the global minimum to be at the input value of $x_{\rm a} = \bar x_{\rm a} = x^{\rm{input}}_{\rm a}$, we modify the last term in $\chi^2_{\rm para}$ (\ref{eq:chi2para}) as follows,
\begin{equation}
\left[ 
  \frac { \left( \sin^2 2 \theta_{\rm a} \right)^{\rm fit} - \overline{\sin^2 2 \theta_{\rm a}}}
				{ \Delta \sin^2 2 \theta_{\rm a} }
\right]^2
\rightarrow 
\left[
  \frac {x^2_{\rm a} - \bar x^2_{\rm a}}
				{0.03}
\right]^2 \,,
\label{eq:xa-constraint}
\end{equation}
which keeps the uncertainty in $\sin^2 2 \theta_{\rm a}$ the same as in (\ref{eq:inputs}) while shifting the mean value to the input, $\sin^2 2 \theta_{\rm a} = 1 - \bar x^2_{\rm a}$. This avoids small but inessential dependence of the $x_{\rm a}$ measurement on the input values of $x_{\rm a}$ in the region $|x_{\rm a}| < 0.2$.

\subsubsection{the Atmospheric Angle}

We compute the minimum of the $\chi^2$ function as a function of the parameter $x_{\rm a}$, by varying all the other 5 parameters with MINUIT2~\cite{minuit}, to obtain the region, $\chi^2_{\rm{min}}(x_{\rm a}) \leq 1$, whose half-width is defined as the expected uncertainty of the $x_{\rm a}$ measurement, $\Delta(x_{\rm a})$ or $\Delta(x^2_{\rm a})$. The results for NH and IH are listed in the left and right panels of \gtable{tab:dxa}, respectively. For each hierarchy, four input values $\bar \delta = 0^\circ, 90^\circ, 180^\circ, 270^\circ$ of the CP phase $\delta$ and three input values $\bar x_{\rm a} = 0.0, \pm 0.2$ for the atmospheric angle $\theta_{\rm a}$ have been tested. Both the uncertainty of $x_{\rm a}$, $\Delta(x_{\rm a})$, and that of $x^2_{\rm a}$, $\Delta(x^2_{\rm a})$, are given in \gtable{tab:dxa}. %We checked that the quadratic approximation of (\ref{eq:xa-constraint}) is valid in the $\pm 3 \sigma$ range.
\begin{table}[h]
\centering
\begin{tabular}{c|c||ccc|ccc}
\multicolumn{2}{c||}{\multirow{2}{*}{$\mu$+e-like}} & \multicolumn{3}{c|}{NH} & \multicolumn{3}{c}{IH} \\
\multicolumn{2}{c||}{} 
 & $\bar x_{\rm a} = - 0.2$ & $\bar x_{\rm a} = 0.0$ & $\bar x_{\rm a} = + 0.2$
 & $\bar x_{\rm a} = - 0.2$ & $\bar x_{\rm a} = 0.0$ & $\bar x_{\rm a} = + 0.2$ \\
\hline\hline
\multirow{4}{*}{$\Delta(x_{\rm a})$} 
& $\bar \delta = 0^\circ$   & 0.013 & 0.033 & 0.012 & 0.014 & 0.035 & 0.012 \\
& $\bar \delta = 90^\circ$  & 0.015 & 0.031 & 0.013 & 0.015 & 0.037 & 0.013 \\
& $\bar \delta = 180^\circ$ & 0.013 & 0.033 & 0.012 & 0.014 & 0.038 & 0.012 \\
& $\bar \delta = 270^\circ$ & 0.015 & 0.031 & 0.013 & 0.015 & 0.038 & 0.013 \\
\hline
\multirow{4}{*}{$\Delta(x^2_{\rm a})$} 
& $\bar \delta = 0^\circ$   & 0.0053 & 0.0011 & 0.0046 & 0.0055 & 0.0013 & 0.0048 \\
& $\bar \delta = 90^\circ$  & 0.0060 & 0.0010 & 0.0051 & 0.0059 & 0.0014 & 0.0053 \\
& $\bar \delta = 180^\circ$ & 0.0052 & 0.0011 & 0.0048 & 0.0055 & 0.0015 & 0.0049 \\
& $\bar \delta = 270^\circ$ & 0.0060 & 0.0010 & 0.0051 & 0.0059 & 0.0014 & 0.0053
\end{tabular}
\caption{The dependence of the uncertainty of $x_{\rm a}$ and $x^2_{\rm a}$, corresponding to $\chi^2_{\rm{min}}(x_{\rm a}) \leq 1$ and $\chi^2_{\rm{min}}(x^2_{\rm a}) \leq 1$, respectively, on the input values $\bar x_{\rm a}$ and $\bar \delta$ after 1-year running of PINGU. The case of both muon- and electron-like events for both NH and IH has been considered with event cuts $E_\nu > 6~\mbox{GeV}$ and $\cos \theta_{\rm z} < -0.4$.}
\label{tab:dxa}
\end{table}

Note that the uncertainty $\Delta(x_{\rm a})$ is larger if the atmospheric mixing angle is maximal, or when $\bar x_{\rm a} = 0$ ($\sin^2 2 \theta_{\rm a} = 1$). 
This is a consequence of relatively large coefficient $N^{(5)}_\mu$ of $x^2_{\rm a}$, shown in \gfig{fig:N45}(b), as compared to the coefficients $N^{(1)}_\mu$ and $N^{(1)}_{\rm e}$ of $x_{\rm a}$, shown in \gfig{fig:Ni1} (a) and (b), respectively. The $x_{\rm a}$-dependence of the muon-like events are expected to be,
\begin{equation}
  x_{\rm a} N^{(1)}_\mu
+ x^2_{\rm a} N^{(5)}_\mu 
+ \cdots \,,
\end{equation}
ignoring the other small terms. The variation of the number of events at $x_{\rm a} = \bar x_{\rm a}$ is then,
\begin{equation}
\left(
  N^{(1)}_\mu
+ 2 \bar x_{\rm a} N^{(5)}_\mu
\right) \delta x_{\rm a} 
+ \cdots \,.
\label{eq:effLinear}
\end{equation}
Since the $N^{(5)}_\mu$ term can dominate over $N^{(1)}_\mu$ and $N^{(1)}_{\rm e}$ for $|\bar x_{\rm a}| = 0.2$, and it vanishes for $\bar x_{\rm a} = 0$, the combined effective coefficient of $\delta x_{\rm a}$ has larger magnitude for nonzero $\bar x_{\rm a}$. This explains the reduced uncertainty $\Delta(x_{\rm a})$ for $x_{\rm a} = \pm 0.2$. The slight difference between the two mirror cases $\bar x_{\rm a} = - 0.2$ and $\bar x_{\rm a} = 0.2$ comes from the first term. Since $N^{(1)}_\mu$ is positive, cancellation happens in the combined effective coefficient when $\bar x_{\rm a}$ is negative, leading to systematically larger $\Delta(x_{\rm a})$. There is some slight dependence on the input value $\bar \delta$ of the CP phase, but not sizable. 

To make a direct comparison with the external constraint (\ref{eq:xa-constraint}), the resolution $\Delta(x^2_{\rm a})$ is also shown in \gtable{tab:dxa}. Since $\sin^2 2 \theta_{\rm a} \equiv 1 - x_{\rm a}^2$, the uncertainty $\Delta(\sin^2 2 \theta_{\rm a})$ is exactly $\Delta(x^2_{\rm a})$. For $x_{\rm a} \approx |0.2|$, it can be roughly estimated as $\Delta(x^2_{\rm a}) \approx (2 x_{\rm a}) \Delta(x_{\rm a}) \approx 0.4 \Delta(x_{\rm a})$ which is typically a factor of $5 \sim 6$ smaller than the uncertainty of $0.03$ given in (\ref{eq:inputs}) and (\ref{eq:xa-constraint}).  If the true value of $x_{\rm a}$ vanishes, its uncertainty should be estimated as $\Delta(x^2_{\rm a}) \approx [\Delta(x_{\rm a})]^2$ which is roughly $0.001$ for NH, which is smaller than the current uncertainty by a factor of $30$. The uncertainties $\Delta(x_{\rm a})$ and $\Delta(x^2_{\rm a})$ are slightly larger for IH due to the smaller coefficients $N^{(1)}_\alpha$ as shown in \gfig{fig:Ni1}. Summing up, a neutrino telescope like PINGU has a potential to resolve the octant degeneracy of the atmospheric mixing angle $\theta_{\rm a}$, and to reduce the uncertainty of $\sin^2 2 \theta_{\rm a}$ by a factor of $5$ to $30$ within one year of running. Although our simulation does not take account of energy and zenith angle resolutions, we expect this high potential to be confirmed in more realistic simulations because the coefficients $N^{(1)}_\mu$ and $N^{(1)}_{\rm e}$ do not oscillate much with $E_\nu$ or $\cos \theta_{\rm z}$. 
 %In other words, PINGU can largely enhance the resolution on the atmospheric angle by roughly one order of magnitude.  

\subsubsection{Octant Sensitivity}

If the linear terms $N^{(1)}_\alpha$ of $x_{\rm a}$ vanish, no difference would
be observed when $x_{\rm a}$ switches its sign. Fortunately, the nonzero 
$N^{(1)}_\mu$ and $N^{(1)}_{\rm e}$ provide us the possibility of determining
the octant of the atmospheric mixing angle. The octant sensitivity can be 
defined as,
\begin{equation}
  \Delta \chi^2_{\tiny \rm octant}
\equiv
  |\chi^2_{\rm min}(x_{\rm a} > 0) - \chi^2_{\rm min}(x_{\rm a} < 0)| \,,
\end{equation}
where the two $\chi^2_{\rm min}$ are obtained by restricting the atmospheric 
mixing angle in the higher or lower octant, respectively.
The results with different input values, $\bar x_{\rm a} = \pm 0.2, \pm 0.1$
and $\bar \delta = 0^\circ, 90^\circ, 180^\circ, 270^\circ$, have been shown 
in \gtable{tab:octant} for both NH and IH.

\begin{table}[h]
\centering
\begin{tabular}{c|c||cccc|cccc}
\multicolumn{2}{c||}{$\Delta \chi^2_{\rm \tiny octant}$} & \multicolumn{4}{c|}{NH} & \multicolumn{4}{c}{IH} \\[1mm]
\hline
\multicolumn{2}{c||}{$\bar x_{\rm a}$} 
 & $- 0.2$ & $- 0.1$ & $+ 0.1$ & $+ 0.2$
 & $- 0.2$ & $- 0.1$ & $+ 0.1$ & $+ 0.2$ \\
\hline\hline
\multirow{4}{*}{$\mu$--like} 
& $\bar \delta = 0^\circ$   & 25.3 & 5.2 & 9.4 & 29.7 & 5.4 & 1.4 & 2.0 & 6.3 \\
& $\bar \delta = 90^\circ$  & 23.9 & 5.1 & 9.4 & 29.6 & 5.7 & 1.6 & 1.7 & 5.8 \\
& $\bar \delta = 180^\circ$ & 25.3 & 6.4 & 7.9 & 27.4 & 5.4 & 1.4 & 1.9 & 5.9 \\
& $\bar \delta = 270^\circ$ & 26.7 & 6.5 & 7.5 & 27.0 & 5.2 & 1.3 & 2.2 & 6.3 \\
\hline
\multirow{4}{*}{$\mu$+e--like} 
& $\bar \delta = 0^\circ$   & 63.3 &  9.2 & 18.9 & 94.7 & 16.7 & 3.8 & 6.2 & 23.5 \\
& $\bar \delta = 90^\circ$  & 62.0 &  9.3 & 18.4 & 98.6 & 18.3 & 4.3 & 5.1 & 20.9 \\
& $\bar \delta = 180^\circ$ & 68.4 & 13.3 & 15.6 & 91.7 & 18.1 & 4.1 & 5.8 & 21.4 \\
& $\bar \delta = 270^\circ$ & 69.9 & 12.8 & 16.2 & 87.1 & 16.4 & 3.5 & 6.8 & 23.8
\end{tabular}
\caption{The dependence of the octant sensitivity on the input values $\bar x_{\rm a}$ and $\bar \delta$ after 1-year running of PINGU. The case of both muon- and electron-like events for both NH and IH has been considered with event cuts $E_\nu > 6~\mbox{GeV}$ and $\cos \theta_{\rm z} < -0.4$.}
\label{tab:octant}
\end{table}

We can see that the octant sensitivity for $\bar x_{\rm a} = \pm 0.1$ is much
smaller than the one for $\bar x_{\rm a} = \pm 0.2$, as expected. With smaller
distance between the two mirrors, the difference due to the linear term is 
much smaller. And we can estimate the significance to scale roughly as 
$\Delta \chi^2_{\tiny \rm octant} \propto x^2_{\rm a}$, according to 
(\ref{eq:chi2-1}), as verified by the 
results in \gtable{tab:octant}. 

Between the two mirrors, the octant sensitivity
is always larger for positive $x_{\rm a}$ due to the same sign between 
$N^{(1)}_\mu$ and $N^{(5)}_\mu$. This makes the effective linear term 
coefficient in (\ref{eq:effLinear}) larger with positive $\bar x_{\rm a}$,
hence enhances the octant sensitivity. This trend is further enhanced by
including the electron-like events which has only a negative linear term 
coefficient $N^{(1)}_{\rm e}$ but not quadratic term. With positive 
$\bar x_{\rm a}$, the event rates become smaller, explaining the further 
enhancement.

The dependence on the neutrino mass hierarchy is much easier to be understood. 
For NH, the octant sensitivity is much larger than that for IH,
because the linear term coefficients $N^{(1)}_\mu$ and $N^{(1)}_{\rm e}$
have much larger magnitude for NH. 

There is small dependence on the CP phase. It comes from the cross term 
$x_{\rm a} \cos \delta'$ whose coefficient $N^{(4)}_\mu$ is mainly positive 
for NH and negative for IH in the considered energy range 
$E_\nu > 6 \, \mbox{GeV}$, especially around $\cos \theta_{\rm z} \approx -0.6$,
as shown in \gfig{fig:N45}(a).
The effective linear term in (\ref{eq:effLinear}) becomes,
\begin{equation}
\left( 
	N^{(1)}_\mu 
+ \cos \bar \delta' N^{(4)}_\mu
+ 2 \bar x_{\rm a} N^{(5)}_\mu 
\right) \delta x_{\rm a} 
+
  \cdots \,.
\end{equation}
With nonzero $\bar x_{\rm a}$, especially when $\bar x_{\rm a} \approx \pm 0.2$,
the quadratic term coefficient $N^{(5)}_\mu$ dominates. Cancellation between 
$\cos \bar \delta' N^{(4)}_\mu$ and $2 \bar x_{\rm a} N^{(5)}_\mu$ happens if
they have opposite signs, leading to smaller octant sensitivity as shown in 
\gtable{tab:octant}. For NH, the octant sensitivity is larger for 
$\cos \bar \delta \lesssim 0$ when $\bar x_{\rm a}$ is negative and
$\cos \bar \delta \gtrsim 0$ when $\bar x_{\rm a}$ is positive. It is the 
opposite for IH. This trend remains when the electron-like events are also 
included in the analysis, since the cross term coefficient $N^{(4)}_{\rm e}$ 
for the electron-like event rates is zero, although the 
sensitivity can be enhanced a lot. Note that the cross term does 
not have sizable effect on the uncertainty of measuring the atmospheric mixing
angle but manifests itself in the octant sensitivity.

Since there is not so much oscillation behavior in $N^{(1)}_\alpha$, 
$N^{(4)}_\mu$, and $N^{(5)}_\mu$, these trends in the octant sensitivity 
$\Delta \chi^2_{\tiny \rm Octant}$, obtained with neutrino events without
full simulation, should survive the smearing effects from the neutrino 
scattering and detector resolutions.

\subsection{Uncertainty of the CP Phase}
\label{sec:CP}

In this section, we will show the capability of PINGU in measuring the CP phase $\delta$ in terms of $\chi^2_{\rm min}(\delta)$. The dependence of $\chi^2_{\rm{min}}(\delta)$ on $\delta$ comes from fixing $\delta$ and fitting the other five parameters to find the minimum. The uncertainty $\Delta(\delta)$ are then obtained according to the condition $\chi^2_{\rm min}(\delta) < 1$ for each case of the input values, $\bar x_{\rm a} = 0, \pm 0.2$, and $\bar \delta = 0, \pm \pi / 2, \pi$, as well as both NH and IH. Since the $\delta$-dependent terms have tiny coefficients, $N^{(2)}_\alpha$ and $N^{(3)}_\alpha$, as shown in \gfig{fig:Ni2N} and \gfig{fig:Ni2I}, it is very challenging to determine the CP phase and can only be possible with a much longer time. We check the results after 10-years running of PINGU with both muon- and electron-like events. 

As explained in \gsec{sec:propagation-basis} and \gsec{sec:rates}, the event rates depend on the CP phase $\delta$ only through the terms proportional to $\cos \delta$ and $\sin \delta$, since the coefficient $N^{(6)}_\mu$ of $\cos^2 \delta$ in the expansion (\ref{eq:Ni-decomposition}) is negligibly small. Therefore, the $\delta$-dependence of $\chi^2_{\rm min}(\delta)$ can be approximated as a quadratic function of $\cos \delta$ and $\sin \delta$,
\begin{subequations}
\begin{eqnarray}
  \chi^2_{\rm{min}}(\delta)
& = &
  \left\lgroup
  \begin{matrix}
    \cos \delta - \cos \bar \delta
  & \sin \delta - \sin \bar \delta 
  \end{matrix}
  \right\rgroup 
  V^{-1}
  \left\lgroup
  \begin{matrix}
    \cos \delta - \cos \bar \delta \\
    \sin \delta - \sin \bar \delta 
  \end{matrix}
  \right\rgroup
+ 
  \mathcal O( (\delta - \bar \delta)^3 )
\\
& = &
  \left\lgroup
  \begin{matrix}
  - \sin \bar \delta
  & \cos \bar \delta 
  \end{matrix}
  \right\rgroup 
  V^{-1}
  \left\lgroup
  \begin{matrix}
  - \sin \bar \delta \\
    \cos \bar \delta 
  \end{matrix}
  \right\rgroup
  (\delta - \bar \delta)^2
+
  \mathcal O( (\delta - \bar \delta)^3 )
\\
& = &
  \left[ \frac {\delta - \bar \delta}{\Delta(\delta)} \right]^2
+
  \mathcal O( (\delta - \bar \delta)^3 ) \,,
\end{eqnarray}
\label{eq:cos-sin-D}
\end{subequations}
\hspace{-2.5mm}
where the covariance matrix $V$ is a $2 \times 2$ real symmetric matrix, and $\Delta(\delta)$ is the uncertainty on the CP phase. For various inputs, the results from exact $\chi^2$ minimization have been shown in \gtable{tab:dCP}.
\begin{table}[h]
\centering
\begin{tabular}{c|c||ccc|ccc}
\multicolumn{2}{c||}{\multirow{2}{*}{$\Delta(\delta)$}} & \multicolumn{3}{c|}{NH} & \multicolumn{3}{c}{IH} \\
\multicolumn{2}{c||}{} & $\bar x_{\rm a} = - 0.2$ & $\bar x_{\rm a} = 0.0$ & $\bar x_{\rm a} = + 0.2$ & $\bar x_{\rm a} = - 0.2$ & $\bar x_{\rm a} = 0.0$ & $\bar x_{\rm a} = + 0.2$ \\
\hline\hline
\multirow{4}{*}{$E_\nu > 6~\mbox{GeV}$} 
& $\bar \delta = 0^\circ$   & $23^\circ$ & $23^\circ$ & $22^\circ$ & $49^\circ$ & $48^\circ$ & $48^\circ$ \\
& $\bar \delta = 90^\circ$  & $20^\circ$ & $20^\circ$ & $18^\circ$ & $41^\circ$ & $41^\circ$ & $38^\circ$ \\
& $\bar \delta = 180^\circ$ & $23^\circ$ & $23^\circ$ & $21^\circ$ & $49^\circ$ & $48^\circ$ & $48^\circ$ \\
& $\bar \delta = 270^\circ$ & $20^\circ$ & $20^\circ$ & $18^\circ$ & $41^\circ$ & $41^\circ$ & $38^\circ$ \\
\hline
\multirow{4}{*}{$E_\nu > 4~\mbox{GeV}$} 
& $\bar \delta = 0^\circ$   & $15^\circ$ & $14^\circ$ & $14^\circ$ & $32^\circ$ & $31^\circ$ & $32^\circ$ \\
& $\bar \delta = 90^\circ$  & $13^\circ$ & $12^\circ$ & $11^\circ$ & $29^\circ$ & $29^\circ$ & $28^\circ$ \\
& $\bar \delta = 180^\circ$ & $15^\circ$ & $14^\circ$ & $14^\circ$ & $32^\circ$ & $31^\circ$ & $32^\circ$ \\
& $\bar \delta = 270^\circ$ & $13^\circ$ & $12^\circ$ & $11^\circ$ & $29^\circ$ & $29^\circ$ & $28^\circ$ 
\end{tabular}
\caption{Dependence of the uncertainty of measuring the CP phase $\delta$ on the input values $\bar x_{\rm a}$ and $\bar \delta$, after 10-years running of PINGU, for the NH (left 3 columns) and for the IH (right 3 columns). Both muon- and electron-like events in the region of $\cos \theta_{\rm z} < -0.4$ are used for $E_\nu > 6~\mbox{GeV}$ (upper 4 rows) and for $E_\nu > 4~\mbox{GeV}$ (lower 4 rows).}
\label{tab:dCP}
\end{table}

Since the coefficients $N^{(2)}_\alpha$ and $N^{(3)}_\alpha$ are much smaller
for IH, as shown in \gfig{fig:Ni2N} and \gfig{fig:Ni2I}, the resultant 
$\Delta(\delta)$ in \gtable{tab:dCP} is expected to be larger. Besides,
there is very slight dependence on the input values of the atmospheric mixing
angle and the CP phase, in contrast to the case at accelerator based neutrino
experiments \cite{degeneracy}. 
Since the $\delta$-dependence of the event rates occurs only through terms of $\cos \delta$ and $\sin \delta$, the $\chi^2_{\rm min}(\delta)$ function is a periodic function of $\delta$. It vanishes at $\delta = \bar \delta$ and has only one period of oscillation in the range of $[0, 2 \pi]$ with peak around $|\delta - \bar \delta| \approx \pi$. With energy cut $E_\nu > 6~\mbox{GeV}$, the maximum value of $\chi^2_{\rm min}(\delta)$ is around $25 \sim 40$ for NH, and just $6 \sim 8$ for IH, depending on the input values of $\bar x_{\rm a}$ and $\bar \delta$. The value increases to around $12 \sim 16$ for IH if we include events down to $E_\nu = 4~\mbox{GeV}$. The function $\chi^2_{\rm min}(\delta)$ can be approximated by a quadratic function (\ref{eq:cos-sin-D}) of $\delta - \bar \delta$ up to more than $3 \sigma$ for all the results quoted in \gtable{tab:dCP} with the only exception of $E_\nu > 6~\mbox{GeV}$ for IH where it holds up only to around $2.5 \sigma$. 

There is some slight $\bar x_{\rm a}$-dependence of the uncertainty 
$\Delta(\delta)$, 
which can then be read off directly by expressing the covariance matrix in terms
of the decomposition coefficients, 
$N^{(2)}_\alpha + \bar x_{\rm a} N^{(4)}_\alpha$ and $N^{(3)}_\alpha$, 
for an input value of $x_{\rm a} = \bar x_{\rm a}$ as follows:
\begin{subequations}
\begin{eqnarray}
  V^{-1}
& = &
  V^{-1}_\mu 
+ V^{-1}_{\rm e} \,,
\\
  V^{-1}_\alpha
& = &
  \int \frac {d E_\nu d \cos \theta_{\rm z}}
						 {N^{(0)}_\alpha + N^{(1)}_\alpha \bar x_{\rm a} + N^{(5)}_\alpha \bar x^2_{\rm a}} (1 - \bar x^2_{\rm a})
\left\lgroup
\begin{matrix}
                 \left[ N^{(2)}_\alpha + N^{(4)}_\alpha \bar x_{\rm a} \right]^2
& N^{(3)}_\alpha \left[ N^{(2)}_\alpha + N^{(4)}_\alpha \bar x_{\rm a} \right]   \\
  N^{(3)}_\alpha \left[ N^{(2)}_\alpha + N^{(4)}_\alpha \bar x_{\rm a} \right]   
&                \left[ N^{(3)}_\alpha \right]^2
\end{matrix}
\right\rgroup \,,
\label{eq:Valpha}
\end{eqnarray}
\label{eq:vinv}
\end{subequations}
\hspace{-2mm}
where we neglect the small $\delta$-dependent terms in the denominator of (\ref{eq:Valpha}). Since $N^{(2)}_\mu$ and $N^{(4)}_\mu$
share the same sign in the considered energy range of $E_\nu > 6 \, \mbox{GeV}$,
especially around $\cos \theta_{\rm z} \approx -0.8 \sim -0.6$, as shown in
\gfig{fig:Ni2N}, \gfig{fig:Ni2I}, and \gfig{fig:N45}, the combination 
$N^{(2)}_\mu + N^{(4)}_\mu \bar x_{\rm a}$ has larger magnitude when 
$\bar x_{\rm a}$ is positive. This explains the reduced uncertainty at 
$\bar x_{\rm a} = + 0.2$.
 Note that there is an overall factor $1 - \bar x^2_{\rm a}$ due to
the modulation $\cos \delta' = \sqrt{1 - x^2_{\rm a}} \cos \delta$ and
$\sin \delta' = \sqrt{1 - x^2_{\rm a}} \sin \delta$. It can increase the CP
phase uncertainty $\Delta(\delta)$ if the atmospheric mixing angle is not
maximal, namely $1 - x^2_{\rm a} < 1$.

The $\bar \delta$ dependence of $\Delta(\delta)$ can be expressed as,
\begin{equation}
  \Delta(\delta)
=
\left[
  (V^{-1})_{11} \sin^2 \bar \delta 
- 2 (V^{-1})_{12} \sin \bar \delta \cos \bar \delta
+ (V^{-1})_{22} \cos^2 \bar \delta
\right]^{-1/2} \,.
\end{equation}
If the elements $(V^{-1})_{11}$, $(V^{-1})_{12}$, and $(V^{-1})_{22}$
have same size, the uncertainty on the CP phase, $\Delta(\delta)$, becomes 
$\bar \delta$-independent. Otherwise, $\Delta(\delta)$ would receive some
variation. We can treat $\cos \delta$ and $\sin \delta$ as
independent functions with nominal uncertainties 
$\Delta(\cos \delta) = \sqrt{V_{11}}$, $\Delta(\sin \delta) = \sqrt{V_{22}}$, 
and $V_{12} = \sqrt{V_{11}} \sqrt{V_{22}} \rho$. The above approximation for 
the covariance matrices gives, for $\bar x_{\rm a} = 0$, 
\begin{subequations}
\begin{eqnarray}
  \Delta(\cos \delta) = 0.65 \,,
\qquad
  \Delta(\sin \delta) = 0.97 \,,
\qquad 
  \rho = -0.29 
& \mbox{for} &
  \mbox{NH ($\mu$-like)} \,,
\\
  \Delta(\cos \delta) = 0.38 \,,
\qquad
  \Delta(\sin \delta) = 0.40 \,,
\qquad 
  \rho = -0.21 
& \mbox{for} &
  \mbox{NH (e-like)} \,,
\\
  \Delta(\cos \delta) = 0.33 \,,
\qquad
  \Delta(\sin \delta) = 0.37 \,,
\qquad 
  \rho = -0.23
& \mbox{for} &
  \mbox{NH ($\mu$+e-like)} \,,
\end{eqnarray}
\label{eq:VNH}
\end{subequations}
\hspace{-2mm} and, 
\begin{subequations}
\begin{eqnarray}
  \Delta(\cos \delta) = 1.43 \,,
\qquad
  \Delta(\sin \delta) = 1.79 \,,
\qquad 
  \rho = -0.27 
& \mbox{for} &
  \mbox{IH ($\mu$-like)} \,,
\\
  \Delta(\cos \delta) = 0.78 \,,
\qquad
  \Delta(\sin \delta) = 0.87 \,,
\qquad 
  \rho = -0.23
& \mbox{for} &
  \mbox{IH (e-like)} \,,
\\
  \Delta(\cos \delta) = 0.69 \,,
\qquad
  \Delta(\sin \delta) = 0.78 \,,
\qquad 
  \rho = -0.24
& \mbox{for} &
  \mbox{IH ($\mu$+e-like)} \,,
\end{eqnarray}
\label{eq:VIH}
\end{subequations}
for the integration region of $6~\mbox{GeV} < E_\nu < 20~\mbox{GeV}$ and $\cos \theta_{\rm z} < -0.4$ with 10 years running of PINGU.

From the results of (\ref{eq:VNH}) and (\ref{eq:VIH}), we find that the uncertainties $\Delta(\sin \delta)$ and $\Delta(\cos \delta)$ are rather large when only the muon-like events of $E_\nu > 6~\mbox{GeV}$ are used in the analysis. This is especially the case of IH, for which the uncertainties are larger than unity even after 10-years of running. In other words, the uncertainty $\Delta(\delta)$ can be rather broad. Therefore, the electron-like events are essential to measure the CP phase. 

Generally speaking, $\Delta(\cos \delta)$ is slightly smaller than $\Delta(\sin \delta)$ due to larger magnitude of $N^{(2)}_\alpha$, resulting in slightly smaller uncertainties $\Delta(\delta)$ at $\bar \delta = \pm 90^\circ$ than those at $\bar \delta = 0^\circ, 180^\circ$. In addition, the correlation turns out to be small $|\rho| \sim 0.2$ and negative for all the cases. From the combined results in (\ref{eq:VNH}) and (\ref{eq:VIH}), we can estimate the smallest and the largest uncertainty of $\Delta(\delta)$ as a function of $\bar \delta$,
\begin{subequations}
\begin{eqnarray}
  \Delta(\delta)_{\rm max} = 23^\circ \quad @ \quad \bar \delta = \phantom{1} 32^\circ, 212^\circ
& \mbox{for} & 
  \mbox{NH ($\mu$+e-like)} \,,
\\
  \Delta(\delta)_{\rm min} = 17^\circ \quad @ \quad \bar \delta = 122^\circ, 302^\circ
& \mbox{for} & 
  \mbox{NH ($\mu$+e-like)} \,,
\end{eqnarray}
\end{subequations}
for NH, and
\begin{subequations}
\begin{eqnarray}
  \Delta(\delta)_{\rm max} = 48^\circ \quad @ \quad \bar \delta = \phantom{1} 32^\circ, 212^\circ
& \mbox{for} & 
  \mbox{IH ($\mu$+e-like)} \,,
\\
  \Delta(\delta)_{\rm min} = 36^\circ \quad @ \quad \bar \delta = 122^\circ, 302^\circ
& \mbox{for} & 
  \mbox{IH ($\mu$+e-like)} \,,
\end{eqnarray}
\end{subequations}
for IH. All the results quoted in \gtable{tab:dCP} lie within the above range.
The largest differences are found to be about a few degrees for the IH case with $E_\nu > 6~\mbox{GeV}$.

\section{Conclusions and Outlook}
\label{sec:conclusions}

In this work, we develop a general decomposition formalism in the propagation
basis, which is extremely useful for the phenomenological study of neutrino 
oscillation. It can analytically separate the contributions of the three
unknown parameters, namely, the neutrino mass hierarchy, the atmospheric mixing
angle, and the CP phase. In this way, the pattern behind $\chi^2$ minimization
can be revealed clearly, especially for atmospheric neutrino which experiences
very complicated Earth matter profile. Hence, it can serve as a complementary
tool to the neutrino oscillogram. The latter is designed for the overall pattern,
especially the resonance behaviors, in the atmospheric neutrino oscillations, 
while our decomposition method can unveil more hidden structures behind the 
oscillogram. In addition, the decomposition method can apply generally to any 
type of neutrino oscillation experiment.

To illustrate the powerfulness of this decomposition formalism, we study in detail the ability of PINGU in determining the neutrino mass hierarchy, the atmospheric angle $\theta_{\rm a}$ and its octant, as well as the CP phase $\delta$ by measuring the oscillation pattern of atmospheric neutrinos. Both muon- and electron-like events have been considered. Our results suggest that PINGU has the potential to determine the mass hierarchy and the octant of the atmospheric mixing angle $\theta_{\rm a}$ within one year of operation if the neutrino energy and the zenith angle can be measured accurately in the region $E_\nu = 6 \sim 20~\mbox{GeV}$ and $\cos \theta_{\rm z} < -0.4$. The uncertainty of measuring the value of $\theta_{\rm a}$ can be reduced by a factor of $5 \sim 30$, while the determination of the CP phase $\delta$ is significantly more challenging. The dependence on the input values of the neutrino mass hierarchy, the atmospheric mixing angle, and the CP phase can be fully understood in our decomposition formalism. Our findings merit a serious investigation of the physics potential of PINGU with realistic detector response which we expect to be underway within the IceCube/PINGU Collaboration.

\section{Acknowledgements}

We thank Naotoshi Okamura for stimulating discussions in the early stage of our investigation. SFG is grateful to Hong-Jian He for kind support and Center for High Energy Physics of Tsinghua University (TUHEP), where part of this work is done, for hospitality. JSPS has provided SFG a generous postdoc fellowship, which is deeply appreciated, to continue the study at KEK. This work is supported in part by Grant-in-Aid for Scientific research (No. 25400287) from JSPS. CR is grateful for support from the Center for Cosmology and AstroParticle Physics (CCAPP) at The Ohio State University, where part of this work was performed. This work is supported in part by the Basic Science Research Program through the National Research Foundation of Korea funded by the Ministry of Education, Science and Technology (2013R1A1A1007068).

\begin{appendix}

\section{Atmospheric Neutrino Oscillation}
\label{sec:setup}
In this appendix, we introduce the input and method that we used to evaluate the
atmospheric neutrino oscillation in this study. They include 
the atmospheric neutrino fluxes, 
interaction cross sections, and an energy-dependent effective detector volume, 
the Earth matter density profile and the numerical procedure.

\subsection{Atmospheric Neutrino Flux, Cross Sections and Effective Volume}
\label{sec:FXV}
\label{sec:fiducial}

The atmospheric neutrino flux depends on many factors. First, it varies with 
the geographic location, mainly due to the earth 
magnetic field at the source regions. In addition, it dependents on the 
neutrino momentum direction, the zenith and the azimuth angles. Seasonal 
effects can also modulate the neutrino flux. For our study, we use an 
annual and azimuth angle averaged neutrino flux computed for the South Pole
\cite{Athar:2012it}. The earth magnetic field effect, which introduces the
largest modification on neutrino fluxes, depending on the position, are
mostly relevant for neutrino energies below the one considered in this
study; therefore our results can be easily transferred to detectors at
other geographic locations. With these factors taken into consideration, 
the neutrino flux is a function of neutrino energy and the zenith angle.

\begin{figure}[h]
\centering
\includegraphics[height=\textwidth,width=7cm,angle=-90]{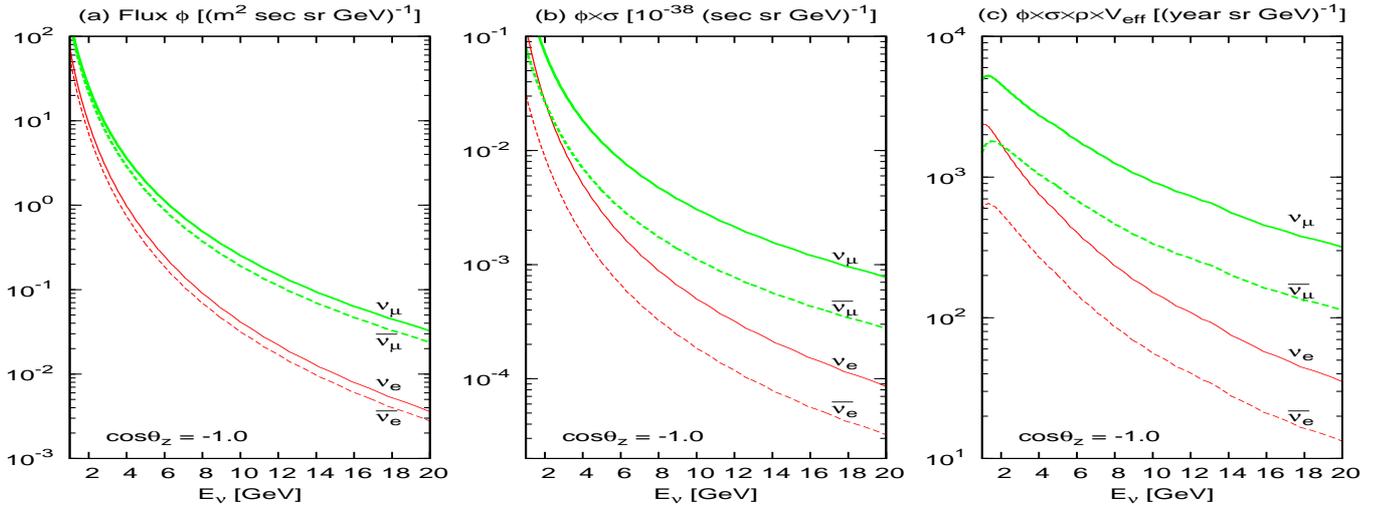}
\caption{Energy dependence of (a) atmospheric neutrino fluxes $\phi$,
                              (b) products of flux and cross section $\phi \times \sigma$,
                              (c) products of flux, cross section and the effective fiducial volume $\phi \times \sigma \times V_{\rm{{\rm eff}}}$, all at $\cos \theta_{\rm z} = - 1.0$.}
\label{fig:flux}
\label{fig:fx}
\label{fig:fxV}
\end{figure}

The energy dependence of the atmospheric neutrino flux is shown in 
\gfig{fig:flux}(a). It can be seen that $\nu_\mu$ and $\bar \nu_\mu$ fluxes 
dominate over the $\nu_{\rm e}$ and $\bar \nu_{\rm e}$ fluxes. 
They drop very quickly with increasing energy, decreasing by four orders of 
magnitude from $E_\nu = 1$~GeV to $E_\nu = 20$~GeV. For both flavors, the 
antineutrino flux is slightly smaller than the neutrino flux due to an 
asymmetry between the $\pi^{+}$ and $\pi^{-}$ production spectra in cosmic ray 
air-showers~\cite{Athar:2012it}. 

The atmospheric neutrino flux will be modulated by the neutrino interaction cross section with nuclei. In this study, we use the charged current (CC) cross sections generated by NEUGEN3 ~\cite{Gallagher:2002sf}. Since the cross sections increase linearly as function of neutrino energy $E_\nu$, higher energy event rates can be enhanced, by roughly an order of magnitude. In addition, the neutrino cross section is a factor of $2-3$ larger than that of the antineutrino. The difference between the neutrino and antineutrino rates becomes even more significant when the atmospheric neutrino flux in \gfig{fig:fx}(a) is multiplied with cross sections, as shown in \gfig{fig:fx}(b). This difference between the neutrino and antineutrino event rates is critical for the mass hierarchy determination with a detector incapable of telling leptons ($\mu$ and e) from antileptons ($\bar \mu$ and $\bar e$), as will be made explicit in \gsec{sec:MSW}. 

The PINGU effective fiducial volume used for our study assume a geometry of 20 
additional strings within a radius of 75~m inside the DeepCore volume~\cite{Veff} 
with an inter-string spacing of approximately 26~m. It was required that at 
least 20 optical sensors would register a Cherenkov photon. At this level, 
events are assumed to be reconstructable. At the relevant energies in this study, 
the effective fiducial volume for electron and muon neutrinos is approximately 
equal \cite{Tang11}. Under this assumption, the fiducial volume is universal 
and will not affect the relative amount of electron and muon 
neutrinos/antineutrino event rates. The fiducial volume increases with energy, 
further increasing the event rates at higher energies. However, due to the 
much larger neutrino flux at lower energies, the highest event rates are 
expected to be detected at lower energies as can be seen in \gfig{fig:fxV}(c).

\begin{figure}[h!]
\centering
\includegraphics[height=8cm,width=5cm,angle=-90]{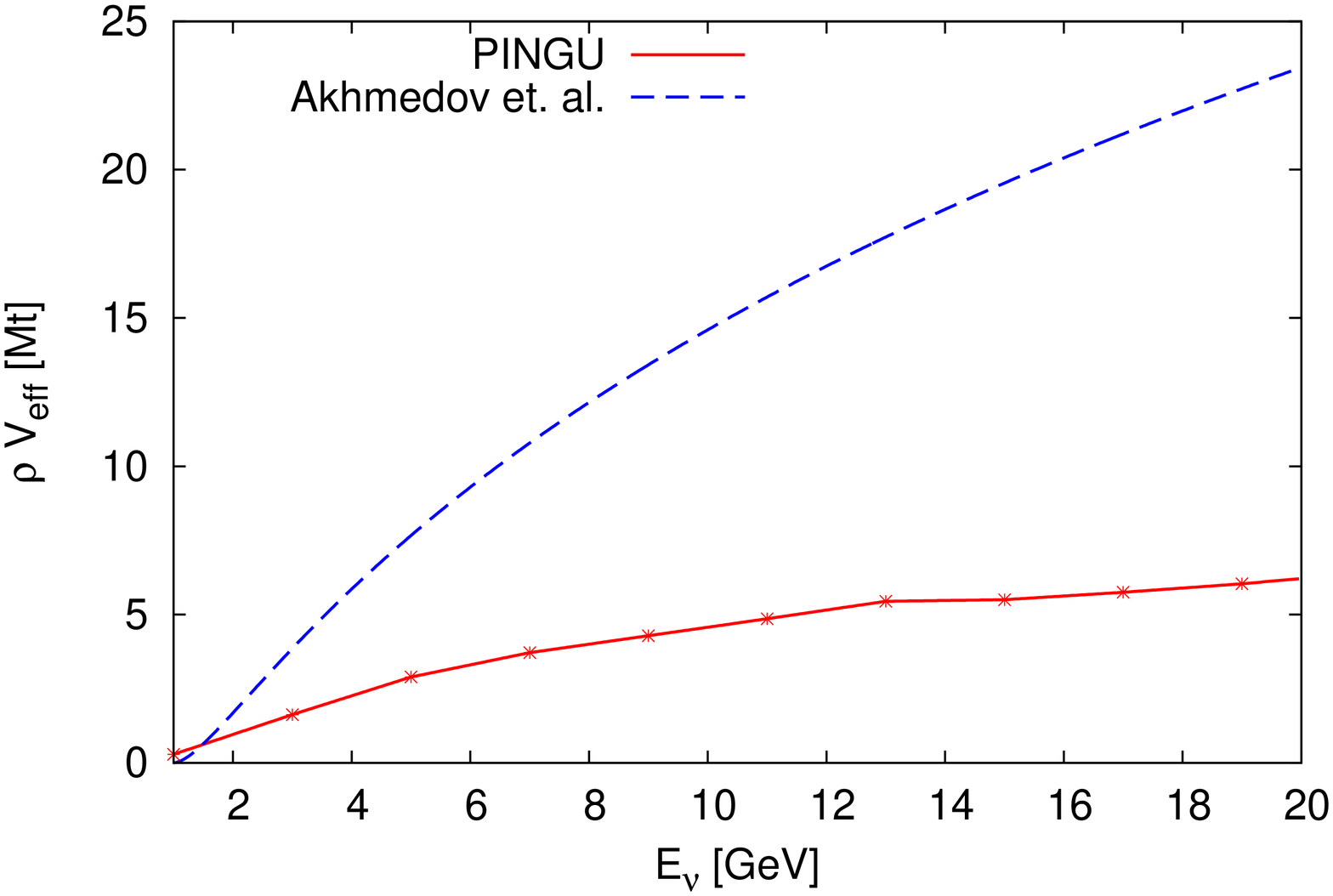}
\includegraphics[height=8cm,width=5cm,angle=-90]{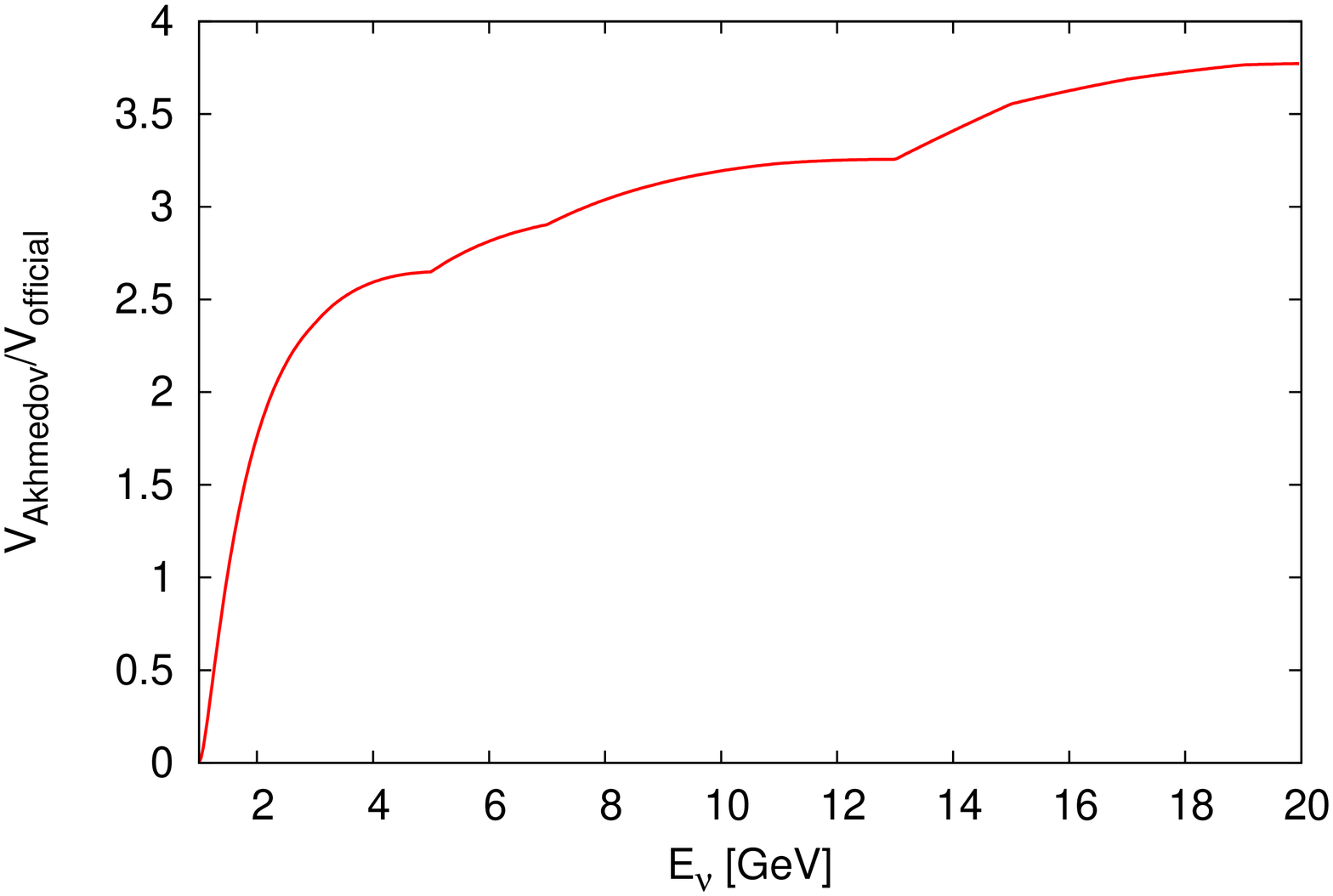}
\caption{Comparison of the effective fiducial volume reported by the PINGU Collaboration~\cite{Veff}, and the one adopted in the work of Akhmedov et. al.~\cite{Akhmedov:2012ah,Ribordy:2013xea}.}
\label{fig:fluxCompare}
\end{figure}

In \gfig{fig:fluxCompare}, we compare the effective fiducial volume reported by the PINGU Collaboration~\cite{Veff}, that we adopt in this study [see \gfig{fig:flux}], and the one adopted in the pioneering work of Akhmedov et. al.~\cite{Akhmedov:2012ah,Ribordy:2013xea}. In the left panel, the absolute values of both PINGU (solid curve) and the Akhmedov et. al.'s paper (dashed curve) are shown while the ratio between the one used in Akhmedov et. al.'s work and the PINGU curve is presented in the right panel. We can see that in most of the regions, the one implemented in the Akhmedov et. al.'s paper is larger than the one from the PINGU Collaboration by a factor of around $3$. It is only in the low energy end that this ratio decreases.

\subsection{Earth matter profile}
\label{sec:PREM}

We use the preliminary Earth reference model (PREM)
\cite{Dziewonski:1981xy} to describe the Earth's matter density distribution. 
It represents the standard framework of interpreting the 
seismological data to determine the Earth matter density, assuming a 
spherically symmetric Earth without describing the chemical composition. 
The Earth radius varies from 6353~km to 6384~km due to rotational flattening, 
which distorts Earth¡¯s shape from that of an ideal sphere. The variation is
 relatively small (0.5\%) and hence it is a reasonable assumption to use a 
spherically symmetric Earth with radius R = 6371~km. As the matter potential
felt by neutrino depends on the electron density, we need to make assumptions 
about the chemical composition.

The matter potential felt by neutrino is proportional to the electron density,
\begin{equation}
  V(x) = \sqrt 2 G_F N_{\rm e}(x) \,,
\label{eq:V}
\end{equation}
while $V(x)$ receives an extra minus sign for antineutrino. The parameter 
$G_F$ is the Fermi constant and $N_{\rm e}(x)$ is the electron number 
density as a function of position, $x \equiv d / L(\theta_{\rm z})$ with $d$ 
being the distance traversed by neutrino and $L(\theta_{\rm z})$ denoting the 
path length corresponding to the neutrino zenith angle $\theta_{\rm z}$.  
The electron number density $N_{\rm e}$ depends on the chemical composition of 
the Earth, which cannot be measured directly. It is approximated as a linear 
function of the matter density $\rho$ as $N_{\rm e} = Y \rho / m_{\rm p}$, 
where $m_{\rm p}$ is the nucleon mass. The coefficient $Y$ is the ratio between 
electron and nucleon number densities, given by 
$n_{\rm p} / (n_{\rm p} + n_{\rm n})$ where $n_{\rm p}$ is the number density 
for proton and $n_{\rm n}$ the number density for neutron. 
Its value can vary significantly from 
light elements $Y\simeq 0.5$ to heavier elements, which increasingly have more 
neutrons per protons (for example $Y=0.466$ for $Fe$). The core is expected to 
be dominated by iron, which takes an $85\%$ share, rendering a smaller 
$Y_{\rm core} = 0.468$. For mantle, $Oxygen$ ($Y = 0.5$), $Magnesium$ 
($Y = 0.494$), and $Silicon$ ($Y = 0.498$) take $44\%$, $23\%$, and $21\%$ of 
the total weight respectively, leading to a larger $Y_{\rm mantle} = 0.497$
\cite{McDonough}. The overall uncertainty in the averaged $Y_{\rm core}$ and 
$Y_{\rm mantle}$ is expected to be small, even if element abundances carry 
uncertainties of about 10\% \cite{McDonough}. This is because the uncertainty 
in the averaged $Y$'s is approximately a product of the 10\% variation in the 
element abundances and the variation in $Y$ for individual element which also has 
around 10\% variation. Hence, the estimated matter potential $V$ would receive 
an uncertainty around 1\%. 
In the current study, we omit this small uncertainty since it would not affect 
the potential of the atmospheric neutrino experiments at a qualitative level.

\begin{figure}[h]
\centering
\includegraphics[height=12cm,width=7cm,angle=-90]{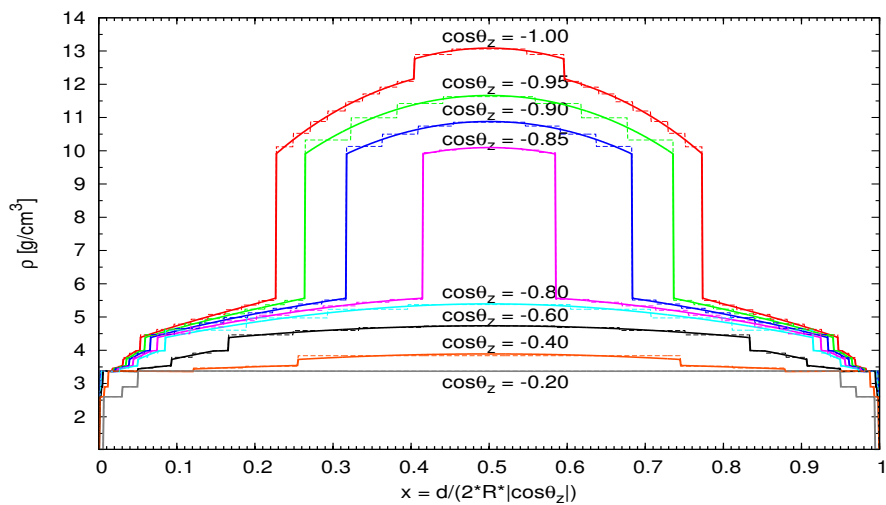}
\caption{Realistic v.s. sectioned PREM matter density profiles along paths with
         solid and dashed lines respectively.
         The Earth's radius is denoted as $R$ while $d$ is the path length of the 
	 neutrino traversing the Earth.
         For the sectioned profile, ten sublayers, from the inside to the outside,
         have been sliced into $n = 2^{2,3,3,0,1,1,0,0,0,0}$ equal steps along the path. See Appendix \ref{sec:numerical} for details.
        }
\label{fig:PREM}
\end{figure}

\gfig{fig:PREM} shows the matter density along different paths, denoted
by the neutrino zenith angle $\theta_{\rm z}$.  For
$\cos \theta_{\rm z} = - 1$, the path runs through the Earth center, 
corresponding to vertically up-going neutrino, with horizontal neutrinos 
denoted by $\cos \theta_{\rm z} = 0$. The path length is related to 
the zenith angle as $L(\theta_{\rm z}) = 2 R |\cos \theta_{\rm z}|$ and is 
symmetric with respect to the central point. For convenience, the horizontal 
axis in \gfig{fig:PREM} is defined as $x \equiv d / L(\theta_{\rm z})$ to make 
comparison between different paths. 
It should be noted that there is a big discontinuity at the boundary between 
the core and mantle. For $\cos \theta_{\rm z} = - 1$, this boundary rests around 
$|x - 0.5| \approx 0.27$ since the core's size ($R_{core} = 3480$~km) is almost 
half of the Earth's radius, $R = 6371~\mbox{km}$. In addition, the core can be 
divided into the inner core and the outer one which are separated by a boundary 
at $r = 1221.5$~km. The inner/outer core and mantle boundaries are very well 
known and have an uncertainty of less than 10~km~\cite{coremantle}. For the 
mantle, there are eight sublayers with boundaries at 
$r = (5701, 5771, 5971, 6151, 6346, 6356, 6368)$~km respectively.  Of these eight 
sublayers, the three outermost layers have constant matter density. Depending on 
the zenith angle $\theta_{\rm z}$, neutrinos pass these various layers 
sequentially.

\subsection{Numerical Method for Oscillation Amplitude Matrix $S'$ in the Propagation Basis}
\label{sec:numerical}

Although the dependences of the oscillation probabilities on the atmospheric 
angle $\theta_{\rm a}$ and the CP phase $\delta$ can be expressed analytically, 
as elaborated in \gsec{sec:propagation-basis}, the other oscillation parameters 
are still entangled with the matter effect inside the oscillation amplitudes 
$S'_{ij}$ in the propagation basis, as explicitly shown in 
(\ref{eq:H'}). Since the matter profile has a very complicated structure, 
depending on the neutrino zenith angle $\theta_{\rm z}$, it is necessary to 
find an accurate and efficient numerical method to evaluate $S'_{\rm ij}$.

\begin{figure}[h!]
\centering
\includegraphics[height=8.8cm,width=5cm,angle=-90]{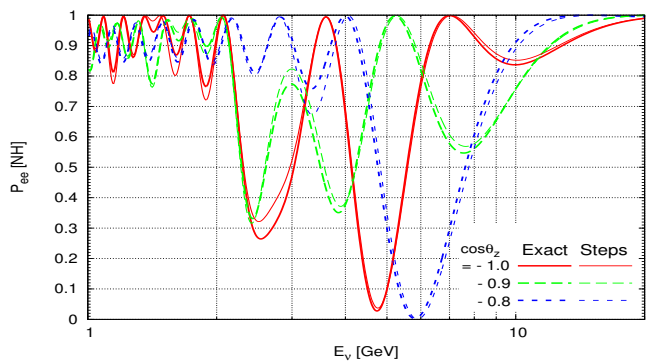}
\caption{The oscillation probabilities $P_{\rm ee}$ obtained with {\it exact solution} [thick lines] v.s. averaged PREM steps for normal hierarchy. The results along three different paths $\cos \theta_{\rm z} = -1.0, -0.9, -0.8$ are shown as solid, dashed and dotted lines respectively.}
\label{fig:Pee:methods}
\end{figure}

As an approximation, we first replace the PREM profile within each 
sublayer, as shown in \gfig{fig:PREM}, by a constant density averaged along the 
path. Within each constant potential, the oscillation amplitude matrix $A_i$ for 
three-neutrino oscillation can be evaluated exactly~\cite{Dziewit:2011pd}. The 
full amplitude matrix is a sequential matrix product of these individual ones.
In this way, we can keep the discontinuity between sublayers, especially the 
periodic mantle-core-mantle structure, which is important for parametric 
resonances. But the slowly varying behavior within each sublayer is averaged out. 
As an illustration, the resultant
oscillation probabilities of the $\nu_{\rm e} \rightarrow \nu_{\rm e}$ channel 
for neutrino zenith angle $\cos \theta_{\rm z} = -1.0, -0.9, -0.8$, respectively, 
are shown in \gfig{fig:Pee:methods} with thin lines, in contrast to the exact 
solutions with thick lines. We can observe that this simple approximation shows 
very good agreements with the exact solution, especially that the peaks and 
troughs appear at almost exactly the same energies. In other words, the resonance
features are maintained. But their amplitude can differ up to 10\% which cannot 
be ignored. A more precise method is needed for a precision analysis such as
$\chi^2$ minimization to obtain the physics potential of atmospheric neutrino
oscillation experiments.

To account for the finer structure, we further divide each sublayer into several 
sections, within each the matter density is approximated by the averaged value 
along the path. Since the matter density has different slopes within different 
PREM layers, the number of sections is chosen accordingly as $n = 2^i$ with 
$i = (2,3,3,0,1,1,0)$ for the density-varying sublayers from the inner core to 
the outer crust. In this way, accuracy and efficiency can be balanced leading to 
an optimized program. In principle, with sublayers divided into more number of 
sections, the result would be closer to the exact solution. This is verified 
for linear potential by comparing with exact solution
\cite{Haxton:1986bc} at various distances. For all the paths 
$(\cos \theta_{\rm z})$ along the PREM matter distribution, we confirm that 
slicing the sublayers into finer sections gives no visible effects on the 
oscillation probabilities as we check by doubling the number of divisions 
with $n = 2^{i+1}$.  Hence, we call the solution with $n = (4, 8, 8, 1, 2, 2, 1)$ 
as the {\it exact solution} in this study. The difference in the probabilities 
is found to be in the order of $10^{-3}$, which can be safely ignored. 
The oscillation probabilities $P_{\rm ee}$ of the exact solution are shown in 
\gfig{fig:Pee:methods} with thick lines.

\subsection{Normal Hierarchy v.s. Inverted Hierarchy}
\label{sec:MSW}

The significant discontinuity in the matter density between the Earth's mantle 
and the core leads to an interesting pattern in the atmospheric neutrino 
oscillation probabilities. Along trajectories with $\cos \theta_{\rm z} < -0.84$,
neutrinos experience a large jump in the potential causing parametric resonance
\cite{parametric} which is also known as oscillation length resonance 
\cite{length}. With periodic matter density profile, the oscillation 
probability is largely enhanced. In addition, there is MSW resonance in the 
wide range of $\cos \theta_{\rm z}$ and $E_\nu$~\cite{MSW} when neutrino 
crosses the mantle region. This makes the 
oscillation pattern of atmospheric neutrinos very sensitive to the neutrino mass 
hierarchy.

\begin{figure}[h]
\centering
\includegraphics[height=8.8cm,width=5cm,angle=-90]{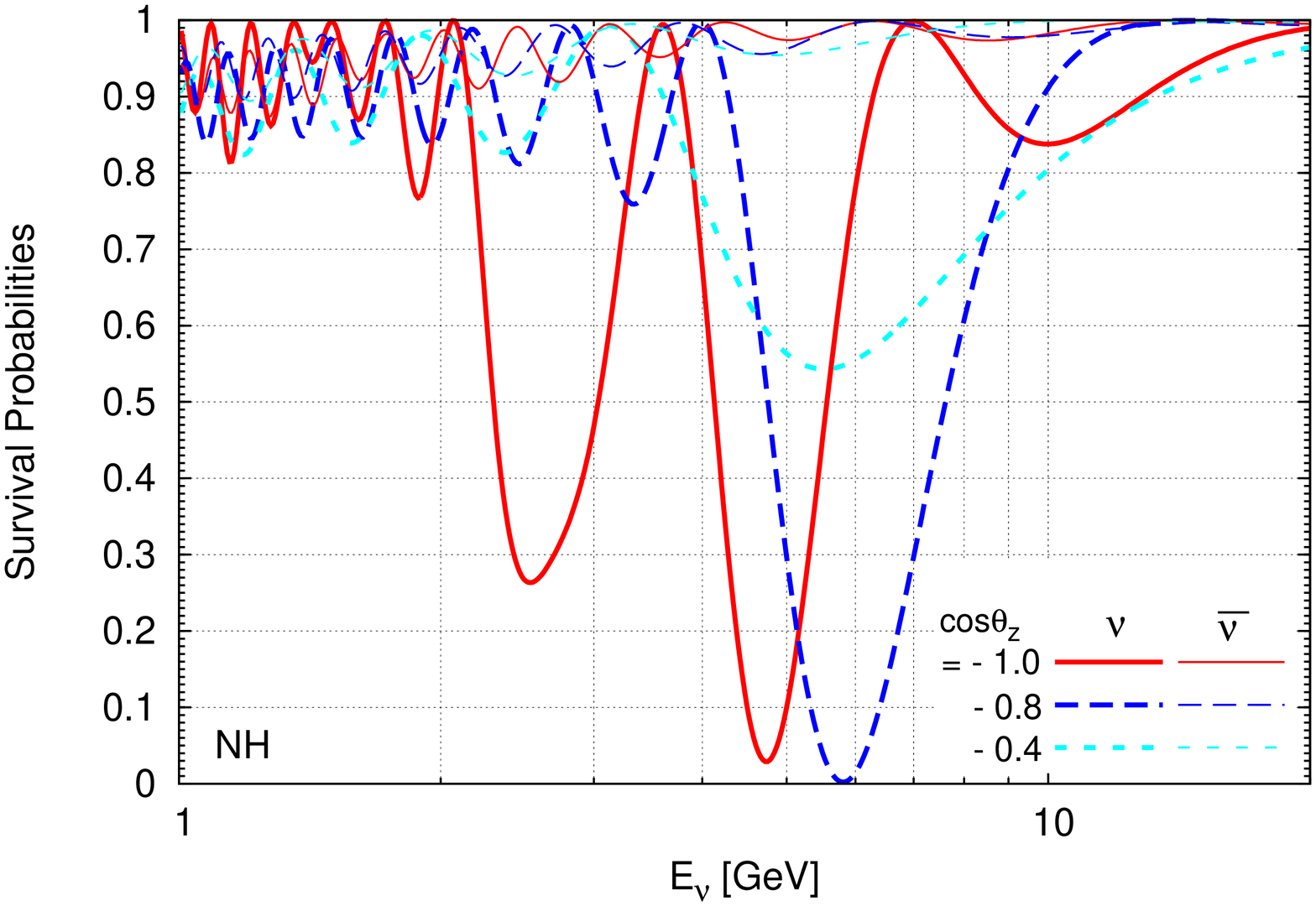}
\includegraphics[height=8.8cm,width=5cm,angle=-90]{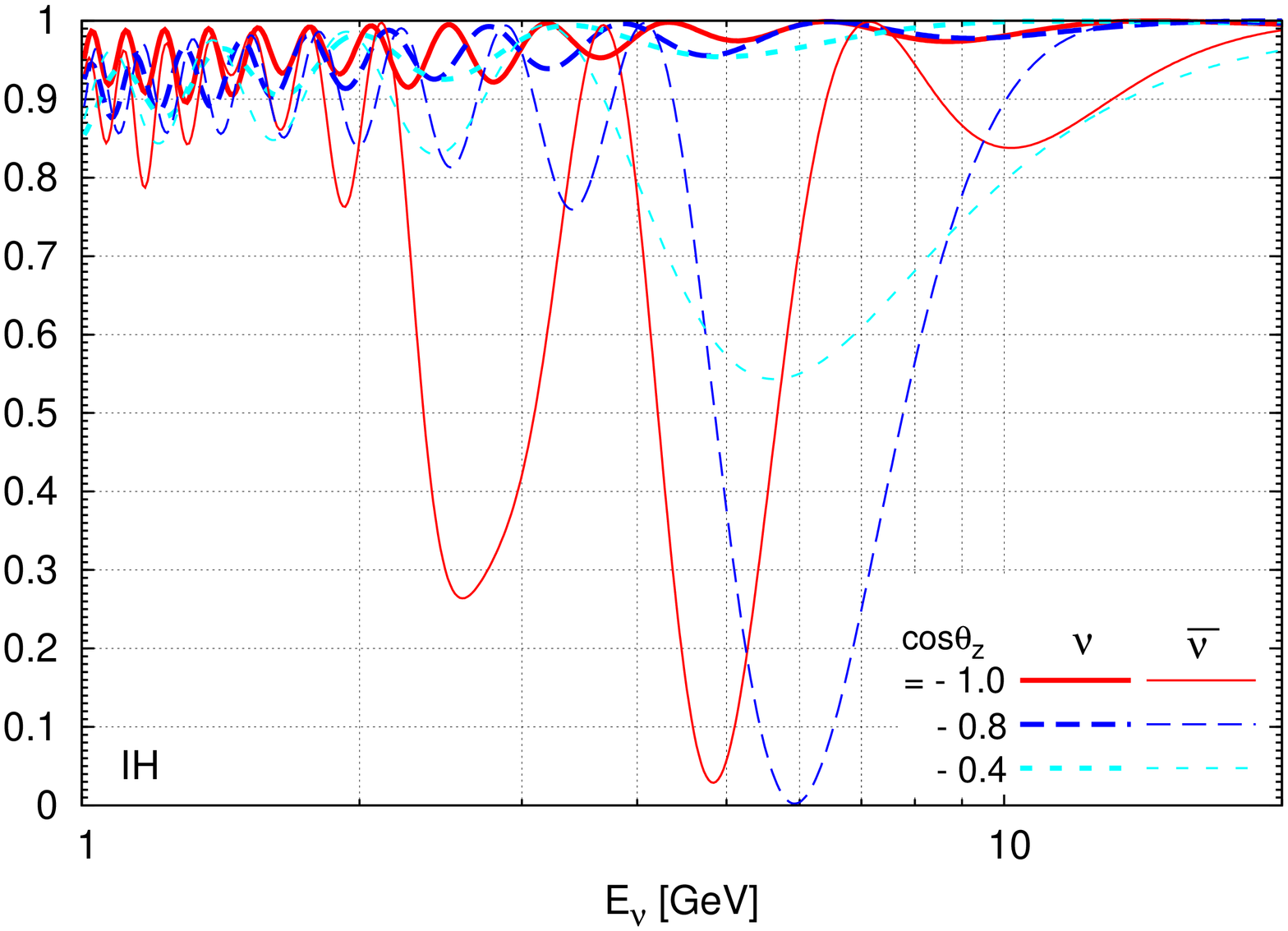}
\caption{Neutrino oscillation probabilities $P_{\rm ee}$ (thick) and 
         $P_{\bar e \bar e}$ (thin)
         v.s. $E_\nu$ for NH (left) and 
         IH (right), along the trajectories $\cos \theta_{\rm z} = -1$, $-0.8$ and $-0.4$.}
\label{fig:Pee:MH}
\end{figure}
We briefly discuss the difference between NH and IH here. In \gfig{fig:Pee:MH}, 
the electron neutrino (antineutrino) survival probabilities, 
$P_{\rm ee}$ ($P_{\bar e \bar e}$), along the paths with neutrino zenith angle
$\cos \theta_{\rm z} = -1$, $-0.8$ and $-0.4$ are plotted for both NH in the 
left and IH in the right panel. For NH, strong oscillation patterns are found 
in the oscillation probability $P_{\rm ee}$ (thick curves) of neutrino, while 
the oscillation probability $P_{\bar e \bar e}$ (thin curves) of antineutrino
has much smaller variation. This significant difference is because of the MSW 
resonance effect which can be demonstrated in the much simpler two-neutrino 
oscillation case. Under this simplified circumstance, the effective mixing angle 
can be analytically expressed as,
$
  \sin 2 \widetilde \theta
=
        {\sin 2 \theta} /
        {\sqrt{\sin^2 2 \theta + \left(\cos 2 \theta - {2 E V}/{\delta m^2} \right)^2}}
$,
where $\theta$ and $\delta m^2$ are the true mixing angle and the true mass
squared difference. The resonance happens at $\cos 2 \theta = 2 E V / \delta m^2$,
leading to a maximal effective mixing angle $\sin 2 \tilde \theta = 1$. With a 
full treatment of three-neutrino oscillation, the probability resonates around 
$E = \cos 2 \theta_{\rm ij} \delta m^2_{\rm ij} / 2 V$, where $\theta_{\rm ij}$
and $\delta m^2_{\rm ij}$ are the relevant true mixing angle and the true mass 
squared difference. For the solar mass squared difference 
$\delta m^2_{12} \equiv m^2_2 - m^2_1$, the resonance energy is around $100$~MeV,
with the typical matter potential in the mantle region, 
which is below the energy region considered in our studies. For the atmospheric 
mass squared difference $\delta m^2_{13} \equiv m^2_3 - m^2_1$, instead, the 
resonance occurs around $4$--$6$~GeV, which is within the accessible region of 
the investigated atmospheric neutrino oscillation in this study. It should be 
noted that the MSW resonance only occurs with $\delta m^2_{13} > 0$ (NH) for 
neutrinos ($V > 0$), and with $\delta m^2_{13} < 0$ (IH) for antineutrinos 
($V < 0$), since $\cos 2 \theta_{13} = 1 - 2 \sin^2 \theta_{13} > 0$. 

It is much simpler to determine the neutrino mass hierarchy if the detector is 
capable of distinguishing neutrinos from antineutrinos. With the MSW resonance 
around $2 \sim 7~\mbox{GeV}$ observed in neutrinos rather than antineutrinos, 
then the mass hierarchy must be normal and vice versa. In other words, the 
existence or the absence of the MSW resonance can serve as a solid discriminator 
of the neutrino mass hierarchy. For a detector without the capability to 
distinguish neutrinos from antineutrinos, the MSW resonance could 
still be used to determine the mass hierarchy. This is made possible by the 
differences in the neutrino and antineutrino fluxes as well as their charged 
current cross sections discussed in \gsec{sec:FXV}. The hierarchy sensitivity 
obtained from the residual difference between NH and IH can still be sizable,
if large enough event rates are collected by a huge underground detector 
such as PINGU.

\end{appendix}

%\newpage
%
%
%\begin{figure}[h]
%\centering
%\includegraphics[height=8cm,width=10cm,angle=-90]{Sij_plot_NH.eps}
%\qquad
%\includegraphics[height=8cm,width=10cm,angle=-90]{Sij_plot_IH.eps}
%\end{figure}
%
%\begin{figure}[h]
%\centering
%\includegraphics[height=14cm,angle=-90]{3nu_plots_t12.eps}
%\includegraphics[height=14cm,angle=-90]{3nu_plots_{\rm s}12.eps}
%\caption{Energy dependence of $\widetilde \theta_{\rm s}$ and $\sin^2 (2 \widetilde \theta_{\rm s})$ in the mantle.}
%\end{figure}
%
%
%\begin{figure}[h]
%\centering
%\includegraphics[height=16cm,angle=-90]{chi_min_error_xa_13.eps}
%\caption{$\chi^2_{\rm{min}}(x_{\rm a})$ obtained from both electron- and muon-like events for NH.}
%\end{figure}
%
%\begin{figure}[h]
%\centering
%\includegraphics[height=16cm,angle=-90]{chi_min_error_dCP_13_6GeV_4y.eps}
%\caption{$\chi^2_{\rm{min}}(\delta)$ obtained from both electron- and muon-like events for NH with 4 years running and energy cut $6 \mbox{GeV} < E_\nu < 20 \mbox{GeV}$.
%         No external constraint on $x_{\rm a}$ is included.}
%\end{figure}

%\newpage
%\begin{figure}[h]
%\centering
%\includegraphics[height=0.9\textwidth,angle=-90]{chi_min_error_xa.eps}
%\\[2cm]
%\includegraphics[height=0.9\textwidth,angle=-90]{chi_min_error_dCP.eps}
%\end{figure}

\end{document}